\documentclass[aps,superscriptaddress,tightenlines,nofootinbib]{revtex4}
\usepackage{amssymb,amsbsy,bm,amsmath,rotating,psfrag}
\usepackage{epsfig}
\usepackage{graphicx}

\newcommand{\be}{\begin{equation}}
\newcommand{\ee}{\end{equation}}
\newcommand{\bea}{\begin{eqnarray}}
\newcommand{\eea}{\end{eqnarray}}
\begin{document}
\title{{\bf Phase transitions in the early and   present Universe.}}
\author{{\bf D. Boyanovsky}}
\email{boyan@pitt.edu} \affiliation{Department of Physics and
Astronomy, University of Pittsburgh, Pittsburgh, Pennsylvania 15260,
USA} \affiliation{Observatoire de Paris, LERMA. Laboratoire
Associ\'e au CNRS UMR 8112.
 \\61, Avenue de l'Observatoire, 75014 Paris, France.}
\affiliation{LPTHE, Universit\'e Pierre et Marie Curie (Paris VI) et
Denis Diderot (Paris VII), Laboratoire Associ\'e au CNRS UMR 7589,
Tour 24, 5\`eme. \'etage, 4, Place Jussieu, 75252 Paris, Cedex 05,
France}
\author{{\bf H. J. de Vega}}
\email{devega@lpthe.jussieu.fr} \affiliation{LPTHE, Universit\'e
Pierre et Marie Curie (Paris VI) et Denis Diderot (Paris VII),
Laboratoire Associ\'e au CNRS UMR 7589, Tour 24, 5\`eme. \'etage, 4,
Place Jussieu, 75252 Paris, Cedex 05,
France}\affiliation{Observatoire de Paris, LERMA. Laboratoire
Associ\'e au CNRS UMR 8112.
 \\61, Avenue de l'Observatoire, 75014 Paris, France.}
\affiliation{Department of Physics and Astronomy, University of
Pittsburgh, Pittsburgh, Pennsylvania 15260, USA}
\author{{\bf D. J.
Schwarz}} \email{dschwarz@physik.uni-bielefeld.de}
\affiliation{Fakult\"at f\"ur Physik, Universit\"at Bielefeld,\\
Postfach 100131, D-33501 Bielefeld, Germany}
\date{\today}

\begin{abstract}
The evolution of the Universe is the ultimate laboratory to study
fundamental physics across energy scales that  span about   25
orders of magnitude: from the grand unification scale  through
particle and nuclear physics scales down to the scale of atomic
physics.  The standard models of cosmology and particle physics
provide the basic understanding of the early and present Universe
and predict a series of phase transitions that occurred in
succession during the expansion and cooling history of the Universe.
We survey these phase transitions, highlighting the equilibrium and
non-equilibrium effects as well as their observational and
cosmological consequences. We discuss the current theoretical and
experimental programs to study phase transitions in QCD and nuclear
matter in accelerators along with the new results on novel states of
matter as well as on multifragmentation in nuclear matter. A
critical assessment of similarities and differences between the
conditions in the early universe and those in ultrarelativistic
heavy ion collisions is presented. Cosmological observations and
accelerator experiments are converging towards an unprecedented
understanding of the early and present Universe.
\end{abstract}


\maketitle

\tableofcontents

\section{Introduction}\label{sec:intro}

The current knowledge of the early and present Universe is
summarized by the standard models of cosmology and of particle
physics. Their symbiosis provides an unprecedented understanding of
the evolution of the Universe solidly based on  a wealth of
observations and experiments. Particle and nuclear physics in a
field theory context provide the fundamental building blocks which
when combined with general relativity and statistical mechanics
yield a description from the early inflationary stage   through a
detailed thermal history of the Universe, to the formation of large
scale structure, galaxies and stars. During the last decade a large
body of observational data has provided a strong evidence in support
of theoretical ideas of an early stage of inflation, during which
the visible size of the Universe grows exponentially. After this
brief, but explosive period of inflation followed by deccelerated
expansion and cooling, the Universe succesively visits the different
energy scales at which particle and nuclear physics predict symmetry
breaking phase transitions. The goals of this article are to review
the equilibrium and non-equilibrium aspects of these phase
transitions, their cosmological imprints and the current theoretical
and experimental efforts to study them with accelerator experiments.
The article begins with an account of both standard models, setting
the stage for a discussion of phase transitions in particle and
nuclear physics. A brief self-contained excursion of cosmology
beginning from inflation follows the early history of the Universe,
visiting the time-marks at which particle physics predicts phase
transitions and exploring their potential consequences. The
\emph{last} phase transition(s) of the standard model of particle
physics took  place when the Universe was about $ 10^{-5} $ secs.
old, they are the deconfinement-confinement and chiral phase
transitions predicted by Quantum Chromodynamics (QCD). The article
presents a summary of the theoretical efforts and the experimental
programs in several accelerator facilities to study the QCD phase
transitions as well as static and \emph{dynamical} aspects of phase
transitions in cold nuclear matter.

\section{The standard models:}


The modern understanding of the early and present Universe hinges
upon \emph{two} standard models: the standard   model of cosmology
and the standard model of particle
physics\cite{bookkolb}-\cite{kamion}. Both have passed stringent
observational and experimental tests. A wealth of cosmological
data is providing confirmation of theoretical ideas in early
Universe cosmology. Measurements of the temperature anisotropy of
the Cosmic Microwave Background  radiation   (CMB) by satellite,
balloon borne and earth based observations, large scale structure
surveys, Lyman $\alpha$ forest, cluster abundance, weak lensing
and measurements of the cosmological expansion and its
acceleration by Type Ia supernovae searches, combined with high
precision measurements of light element abundances provide an
impressive body of high quality data that yield an unprecedented
understanding of cosmology.

\subsection{Observational ingredients} The main observational
pillars that support  the standard   model of cosmology
are\cite{bookkolb}-\cite{kamion}:
\begin{itemize} \item{Homogeneity and isotropy: on scales larger than
$\geq 100\,\textrm{Mpc}$ the Universe looks homogeneous and
isotropic. This is confirmed by large scale surveys and by the
almost isotropy of the CMB.}

\item{The Hubble expansion: objects that are separated by a comoving distance
$d$  recede from each other with a velocity $ v=H\,d $, with $ H $
the Hubble parameter, whose value today is $ H_0 \approx
72\,\textrm{km}/s/\textrm{Mpc} $. The Hubble expansion law
determines the size of our causal horizon: objects separated by a
comoving distance \be d_H = 3000\,h^{-1}\,\textrm{Mpc} \quad , \quad
h = H_0/100\,\textrm{km}/s/\textrm{Mpc} \ee recede from each other
at the speed of light and are therefore causally disconnected.}

\item{The Cosmic Microwave Background   radiation (CMB): a bath of
thermal photons with an \emph{almost} perfect Planck distribution
at a temperature $ T_0= 2.725 \pm 0.001 \,K$. Temperature
anisotropies $ \Delta T/T_0 \sim 10^{-5} $ were first measured in
1992 by the COBE satellite\cite{COBE}, their detection represents
a triumph for cosmology. This small temperature anisotropy, whose
existence is predicted by cosmological models, provide the clue to
the origin of structure. It is an important confirmation of
theories of the early Universe. }

\item{The abundance of light elements: observations of the
abundance of elements in low metallicity regions reveals that
about $76\%$ of ordinary matter is in the form of hydrogen, about $24\%$
(by mass) in ${}^4He$ and trace abundances of ${}^3He \, (\sim
10^{-5})$,  deuterium ($\sim 10^{-5}$) and ${}^7Li\,(\sim
10^{-10}$), all relative to hydrogen\cite{steigman,turnerBBN}.
These elements were formed during the first three minutes of the Universe,
while heavier elements (metals) are produced in the
interior of stars and in astrophysical processes during supernovae
explosions. }

\item{The concordance model: dark matter and dark energy. In the
last few years there has been a wealth of observational evidence
from CMB, large scale structure and high redshift supernovae
Ia\cite{SN1} data that leads to the remarkable conclusions that i)
the spatial geometry of the Universe is flat, ii) the Universe is
accelerating today, and iii) most of the matter is in the form of
dark matter. Current understanding of cosmology is based on the
\emph{concordance} or $\Lambda CDM$ model in which the total energy
density of the Universe has as main ingredients: $5\%$ of baryonic
matter, $25\%$ of \emph{dark matter} and $70 \%$ of \emph{dark
energy}\cite{WMAP1,kogut,spergel,peiris}. The present observations
indicate that the dark energy can be described by a cosmological
constant\cite{negra}.}

\end{itemize}

\subsection{The building blocks}\label{ladrillos}
The main building blocks for a theory of the standard cosmology are:
\begin{itemize}
\item{ {\bf Gravity}: {\em Classical} general relativity provides a good
description of the geometry
of space time for scales $ l \gg l_{Pl} \sim 10^{-33}\mbox{cm}$ or
time scales $ t \gg t_{Pl} \sim 10^{-43}\mbox{s} $, or equivalently
energy scales well below the Planck scale $ M_{Pl} \sim
10^{19}\mbox{Gev} $. A consistent quantum theory of gravity unified
with matter describing the physics at the Planck scale is yet to emerge.

Homogeneity and isotropy for a spatially flat space-time lead to
the Friedmann-Robertson-Walker (FRW) metric
\begin{equation}\label{FRWmetric}
ds^2 = dt^2- a^2(t) \; d\vec{x}^2
\end{equation} where $ t $ is the comoving time (the proper time of a comoving
observer). Physical scales are stretched
by the scale factor $ a(t) $ with respect to the comoving scales
\begin{equation}\label{scale}
l_{phys}(t)= a(t) \; l_{com} \; .
\end{equation}
A physical wavelength redshifts proportional to the scale factor
[eq.(\ref{scale})], therefore its time derivative obeys the Hubble
law $ \dot{l}_{phys}(t) = H(t) \; l_{phys}(t)= l_{phys}(t)/d_H(t)
$. At equilibrium temperature decreases as the universe expands as
\be \label{Ta} T(t) = \frac{T_0}{a(t)} \ee In the   homogeneous
and isotropic FRW universe described by eq.(\ref{FRWmetric}), the
matter distribution must be homogeneous and isotropic, with an
energy momentum tensor with the fluid form \be\label{fluido}
\langle T^{\mu}_{\nu} \rangle={\rm diag}[\rho, -p,-p,-p ] \; , \ee
where $ \rho, \; p $ are the energy density and pressure,
respectively. In such geometry the Einstein equations of general
relativity reduce to   the Friedmann equation, which determines
the evolution of the scale factor from the energy density
\begin{equation}\label{friedeqn}
\left[\frac{\dot{a}(t)}{a(t)} \right]^2\equiv H^2(t) = \frac{
\rho}{3 M^2_{Pl}} \,.
\end{equation}
where $ M_{Pl}= 1/\sqrt{8\pi G} = 2.4\times 10^{18}$ GeV
$ = 0.434\times 10^{-5}$ g.
A spatially flat Universe has the critical density
\begin{equation}\label{rhocrit}
\rho_c = 3 \; M^2_{Pl} \; H^2_0= 1.88 \; h^2 \;
10^{-29}\textrm{g}/\textrm{cm}^3\,.
\end{equation}
where $H_0$ is the Hubble constant today.
The energy momentum tensor conservation reduces to the single
conservation equation,
\begin{equation}\label{conener}
\dot{\rho}+3\left( \rho+p\right) \frac{\dot{a}}{a} =0
\end{equation}
\noindent The two equations (\ref{friedeqn}) and (\ref{conener}) can
be combined to yield the acceleration of the scale factor,
\begin{equation}\label{accel}
\frac{\ddot{a}}{a}= -\frac1{6 \; M^2_{Pl}}(\rho + 3 \; p)
\end{equation}
\noindent which will prove useful later.
In order to provide a close set of equations we must append an
equation of state $ p=p(\rho) $ which is typically written in the form
\begin{equation}\label{eqnofstate}
p=w(\rho) \; \rho
\end{equation}
The following are  important cosmological solutions:
\begin{eqnarray}\label{scalefac}
&&  {\rm Cosmological ~ Constant } \; \Rightarrow
w=-1 : {\rm de~Sitter~ expansion}  \Rightarrow \rho = {\rm
constant}~;~ a(t) \propto e^{Ht} ~;~H= \sqrt{\frac{
\rho}{3M^2_{Pl}}} \cr \cr
&& {\rm Radiation}\; \Rightarrow w=\frac{1}{3} :
{\rm Radiation ~ domination}  \Rightarrow \rho \propto a^{-4}~;~
a(t) \propto t^{\frac{1}{2}} \label{RD} \\
&& {\rm Non-relativistic (cold) ~ Matter} \; \Rightarrow w=0 :
{\rm Matter ~ domination}  \Rightarrow \rho \propto
a^{-3}~;~ a(t) \propto t^{\frac{2}{3}} \nonumber
\end{eqnarray}
Furthermore, we see from eqs.(\ref{accel}) and (\ref{RD}) that
accelerated expansion takes place if $ w < -1/3 $. }

\item{ {\bf The Standard Model of Particle Physics:}\cite{SM} the
current standard model of particle physics, experimentally tested
with remarkable precision describes the theory of strong (QCD),
weak and electromagnetic interactions (EW) as a gauge theory based
on the group $SU(3)_c \otimes SU(2) \otimes U(1)_Y$. The particle
content is: three generations of quarks and leptons:   }
\[
\left( \begin{array}{c} u\\d
\end{array}\right)
\left(\begin{array}{c} c\\s
\end{array}\right)
\left(\begin{array}{c} t\\b
\end{array}\right)\; ; \;
\left(\begin{array}{c} \nu_e\\e
\end{array}\right)
\left(\begin{array}{c} \nu_{\mu}\\\mu
\end{array}\right)
\left(\begin{array}{c} \nu_{\tau}\\\tau
\end{array}\right)
\]
vector Bosons: 8 gluons (massless) , $ Z^0 $, $ W^{\pm} $   with
masses $ M_{Z} = 91.18 \pm 0.02$ GeV and $ M_{W} = 80.4 \pm 0.06$
GeV, the photon (massless) and the scalar Higgs, although the
experimental evidence for the Higgs bosons is still inconclusive.

\end{itemize}

Current theoretical ideas supported by the renormalization group
running of the couplings propose that the strong, weak and
electromagnetic interactions are unified in a grand unified theory
(GUT) at the scale $ M_{GUT} \sim 10^{16}$ Gev. Furthermore
the ultimate scale at which Gravity is eventually unified with the
rest of particle physics is the Planck scale $ M_{Pl} \sim
10^{19}\mbox{Gev} $. Although there are proposals for the total
unification of forces within the context of string theories, their
theoretical understanding as well as any experimental confirmation
is still lacking. However, the physics of the standard model of
the strong and electroweak interactions that describes phenomena
at energy scales below $ \sim 100 $ GeV is on solid experimental
footing.

The connection between the standard model of particle physics and
early Universe cosmology is through Einstein's equations that
couple the space-time geometry to the matter-energy content. As
argued above, gravity can be studied semi-classically at energy
scales well below the Planck scale. The standard model of particle
physics is a {\em quantum field theory}, thus the space-time is
classical  but with sources that are quantum fields. Semiclassical
gravity is defined by the Einstein's equations with the
expectation value of the energy-momentum tensor $ \hat{T}^{\mu
\nu} $ as sources
\begin{equation}\label{einstein}
G^{\mu \nu} = R^{\mu \nu} -\frac{1}{2}\, g^{\mu \nu} R = \frac{
\langle \hat{T}^{\mu \nu} \rangle}{M^2_{Pl}} \; .
\end{equation}
The expectation value of  $ \hat{T}^{\mu \nu} $ is taken in a
given quantum state (or density matrix) compatible with
homogeneity and isotropy which must  be translational   and
rotational invariant. Such state yields  an expectation value of
the energy momentum tensor with the fluid form eq.(\ref{fluido}).

Through this identification the standard model of particle physics
provides the sources for Einstein's equations. All of the elements
are now in place to understand the evolution of the early Universe
from the fundamental standard model. Einstein's equations
determine the evolution of the scale factor, the standard model
provides the energy momentum tensor and statistical mechanics
provides the fundamental framework to describe the thermodynamics
from the microscopic quantum field theory of the strong,
electromagnetic and weak interactions.

\subsection{Energy scales, time scales and phase transitions}\label{2C}

{\bf\underline{Energy Scales}:}

While a detailed description of early Universe cosmology is
available in several books~\cite{bookkolb}-\cite{liddlerev}, a broad- brush picture of
the main cosmological epochs can be obtained by focusing on the
energy scales of particle, nuclear and atomic physics.

\begin{itemize}

\item{{\bf Total Unification:} Gravitational, strong and
electroweak interactions are conjectured to become unified and
described by a single quantum theory at the Planck scale $\sim
10^{19}$ Gev. There are currently many proposals that seek to
provide such fundamental description such as string theories,
however, their theoretical consistency is still being studied and
experimental  confirmation is not yet available.}


\item{{\bf Grand Unification:} Strong and electroweak interactions
(perhaps with supersymmetry) are expected to become unified at an
energy scale $\sim 10^{16}$ Gev corresponding to a temperature $ T
\sim 10^{29}$ K under a larger gauge group $ G $, for example $
SU(5), \; SO(10), \; E_8 $,  which breaks spontaneously  $ G
\rightarrow SU(3)_c\otimes SU(2)\otimes U(1)_Y$ at a scale below
grand unification. There are very compelling theoretical reasons
for the existence of the GUT scale such as the merging of the
running coupling constants of the strong, electromagnetic and weak
interactions, shown in fig. \ref{unif} for the minimal
supersymmetric standard model (MSSM). Yet another reason is the
explanation of the small neutrino masses via the see-saw mechanism
in terms of the ratio between the weak and the grand unification
scale.}

\begin{figure}[h]
\includegraphics[height=3in,width=3in,keepaspectratio=true]{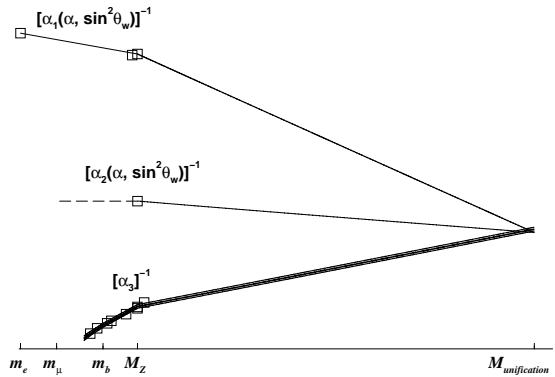}
\caption{Evolution of the weak $ \alpha_1 $, electromagnetic $
  \alpha_2 $ and strong $ \alpha_3 $ couplings with energy in the MSSM
  model. Notice that only  $ \alpha_3 $ decreases with energy
 (asymptotic freedom) extracted from ref.\cite{unifica}} \label{unif}
\end{figure}


\item{{\bf Electroweak :} Weak and electromagnetic interactions
become unified in the electroweak theory based on the gauge group
$SU(2)\otimes U(1)_Y $. The weak interactions become short ranged
after a symmetry breaking phase transition $ SU(2)\otimes U(1)_Y
\rightarrow U(1)_{em}$ at an energy scale of the order of the
masses of the $ Z^0,W^{\pm} $ vector bosons,
corresponding to a temperature $ T_{EW} \sim 100$ GeV $ \sim
10^{15}$ K. At temperatures $ T>T_{EW} $ the symmetry is
restored and  all  vector bosons are (almost) massless (but for
plasma effects that induce screening masses). For $ T<T_{EW} $ the
vector bosons that mediate the weak interactions (neutral and
charged currents) $W^{\pm},Z^0$ acquire masses through the Higgs
mechanism while the photon remains massless, corresponding to the
unbroken $U(1)$ abelian symmetry of the electromagnetic
interactions. Thus $T_{EW}$ determines the temperature scale of
the electroweak phase transition in the early Universe and is the
{\em earliest} phase transition that is predicted by the standard
model of particle physics. The understanding of the symmetry
breaking sector of the standard model is one of the primary goals
of the Large Hadron Collider at CERN. The standard model has the
necessary ingredients to explain the origin of the baryon
asymmetry, leading to the possibility that the asymmetry between
matter and antimatter was produced at the electroweak scale (see
section \ref{sec:EWPT}). }


\item{{\bf QCD:} The strong interactions have a typical energy
scale $ \Lambda_{QCD} \sim \,200$ MeV at which the coupling
constant becomes strong [$ \alpha_s \sim \mathcal{O}(1) $]. This
energy scale corresponds to a temperature scale $T_{QCD} \sim
10^{12}$K. QCD is an asymptotically free theory, the coupling
between quarks and gluons becomes smaller at large energies, but
it diverges at the scale $ \Lambda_{QCD} $. For energy scales
below $ \Lambda_{QCD}$ QCD is a strongly interacting theory and
quarks and gluons are bound into mesons and baryons. This
phenomenon is interpreted in terms of a phase transition at an
energy scale $ \Lambda_{QCD} $ or  $ T_{QCD} $. For $ T>T_{QCD} $
the relevant degrees of freedom are weakly interacting quarks and
gluons, while below are hadrons. This is the quark-hadron or
deconfinement-confinement phase transition. In the limit of
massless up and down quarks, QCD features an $ SU(2)_L\otimes
SU(2)_R $ chiral symmetry, which is spontaneously broken at  about
the same temperature scale as the confinement-deconfiment
transition. Pions are the (quasi) Goldstone bosons emerging from
the breakdown of the chiral symmetry $ SU(2)_L\otimes SU(2)_R
\rightarrow SU(2)_{R+L} $. The QCD phase transition(s) are the
{\em last} phase transition predicted by the standard model of
particle physics. The high temperature phase above $T_{QCD}$, with
almost free quarks and gluons (because the coupling is small by
asymptotic freedom) is a {\em quark-gluon plasma} or QGP.
Experimental programs at CERN (SPS-LHC) and BNL (AGS-RHIC) study
the QCD phase transition via ultrarelativistic heavy ion
collisions (URHIC) and a systematic analysis of the data gathered
at SPS and RHIC during the last decade has given an optimistic
perspective of the existence of the QGP\cite{phystoday,heinz} (see
section 4 below).}


\item{{\bf Nuclear Physics:}~Low energy scales that are relevant
for cosmology are determined by the binding energy of light
elements, in particular deuterium, whose binding energy is $\sim
2$~Mev corresponding to a temperature $T \sim 10^{10}$~K. This is
the energy scale that determines the onset of primordial
nucleosynthesis. The first step in the network of nuclear
reactions that yield the primordial elements is the formation of
the deuteron via $n+p \leftrightarrow d + \gamma$. The large
number of photons per baryon results in that high energy photons
in the blackbody tail dissociate the deuterons formed in the
forward reaction until the temperature becomes of the order of
$T_{NS} \sim 0.1\,\textrm{MeV}$. Once deuterons are formed, a
network of nuclear reactions results in that all neutrons end up
in nuclei, mainly helium, resulting in a helium abundance of about
$25\%$\cite{steigman,turnerBBN}. The nature of the nuclear forces
suggests the possibility of a   liquid-gas phase transition in
cold nuclear matter at an energy $\sim 15-20\,\textrm{MeV}$
discussed in section \ref{nuclearpt}.}


\item{{\bf Atomic Physics:} A further very important low energy
scale relevant for cosmology corresponds to the binding energy of
hydrogen $ \sim 10 ~\textrm{eV} $. This is the energy scale at
which free protons and electrons combine into neutral hydrogen or
`recombination'. The large number of photons per baryon results
in that recombination actually takes place at an energy scale of
order $0.3~\textrm{eV}$, at about $400000$ years after the
beginning of the Universe. At this time when neutral hydrogen is
formed the Universe becomes transparent. This event determines the
last scattering surface, after neutral hydrogen is formed photons
no longer scatter and travel freely. These are the photons
measured by CMB experiments \emph{today}.  }

 \end{itemize}

{\bf\underline{Time Scales}:} An important ingredient of modern
standard cosmology is a brief but explosive early period of
\emph{inflation} during which the scale factor   grows
exponentially as $a(t)=e^{Ht}$ [see eq. (\ref{scalefac})].
WMAP\cite{WMAP1,kogut,spergel,peiris} yields an upper bound on the
energy scale of the inflation [for a detailed discussion see sec.
\ref{sri}] $ H  \lesssim 10^{13}$ GeV. In order to solve the
entropy and horizon problems, the inflationary stage must last a
time interval $ \delta t $ so that $ H \; \delta t \sim 60 $,
hence the inflationary stage lasts a time scale \be \Delta t_{inf}
\sim 10^{-34} \,\textrm{secs}\; . \label{inflatime} \ee Field
models of inflation are discussed in sec. \ref{sri}. The
inflationary stage is followed by a radiation dominated era
(standard hot big bang) after a short period of reheating during
which the energy stored in the   field that drives inflation
decays into   quanta of   many other fields, which through
scattering processes reach a state of local thermodynamic
equilibrium.

Once local thermodynamic equilibrium is reached, a very detailed
picture of the thermal history of the Universe emerges combining
statistical mechanics with the basic ingredients described above
~\cite{bookkolb}-\cite{liddlerev}: during the first $\sim 10000$
years of the Universe and after the inflationary stage that lasted
$ \sim 10^{-34} $secs, the Universe was radiation dominated
expanding and cooling (almost) adiabatically. As a consequence the
entropy $ S \propto V(t) \; T^3(t) \propto V_0 \; \left[a(t) \;
T(t)\right]^3 $ is almost constant according to eq.(\ref
{Ta}) but for the change in the number of relativistic degrees of freedom.
Therefore, for a relativistic equation of state $ p=\rho/3
$ with $ p $ being the pressure and $ \rho $ the energy density, $
\rho(t)\propto T^4(t) $. Such equation of state yields the
following evolution of the scale factor as a function of time \be
\label{HofT} H=\frac{1}{2t}= 0.33 \, g^{\frac{1}{2}}
\frac{T^2}{M_{Pl}} \ee where $g$ is the number of relativistic
degrees of freedom, which is also a function of temperature $ 10
\lesssim g \lesssim 100 $ for $ 0.5\,\textrm{MeV}\lesssim T
\lesssim 300\,\textrm{GeV} $. The above expression yields a simple
dictionary that allows to translate temperature (or energy scale)
into time scales, namely \be \label{Tt} T  \sim
\frac{\textrm{MeV}}{\sqrt{t(\textrm{sec})}}\sim
\frac{10^{10}\,\textrm{K}}{\sqrt{t(\textrm{sec})}} \; . \ee This
simple dictionary allows to establish the time scales at which the
standard model of particle physics predicts phase transitions as
well as an estimate of whether the transitions are likely to occur
in LTE or not. The electroweak transition would have ocurred at $
T \sim 100 \,\textrm{GeV} $ at a time scale $ t_{EW} \sim
10^{-12}\,\textrm{secs} $ and the QCD phase transition at $ T\sim
170\,\textrm{MeV} $ at $ t_{QCD}\sim 10^{-5}\,\textrm{secs} $.

{\bf\underline{Local Thermal Equilibrium (LTE) or Nonequilibrium}:}

Whether a phase transition occurs in or out of local thermodynamic
equilibrium (LTE) depends  on the comparison of  two time scales:
the cooling rate and the rate of equilibration. In the early
Universe or in ultrarelativistic heavy ion collisions, the rate of
change of temperature is determined by the expansion rate of the
fluid. The rate of cooling by cosmological expansion follows from
eq.(\ref{Ta}) $ \dot{T}(t)/T(t) =-\dot{a}(t)/a(t)= -H(t) $.
Collisions as well as
non-collisional processes  contribute to establish equilibrium
with a rate $ \Gamma $.   Local thermodynamic equilibrium ensues
when $ \Gamma > H(t) $, in which case the evolution is adiabatic
in the sense that the thermodynamic functions depend slowly on
time through the temperature. When the cosmological expansion is
too fast, namely $ H(t) \gg \Gamma $, local thermodynamic
equilibrium \emph{cannot} ensue, the temperature drops too fast
for the system to have time to relax to LTE and the phase
transition occurs via a \emph{quench} from the high into the low
temperature phase.

While a detailed understanding of the relaxational dynamics
requires an analysis via quantum Boltzmann equations, a simple
order of magnitude estimate for a collisional rate is given by the
ensemble average $\Gamma \sim \langle\sigma\,n\,v\rangle$, where $
\sigma $ is a scattering cross section, $ n $ is the density of
scatterers and $ v $ the average velocity. For electromagnetic
scattering a typical cross section is of order $ \sigma_{em} \sim
\alpha^2/Q^2 $ with $ Q^2 $ the transferred momentum, at high
temperature single photon exchange yields the estimate $
\sigma_{em}\sim \alpha^2/T^2 $, the density of ultrarelativistic
degrees of freedom $ n\sim T^3 $ and $ v\sim 1$  yields $
\Gamma_{em} \sim \alpha^2  \; T $. In QCD, simple gluon exchange
yields the estimate $ \Gamma_{QCD} \sim \alpha^2_s \; T $. Comparing
with $H$ given by eq.(\ref{HofT}), it is found that the strong
interactions are in LTE for $ T \lesssim 10^{16}$ GeV
and electromagnetic interactions are in LTE for $ T \lesssim
10^{14}$ GeV. A similar estimate is obtained for the
weak interactions: a typical scattering process with energy
transferred $ E\ll M_W $ has a scattering cross section $\sigma
\sim G^2_F \;  E^2 $ whereas if $ E>>M_W, \; \sigma \sim g^4/E^2 $,
therefore in a thermal medium with $E\sim T$ and density of
relativistic particles $ n \sim T^3 $ a typical weak interaction
reaction rate is $ \Gamma_{EW} \sim g^4 \; T $ for $ T \gg M_{W,Z}
$, and $ \Gamma_{EW}\sim G^2_F \;  T^5 $  for $ T\ll M_{W,Z} $. In
this latter temperature regime the ratio $ \Gamma_{EW}/H \sim
(T/\textrm{MeV})^3 $, hence the weak interactions fall out of LTE
for $ T \lesssim 1$ MeV. This simple analysis, while
providing an intuitive order of magnitude estimate for the
relaxation time scales, neglects several subtle but important
aspects that must be studied on a case-by-case basis:
\begin{itemize}
\item{\textit{Screening and infrared phenomena}: the estimate for
the relaxational rates $\Gamma$ invoked the exchange of a vector
boson or relativistic degrees of freedom. In a medium at high
temperature and or density there are important screening effects
and infrared phenomena that change these assessments both
quantitatively and qualitatively and depend on whether the gauge
symmetry is abelian or not\cite{tembooks}.}

\item{\textit{Critical slowing down}: condensed matter experiments
reveal that systems that undergo second order (or in general
continuous) transitions exhibit a slowing down of relaxational
dynamics of long wavelength fluctuations near the critical point
of a phase transition. Such is the case in ferromagnets,
superconductors and superfluids. While short wavelength
fluctuations remain in LTE through the transition through
microscopic scattering mechanisms, long wavelength fluctuations
feature   slower relaxational dynamics and even cooling rates far
smaller than microscopic relaxational rates produce quenched
transitions and departures from equilibrium. Critical slowing down
in \emph{classical} models of critical phenomena is fairly well
understood\cite{golden,langer,guntonmiguel}, but a similar level
of understanding in \emph{quantum field theories} at extreme
temperature and density is now emerging\cite{boycrit}.}

\item{\textit{Strong first order transitions and metastable
states:} In a strong first order transition (see below) the system
is trapped in a metastable state which is a local minimum of the
free energy but not the global one. Within this local, metastable
minimum, collisions can bring the system to LTE, but the
metastable state will eventually decay and fall out of LTE by the
non-perturbative process of nucleation, which is described below
in detail. }

\end{itemize}

\subsection{Phase transitions: early Universe vs. accelerator
experiments}

The control variables in a collision experiment are the beam
energy and the luminosity.  For a   phase transition to be
achieved in an accelerator experiment  an environment with a
temperature close to the transition value $T_c$ must be formed in
the collision region. The blackbody relation, valid for
ultrarelativistic particles in equilibrium yields the energy
\emph{density}-temperature relation \be \varepsilon = C \, T^4 \;
, \ee where the constant $ C $ depends on the number of degrees of
freedom.  For the electroweak phase transition with $ T_c \sim
100\,\textrm{GeV} $, the energy \emph{density} that must be
deposited in the collision region to achieve the conditions for a
phase transition is $ \varepsilon \sim 10^{10}
\,\textrm{GeV}/\textrm{fm}^3 $. This energy density is about $
10^{11} $ times \emph{larger} than that of nuclear matter. For the
QCD phase transition with $ T_c \sim 177\,\textrm{MeV} , \;
\varepsilon \sim \textrm{GeV}/\textrm{fm}^3 $ which is achieved in
ultrarelativistic heavy ion collisions at SPS-CERN and RHIC-BNL.
Therefore, from the perspective of studying particle physics phase
transitions with \emph{accelerator experiments} the only realistic
possibility for the foreseeable future is the QCD transition with
URHIC at RHIC and the forthcoming LHC. Hence the potential
observables from phase transitions in the early Universe before
the QCD scale must be inferred \emph{indirectly} from the
aftermath. In section \ref{PUN} we compare the conditions for the
QCD phase transition both in the early Universe and in URHIC to
assess whether current experiments reproduce the conditions that
prevailed about $10 \,\mu\,\textrm{secs}$ after the beginning of
the Universe.

\section{Phase Transitions: equilibrium and non-equilibrium aspects:}
\label{sec:phases}

\subsection{Equilibrium aspects: free energy, effective potentials and critical
phenomena.}\label{crit}

Phase transitions are broadly characterized as either second or
first order\cite{golden,langer,guntonmiguel}. In the Landau theory
of phase transitions the order parameter plays a central role. In
particle physics the order parameter is the expectation value of a
spin zero field in the state that extremizes the free energy. A
non-zero expectation value for a non-zero spin field entails the
breakdown of rotational symmetry. A particularly illuminating
example is that of ferromagnetic materials where the order
parameter is the total magnetization of the sample. For these
materials the magnetization vanishes above the Curie temperature
while a nonzero spontaneous magnetization emerges below such
critical temperature. In the standard model, the order parameter
is the expectation value of the neutral component of the Higgs
$\langle \varphi_0 \rangle$, in QCD the chiral order parameter is
the expectation value of the pseudoscalar density $\langle
\bar{\psi} \gamma^5 \psi\rangle$. For the
confinement-deconfinement phase transition lattice studies show a
rapid variation of the free energy as a function of temperature
and the Polyakov loop may play the role of order parameter (see
below).

At zero temperature and chemical potential the state of lowest free
energy is the vacuum state of the theory. An important concept to
study the nature of the phase transition is that of the effective
potential or Gibbs free energy which is a function of the expectation value
of the scalar field.

Consider adding to the Hamiltonian of the theory $ H $ a term   $
J\,\varphi $ where $ J $ is a constant and $ \varphi $ is the
scalar field whose expectation value  is the order parameter. The
total Hamiltonian is $ H[J]=H+J\,\varphi $. The partition function
is given by \be Z[J,T]= \textrm{Tr} \,e^{-\frac{H[J]}{T}}\, \ee
introducing the Helmholtz free energy \be W[T,J] = -T\, \ln Z[J,T]
\; , \ee the order parameter $\Phi$ is given by \be \frac{\delta
W[T,J]}{\delta J} = \Phi \; , \ee inverting this relation one
finds $ J(\Phi) $. The Gibbs free energy or effective potential $
V[\Phi] $ follows from a Legendre transform \be V[\Phi] =
W[J(\Phi)]-J(\Phi)\,\Phi \; , \ee and it is a function of the
order parameter and other intensive thermodynamic variables such
as temperature, chemical potential, etc. It is a very powerful
concept that provides information on the \emph{equilibrium
thermodynamic} aspects and the phase structure of the theory and
the possible transitions between them.

\begin{figure}[h]
\begin{center}
\epsfig{file=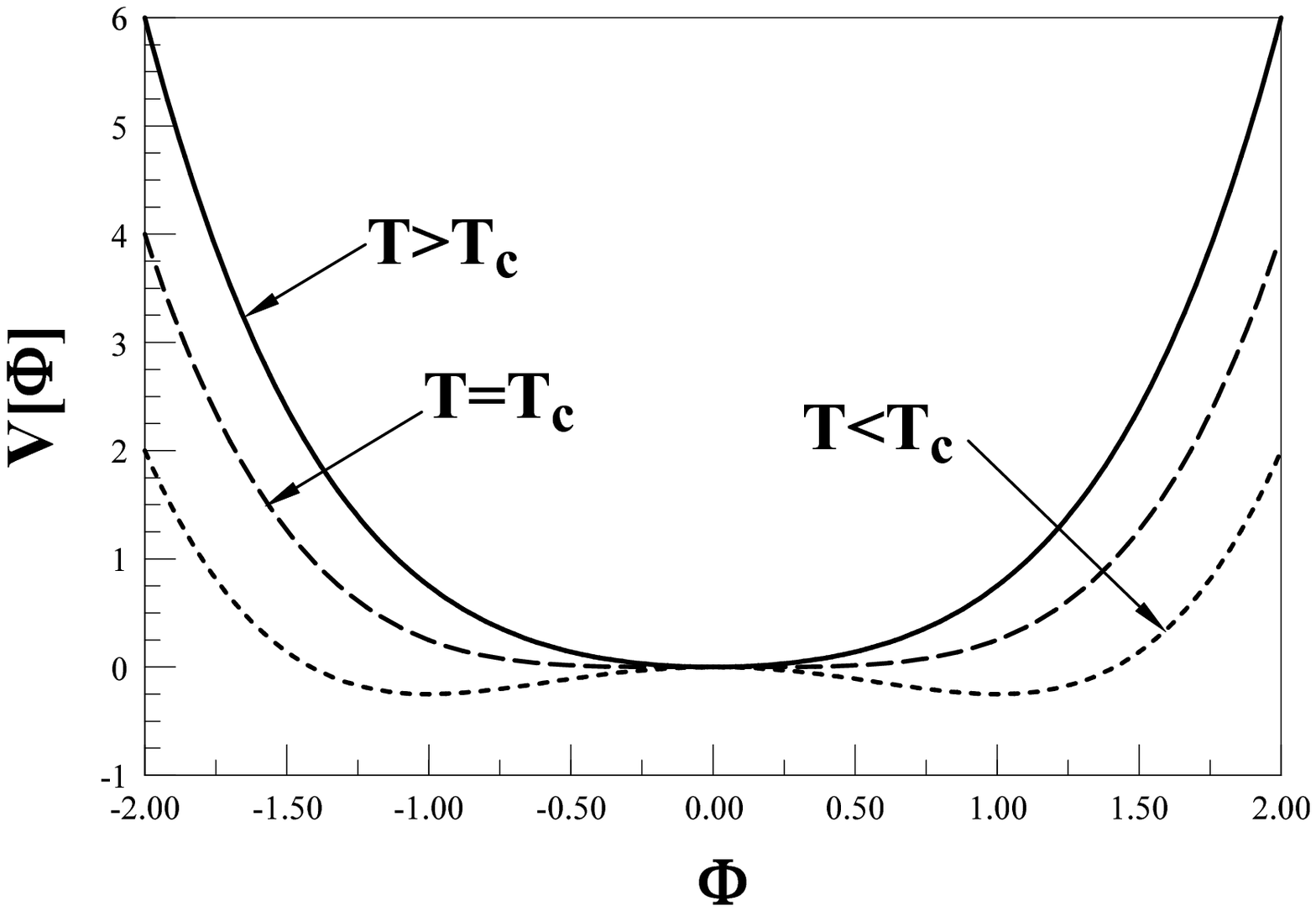,height=2in,width=2in,keepaspectratio=true}
\epsfig{file=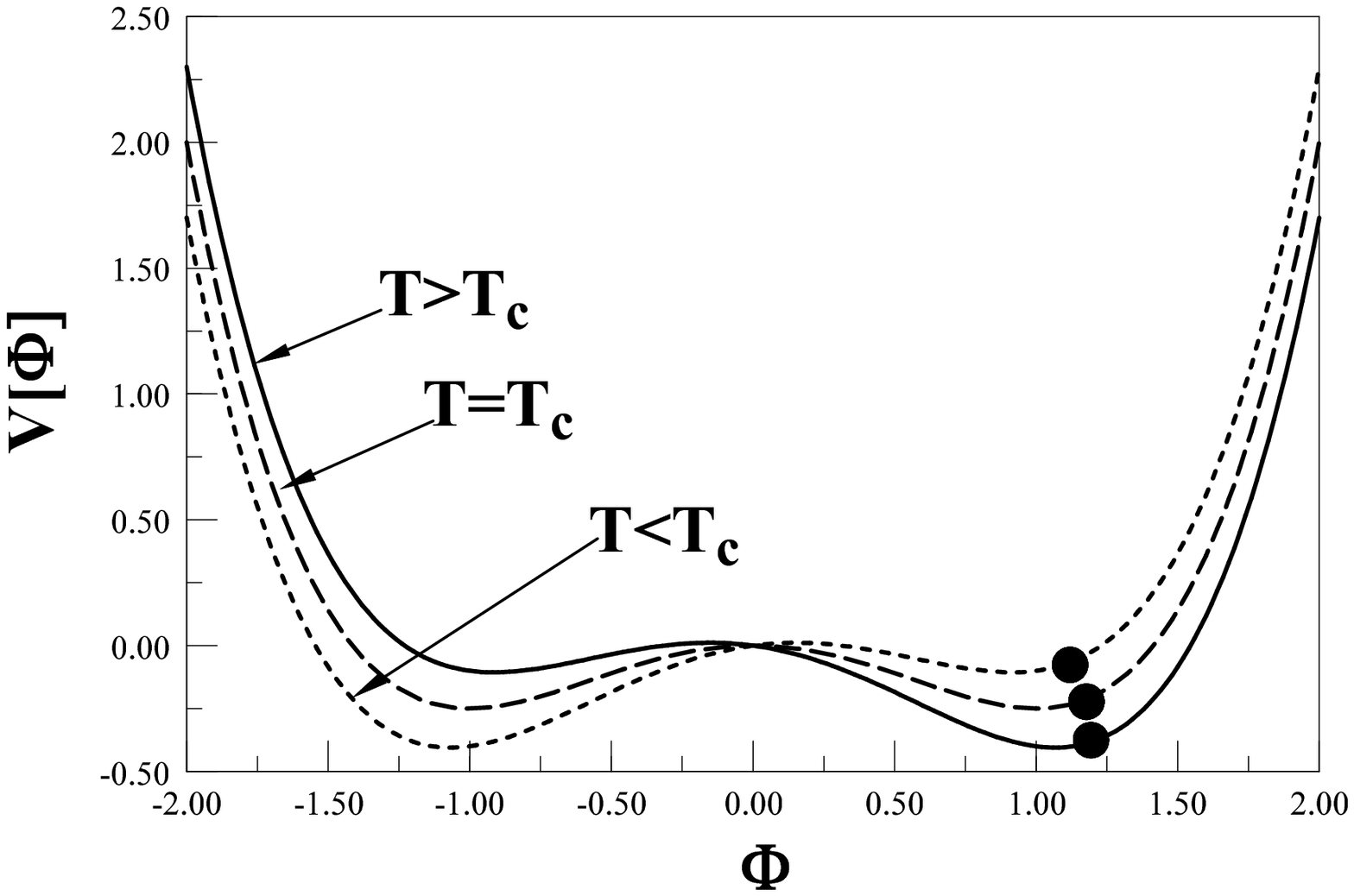,height=2in,width=2in,keepaspectratio=true}
 \caption{Effective potentials for a second order phase transition with
the breakdown of a discrete symmetry (left panel)
  and for a first order phase transition (right  panel).
 The filled circles in the right panel figure display the equilibrium minima
for $T>T_c$ and $T=T_c$ and the metastable minimum for $T<T_c$. }
 \label{fig:Veff}
\end{center}
\end{figure}

Fig. \ref{fig:Veff} displays the typical effective potentials
for a second order (left panel) and first order (right panel) phase
transitions. In the case of a second order transition, the order
parameter  in the state of minimum free energy  vanishes
continuously as the temperature approaches the critical. In second
order phase transitions the \emph{second} derivative of the free
energy with respect to temperature is divergent at the critical
temperature. Near the critical temperature the free energy per unit
volume, or equivalently the pressure varies as $ p(T)\sim
|T-T_c|^{2-\alpha} $ with $ \alpha $ the thermal critical exponent. The order
parameter itself vanishes as $ |T-T_c|^\beta $ for $ T\rightarrow T_c $
from below, with $ \beta >0 $ and response functions feature power-law
singularities. In contrast to this case, in a first order phase
transition the value of the order parameter and the position of the
minimum of the free energy jump discontinuously at the critical
temperature. In fig. \ref{fig:Veff} the black circles in the right panel
 show the behavior of the order parameter as the temperature
is \emph{lowered} from above to below $ T_c $. The state that
minimizes the effective potential changes suddenly from the right to
the left well as the temperature is lowered below $ T_c $. For
$ T>T_c $ the right well in the right panel remains a
\emph{local} minimum of \emph{higher} free energy, thus a metastable
state. The difference in free energies between the local and the
global minima yields a latent heat which is released upon the decay
of the metastable state.

The effective potential also provides information on the
excitation spectrum. Consider a state characterized by a given
value of the order parameter $\Phi$, the frequency of harmonic
excitations of wavevector $k$ around this state is given by
\be\Omega(k;\Phi)= \sqrt{k^2+V''(\Phi)}\,~~,~~V''(\Phi)=\frac{d^2
V(\Phi)}{d\Phi^2}\,.\label{freqs}\ee \emph{Most} of the phase
transitions in particle physics models involve the spontaneous
breakdown of a global symmetry and the order parameter transforms
covariantly under this symmetry. An expectation value of the order
parameter in the state of lowest free energy indicates the
spontaneous breakdown of the symmetry. Different phase transitions
in particle physics involve different order parameters: for the
electroweak transition the order parameter is the expectation
value of the Higgs doublet along the neutral component, $
\Phi=\langle \varphi_0 \rangle $. This expectation value breaks
the symmetry $ SU(2)\otimes U(1)_Y \rightarrow U(1)_{em} $, the
three Goldstone bosons emerging from the \emph{global} broken
symmetry make up the longitudinal components of the $ W^{\pm},Z^0
$ vector bosons via the Higgs mechanism. In QCD with massless
quarks the expectation value of the chiral density $
\langle\bar{\Psi}\gamma^5 \Psi\rangle $ with $ \Psi $ an up-down
quark doublet breaks the $ SU(2)_L\otimes SU(2)_R $ symmetry of
the QCD Lagrangian       down to $ SU(2)_{L+R} $ with pions
emerging as the triplet of Goldstone bosons. Because up and down
quarks are not exactly massless, pions are \emph{pseudo} Goldstone
bosons. The QCD confinement-deconfinement transition does not have
an obvious order parameter. However, in absence of dynamical
quarks, for example in  an $ SU(N) $ Yang-Mills theory of gluons,
the \emph{Polyakov loop}\be L(x) = \frac{1}{N}\left\langle
\,\textrm{Tr}\, \mathcal{P}
\exp\left(ig\int_0^{\frac{1}{T}}\textbf{A}_0(\vec x,\tau) \; d\tau
\right) \right\rangle \ee has been shown to be a suitable gauge
invariant order parameter. In the above expression $ g $ is the
gauge coupling, $ T $ is the temperature $ \textbf{A}_0 $ is the
temporal component of the $ SU(N) $ gauge potential, and $
\mathcal{P} $ is the path ordering symbol, the line integral is
carried out in Euclidean time. The Polyakov loop is usually
interpreted as the free energy of an infinitely heavy test quark.
It vanishes in the confined phase for $ T<T_c $ and is
non-vanishing in the deconfined phase $ T>T_c $. An effective
potential in terms of the Polyakov line has recently been proposed
to study the confining-deconfining phase
transition\cite{polypisa}. When light dynamical quarks are
included there is no longer an obvious order parameter for the
confinement-deconfinement phase transition because light dynamical
quarks  screen the potential from gluon exchange.

\textbf{Second order phase transitions in equilibrium and critical
phenomena: }  Consider   a continuous (second order) phase
transition\cite{golden} described by an effective potential akin to
the left panel in fig. \ref{fig:Veff} in  the case in which the
temperature falls from a value larger than the critical value $ T_c
$. If the cooling rate is much smaller than the relaxation rate the
evolution will be in LTE, but as the temperature approaches the
critical, collective long-wavelength fluctuations develop and become
strongly correlated. Near $ T_c $ these collective long-wavelength
fluctuations become massless (critical), their correlation function
becomes  scale invariant and large regions behave coherently with
strong correlations over arbitrarily large scales. This is the
hallmark of critical phenomena at second order phase transitions
which are characterized by response functions that feature
singularities in $ |T-T_c| $ in terms of power laws with
\emph{critical exponents}. The correlation function of the order
parameter field $ \Phi $ is given by $ \langle \Phi(\vec x)\Phi(\vec
0)\rangle \propto e^{-x/\xi(T)} $. The correlation length $ \xi(T) $
diverges at the critical temperature as $ \xi(T) \sim |T-T_c|^{-\nu}
$, with $ \nu
> 0 $ being a critical exponent. At this point the system becomes
correlated over large distances and the spectrum of
long-wavelength fluctuations becomes \emph{scale invariant}. This
is a very important aspect of critical phenomena associated with
second order phase transitions. Near the critical point where the
correlation length diverges the long distance physics is
\emph{universal} and the \emph{static aspects} can be described in
terms of a Landau-Ginzburg low energy effective theory. The
different \emph{universality classes} that characterize the
different critical exponents, are determined by few properties of
the system such as the dimensionality, the symmetry of the order
parameter, the number of independent fields, etc. The description
of static critical phenomena in terms of the Landau-Ginzburg
approach has been confirmed by a wealth of experiments in
condensed matter systems. A relevant example of a Landau-Ginzburg
description for the case of a scalar field, whose expectation
value breaks a discrete symmetry $ \varphi\rightarrow -\varphi $
is described by the finite temperature effective potential given
by \be\label{LGveff}
V[\Phi;T] = \frac{a}{2} \; (T-T_c) \; \Phi^2 +\frac{b}{4} \; \Phi^4
\ee and featured in fig.\ref{fig:Veff}. The
concept of \emph{universality classes} that describes critical
phenomena for a wide variety of systems is of fundamental
importance. Relevant to the discussion below is the case of the
chiral phase transition in QCD, which for massless up and down
quarks, is described by the same universality class as the
Heisenberg ferromagnet with $ O(4) $ symmetry\cite{raja}.

An important \emph{dynamical} aspect of second order phase
transitions is that near the critical region the relaxation time
scale  of long-wavelength fluctuations  feature \emph{critical
slowing down}\cite{langer,guntonmiguel,golden}. Critical slowing
down has been studied theoretically and experimentally in condensed
matter systems, where a large body of experimental work confirms the
critical slowing down of these fluctuations near second order phase
transitions in \emph{classical systems} in agreement with theory.
The study of critical slowing down in \emph{quantum field theory}
has recently began  to receive attention \cite{boycrit}. For any
finite cooling rate, long-wavelength modes will be \emph{quenched}
through the phase transition. The effective potential  provides
\emph{only} an \emph{equilibrium} description but does not address
the issue of the \emph{dynamics} of the transitions between
different phases.

\subsection{Non-equilibrium aspects:  spinodal
decomposition and nucleation. }\label{noneqPT}

   As discussed   in sec.\ref{2C} whether a
phase transition occurs in LTE or not depends on  whether, $ \Gamma
> H(t) $ or   $ \Gamma < H(t) $, respectively. In the first case the
transition occurs adiabatically (similar to the Minkowski case)
while in the latter case, the phase transition occurs via a
\emph{quench} from the high into the low temperature phase. The
dynamics are different depending on whether the \emph{equilibrium}
effective potential describes a second or first order transition.

\subsubsection{ {Second order case: spinodal decomposition}}
\begin{figure}[h]
\begin{center}
\epsfig{file=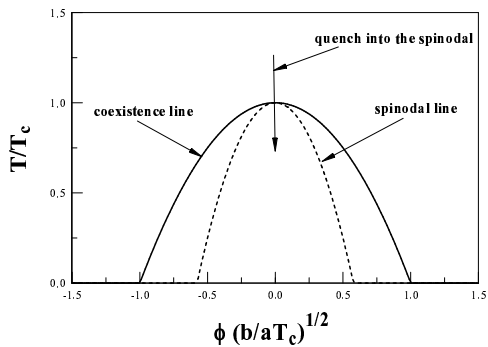,height=2in,width=2.5in,keepaspectratio=true}
\caption{A quench into the spinodal region.}
 \label{fig:quench}
\end{center}
\end{figure}

Let us first consider the simpler case of a symmetry breaking
second order transition where the symmetry that breaks
spontaneously is discrete. To focus on a simple, yet relevant
example, consider the   form for the finite temperature effective
potential given by eq.(\ref{LGveff}). For $ T>T_c $ the state of
minimum free energy corresponds to $ \Phi=0 $, whereas for $ T<T_c $
the state of minimum free energy corresponds to either the right
or left equilibrium minima in fig. \ref{fig:Veff}, namely
$ \Phi=\pm \Phi_e $ with $ \Phi_e = \pm \left[\frac{a
T_c}{b}(1-\frac{T}{T_c})\right]^{\frac12} $. The inflection
points at which $ V''(\Phi,T)=0 $ are at $ \Phi= \pm\Phi_s $ with
$ \Phi_s = \left[\frac{a
T_c}{3b}(1-\frac{T}{T_c})\right]^{\frac{1}{2}} $. The states with $
 -\Phi_s \leq \Phi \leq + \Phi_s $ are \emph{thermodynamically unstable}. In
 order to understand the non-equilibrium aspects of a rapid phase
 transition it is convenient to plot the coexistence lines at which
 thermodynamic equilibrium states coexist and the spinodal line that
limits the region of
 thermodynamic stability in the $ T-\phi $ plane. These are shown in
 fig. \ref{fig:quench}. The coexistence line is given by the
 relation between $ \Phi $ and $ T $ for the minima of the potential,
 namely $ T_{coex}(\Phi)= -b \; \Phi^2/a + T_c $, and similarly the
 spinodal line is determined by the condition $V''(\Phi,T)=0$,
 namely the inflexion points of the effective potential, which
 yields $ T_{sp}(\Phi) = -b  \; \Phi^2/(3 \; a) + T_c $. Fig. \ref{fig:quench}
 depicts a quench from a high temperature state with vanishing order
 parameter to a low temperature state when the equilibrium state
 corresponds to a broken symmetry.  In \emph{classical}
 statistical mechanics out of equilibrium the \emph{dynamics} of the
 evolution out of equilibrium is studied via purely dissipative
 equations of motion, the Cahn-Hilliard equations\cite{langer,guntonmiguel,golden}.
 The initial stages of spinodal decomposition in \emph{classical}
 systems are characterized by growth of unstable long-wavelength
 perturbations which lead to the formation of domains and
 coarsening\cite{langer,guntonmiguel}. Beautiful experiments in
 phase separation and spinodal decomposition in binary fluids have provided a
 spectacular confirmation of  this mechanism as well as the
 dynamical aspects predicted by the Cahn-Hilliard
 theory\cite{goldburg}.   In \emph{quantum
 field theory} the dynamics is determined by the unitary time
 evolution of an initially   quantum state or density matrix.
 \emph{Quantum spinodal decomposition} is studied in
 ref.\cite{calzetta,danspino} and the non-linear backreaction of
 unstable fluctuations was studied in \cite{danspino}  by implementing
a non-perturbative self-consistent method.  States between the
 spinodal lines are thermodynamically unstable to small amplitude
long-wavelength  perturbations, since $ V''(\Phi)<0 $ in the spinodal region
$ -\Phi_s\leq \Phi \leq +\Phi_s $,
 the frequency of small amplitude fluctuations around this state  become
\emph{imaginary} for wavevectors $ k< |V''(\Phi)| $, while
fluctuations with $k>|V''(\Phi)|$ remain stable.
 Fluctuations with wavevectors in the spinodal band $ k < \sqrt{|V''(0)|} $
grow as $ e^{\gamma_k\,t} $ with $ \gamma_k =
 \sqrt{|V''(0)|-k^2} $. These unstable modes lead to the formation of
 domains whose size is determined by a time dependent correlation length
 $ \xi(t) $.   Inside these domains
 the expectation value attains the equilibrium values $ \pm \Phi_e $
 at the final temperature. The formation of correlated domains of each phase
is the hallmark of the
 process of phase separation via  \emph{spinodal decomposition}. The dynamics
of this process is
  non-perturbative: initially long wavelength fluctuations grow exponentially
until
 they become non-linear and react back   on the dynamics
shutting off the instabilities.
  In scalar field theories it is  found that during the initial stages of
phase separation  $ \xi(t) \propto t $.
The above discussion focused on the case of discrete symmetry that
is spontaneously broken. When the broken symmetry is continuous a
wealth of fascinating non-perturbative topological excitations
emerge, such as vortices, monopoles, and textures\cite{topo}.

\subsubsection{{First order phase transitions: nucleation}}

\begin{figure}[h]
\begin{center}
\epsfig{file=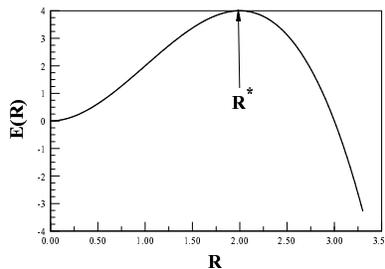,height=2in,width=2in,keepaspectratio=true}
\caption{$E(R)$ vs. $R$, arbitrary units.}
 \label{fig:bub}
\end{center}
\end{figure}

The dynamics of first order phase transition is   different: in
this case the thermodynamic state of the system is a trapped
\emph{metastable state}, locally stable under small thermodynamic
perturbations, but with a higher free energy than the global
minimum\cite{langer,guntonmiguel}. This is the situation of the
liquid-gas phase transition for example in water. The metastable
state is stable under small amplitude perturbations (oscillations
in the right-most well in fig. \ref{fig:Veff}) but decays via
spontaneous large amplitude fluctuations corresponding to bubbles
of the stable phase immersed in a host of the metastable state.
Consider a spherical such bubble or radius $R$, where   inside of
the bubble the value of the order parameter is that of the
globally stable state, but outside is that of the metastable
state. Since the globally stable state has a lower free energy
than the metastable state, there is a gain in the free energy
given by $-4\pi R^3 \Delta V/3$ where $\Delta V$ is the difference
in the effective potential between the stable and the metastable
state. Because the order parameter is inhomogeneous in this
configuration there is an elastic contribution to the free energy
from the gradients of the order parameter field which is
proportional to the \emph{surface} of the bubble, because this is
the region in which the spatial derivatives of  the order
parameter are non-vanishing. This elastic contribution is positive
and given by $4\pi R^2\,\sigma$ where $\sigma$ is the surface
tension, thus the total change in the free energy for such
inhomogeneous bubble configuration is \be E[R] = 4\pi R^2 \sigma -
\frac{4\pi}{3}\,R^3 \Delta V \,.\ee  This change in the free
energy is depicted in fig. \ref{fig:bub}.

Bubbles with radii $R < R^*= 2\sigma/\Delta V$ shrink while those
with $R > R^*= 2\sigma/\Delta V$ grow and convert the metastable
phase into the globally stable phase releasing the latent heat.
The phase transition completes via percolation of these growing
bubbles. The probability of a bubble to appear spontaneously in
the heat bath in equilibrium at temperature $T$ is given by \be
P(R^*) = D\, e^{-\frac{E[R^*]}{T}}~~;~~ E[R^*] = \frac{16\pi}{3}
\frac{\sigma^3}{(\Delta V)^2}\ee where the prefactor $D$ depends on
the spectrum of small amplitude fluctuations around the critical
bubble configuration. The rate of decay per unit volume and per
unit time of the metastable state is given
by\cite{langer,guntonmiguel,golden} \be \Gamma = \Gamma_0 \,
e^{-\frac{E[R^*]}{T}} \ee where $\Gamma_0$ depends on the
prefactor $D$ as well as in general on transport properties such
as the heat conductivity\cite{langer,guntonmiguel}. These
processes which require large amplitude fluctuations to overcome a
potential barrier, which in this case is determined by $E[R^*]$
are called thermally activated and the decay rate features the
Boltzmann suppression factor corresponding to the  energy of the
configuration at the top of the barrier. Since the metastable
state has a higher free energy than the globally stable state, the
nucleation process that results in the conversion of the
metastable into the stable phase must release the latent heat.

\subsubsection{{Liquid-gas phase transition: phase coexistence}}
\label{sec:liqgas} 
\begin{figure}[h]
\begin{center}
\epsfig{file=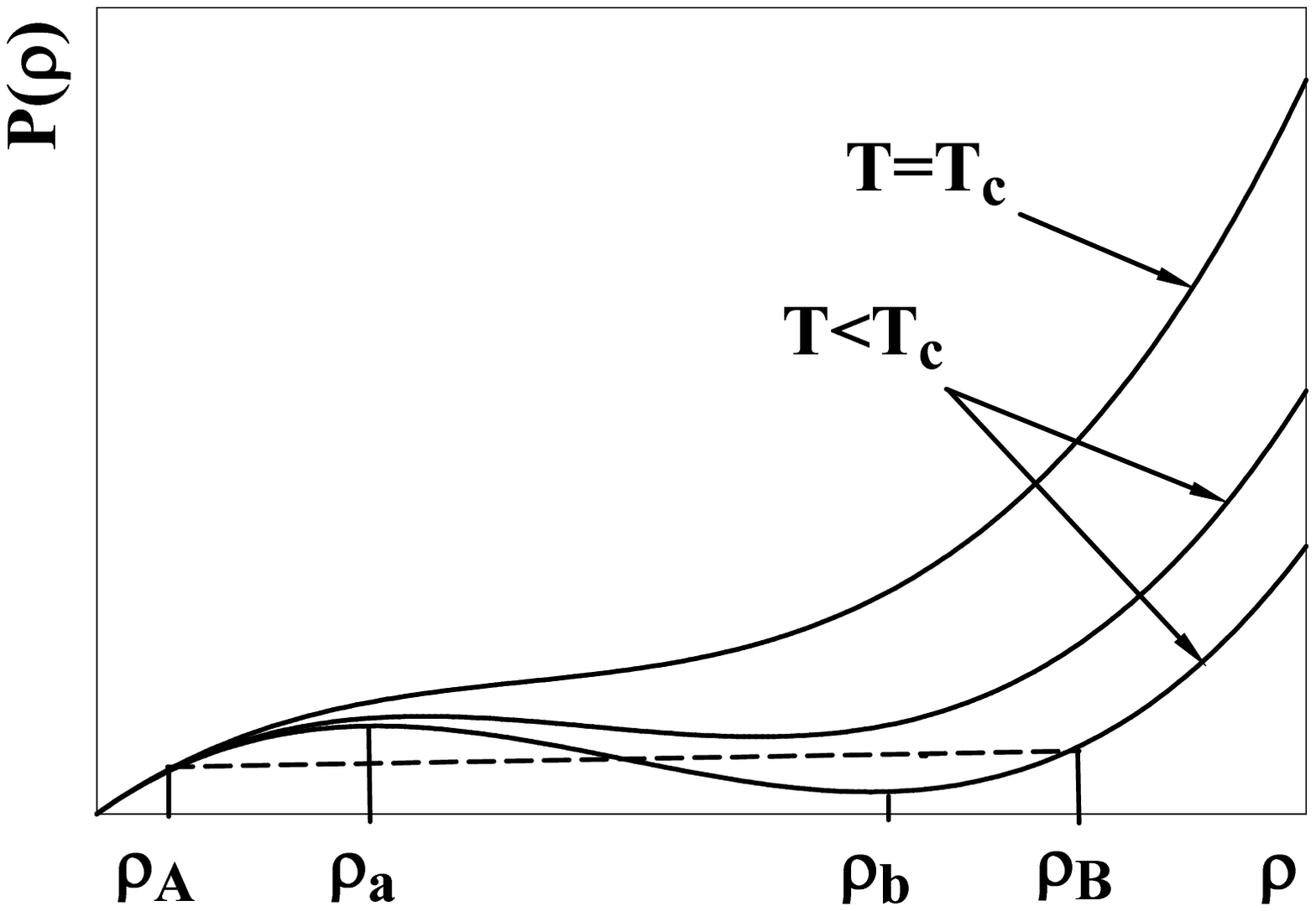,height=2in,width=2in,keepaspectratio=true}
\epsfig{file=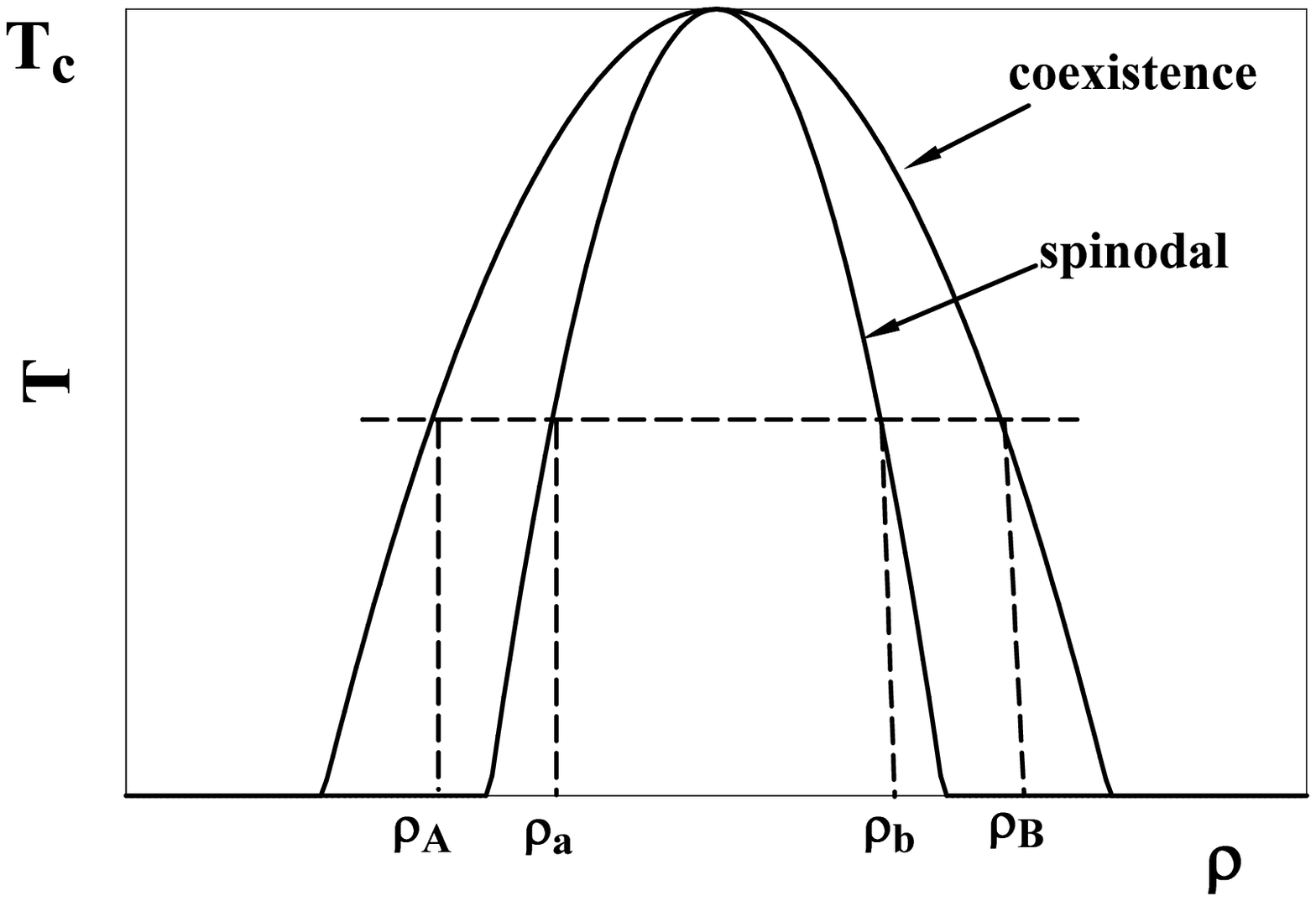,height=2in,width=2in,keepaspectratio=true}
\caption{Left panel: equation of state for a liquid gas transition.
The dashed line is the coexistence line (Maxwell construction),
$\rho_{a,b}$ determine the position of the spinodal line. Right
panel: coexistence and spinodal lines in the $T-\rho$ plane. }
\label{fig:liqgas}
\end{center}
\end{figure}

Another important and very relevant example of
a first order phase transition is the liquid-gas transition. This
transition is exemplified by an equation of state of the Van der
Walls form and results in general when the microscopic
interactions have a short range repulsive and a long range
attractive components. Such is the case of nuclear forces between
nucleons (see section \ref{nuclearpt}).  Fig. \ref{fig:liqgas}
displays a few isotherms for a typical equation of state in terms
of pressure $ P $  vs. density $ \rho $ for a liquid-gas  type
transition. For $ T>T_c $ the isotherms are single valued, there
is only one value of density $\rho$ for a given pressure  $P $ and
the system is in one phase. However, for $ T<T_c $, for a given
value of the pressure there are three values of the density. The
derivative $
\partial P/\partial \rho|_T = c^2_{s,T} $, where $ c_{s,T} $ is the
isothermal speed of sound for hydrodynamic modes. Obviously, we
see that $ c^2_{s,T} < 0 $ in the region between $ \rho_a $ and $
\rho_b $ in fig.\ref{fig:liqgas}. This corresponds to an imaginary
speed of sound, therefore to unstable density fluctuations that
grow exponentially with time as $ e^{k \; | c_{s,T}| \; t} $.
Therefore, such unstable thermodynamic states and the value of the
pressure in this region do not correspond to a thermodynamically
stable equilibrium state. The two different values $ \rho_{A,B} $
describe two different phases: the phase corresponding to $ \rho_A
$ features a small speed of sound, hence a highly compressible
medium, this is the gas phase, while the phase described by $
\rho_B $ features a large speed of sound, hence a rather small
compressibility, namely a liquid. The dashed line at constant
pressure in fig. \ref{fig:liqgas} describes the coexistence of
these phases in thermodynamic equilibrium at the same pressure,
temperature and chemical potentials. The position of this line is
determined by the equal area law or \emph{Maxwell construction}.
Along the coexistence curve the state of the system is a
\emph{heterogeneous} mixture of both phases with a composition
given by the lever rule: a state with global density $\rho$ is a
mixed phase with islands of gas with density $\rho_A$ in a host of
liquid with density $\rho_B$ with $\rho = \zeta \rho_A +
(1-\zeta)\rho_B$ where $\zeta = (\rho_A-\rho_B)/(\rho-\rho_B)<1$
is the proportion of the phases. A change of the global density in
this region results in a conversion of part of one of the phases
into the other at constant pressure and temperature. Therefore in
the coexistence region or mixed phase small variations in the
density do \emph{not} result in a change in the pressure or
temperature, just in the isothermal and isobaric conversion of
phases, hence the isothermal speed of sound \emph{vanishes} along
the coexistence curve or mixed phase. This flat portion of the
isotherm in the mixed phase is a \emph{soft} part of the equation
of state since the isothermal speed of sound vanishes. This
property of the mixed phase or coexistence region for first order
phase transitions has important consequences both in cosmology as
well as in ultrarelativistic heavy ion collisions to be discussed
below.

The region $ \rho_a < \rho < \rho_b $ corresponding to $ \partial
P/\partial \rho < 0 $ is the \emph{spinodal} region: small
pressure (or density) perturbations of a \emph{homogeneous phase}
are unstable and grow. The regions $ \rho_A < \rho < \rho_a\,;\,
\rho_b < \rho < \rho_B $ are actually \emph{metastable}. Consider
the following experiments: a) begin at $ T>T_c $ with a state in
thermodynamic equilibrium with a single phase at density $\rho_a <
\rho < \rho_b$\footnote{Since this state  is continuously
connected to the low density region, this single phase describes a
gas.} and quench the system in temperature by lowering the
temperature to $ T<T_c $ on a very short time, b) repeat the
experiment but now quenching the system from $ T >T_c $ to $ T<T_c
$ at constant density $ \rho_b < \rho < \rho_B $, what is the
dynamics in both cases? In the case (a) the initial state is a
single (gas) phase, homogeneous of density $\rho$ which is
quenched inside the spinodal region where small amplitude density
perturbations of a homogeneous phase are unstable. The spinodal
instabilities lead to a fragmentation of the homogeneous phase and
the formation of domains of the liquid phase, these domains grow
and separate, leading to the process of phase separation and an
heterogeneous mixed phase in coexistence with proportions of
phases given by the lever rule. In the case (b) the initial
homogeneous state (gas) with global density $ \rho_b < \rho <
\rho_B $ is quenched into a region in which the homogeneous phase
is \emph{mechanically stable} since $ \partial P/\partial \rho
>0 $, however at this temperature the free energy of the liquid is
smaller, hence the state is \emph{metastable}. It will decay to a
mixed state of gas and liquid along the coexistence line by
\emph{nucleation}: a large amplitude fluctuation corresponding to a
droplet larger than the critical, the growth and percolation of
these droplets eventually leads to the mixed phase on the
coexistence curve. Spontaneous nucleation of a liquid droplet in the
homogenous gaseous host is referred to as \emph{homogeneous
nucleation}, but also the presence of impurities can act as
nucleation seeds, in which case the situation is referred as
\emph{heterogeneous} nucleation.

This is in fact the principle of the cloud chamber, one of the first
particle detectors where a gas was supercooled and cosmic rays (or
other particles) passing through the gas induced the nucleation of
droplets, namely these particles triggered the \emph{heterogeneous
nucleation of the metastable phase}.

The spinodal and coexistence regions end at the critical point, and
for $ T>T_c $ the system can only be in a single thermodynamically
stable phase. Any path taking the system from a liquid into the gas
phase above $T_c$ is continuous. The critical point itself is of
second order. If the system is brought near the critical temperature
from above but always in LTE, the system develops large density
fluctuations with a correlation length that diverges. These strong
fluctuations in liquid-gas systems gives rise to the phenomenon of
\emph{critical opalescence} which  is observed in the familiar
example of water near its boiling point when bubbles of vapor are
beginning to be formed and the liquid becomes opaque as a result of
light scattering from the large vapor domains.

A liquid-gas type transition is ubiquituous in systems that
feature microscopic short range repulsion and a long range
attraction. Such is indeed the case of nucleons in nuclei,
semi-phenomenological effective nucleon interactions yield
equations of state of the Van der Walls type and predict that
nuclear matter undergoes a liquid-gas phase transition at a
critical temperature $ T_c \sim 15-20$ MeV. The theoretical and
experimental aspects of such phase transition in nuclear matter
are discussed further in section \ref{nuclearpt}. \emph{If} QCD
has a first order confinement-deconfinement (quark hadron) phase
transition the soft-part of the equation of state corresponding to
a mixed phase gives rise to an anomalously small speed of sound
with important consequences both in the early Universe as well as
in URHIC. A mixed phase during the quark-hadron transition in the
early Universe would result in an almost vanishing speed of sound,
as a result there are no pressure gradients that can
hydrostatically balance the gravitational pull and gravitational
collapse would ensue, perhaps resulting in the formation of
primordial black holes. In URHIC, if the quark-hadron gas reaches
coexistence as a mixed phase in equilibrium, again the small speed
of sound (soft point) hinders the pressure gradients that drive
the expansion and hydrodynamic flow, with potential experimental
observables.

\textbf{Crossover:} An alternative to a second or first order
phase transition is a simple crossover in thermodynamic behavior
without discontinuities or singularities in the free energy or any
of its derivatives. If the crossover is smooth, then no
out-of-equilibrium aspects are expected as the system will evolve
in  LTE. However, if the crossover is relatively sharp
  the situation may not be too different from a phase
transition. There is now evidence suggesting that the
\emph{standard model} does not feature a sharp electroweak phase
transition (either first or second order) but is a smooth
crossover. Furthermore, current lattice studies indicate that for
two light quarks (u,d) and one heavier quark (s) with a mass of
the order of the QCD scale the confinement-deconfinement
transition \emph{maybe} a sharp crossover rather than either a
first or second order transition. These possibilities are
discussed further below.

\section{Inflation and  WMAP }\label{inflation}

\subsection{Inflationary dynamics}

Inflation was originally proposed to solve several outstanding
problems of the standard big bang model \cite{guthsato} thus
becoming an important paradigm in cosmology. At the same time, it
provides a natural mechanism for the generation of scalar density
fluctuations that seed large scale structure, thus explaining the
origin of the temperature anisotropies in the cosmic microwave
background (CMB), as well as that of  tensor perturbations
(primordial gravitational waves). Inflation is the statement that
the cosmological scale factor $a(t)$ in eq.(\ref{FRWmetric}) has a
\emph{positive acceleration}, namely,   $ \ddot{a}(t)/a(t) > 0 $.
Hence, eq.(\ref{accel}) requires the equation of state $ w= p/\rho
< -1/3 \, $.

\begin{figure}[h]
\begin{center}
\epsfig{file=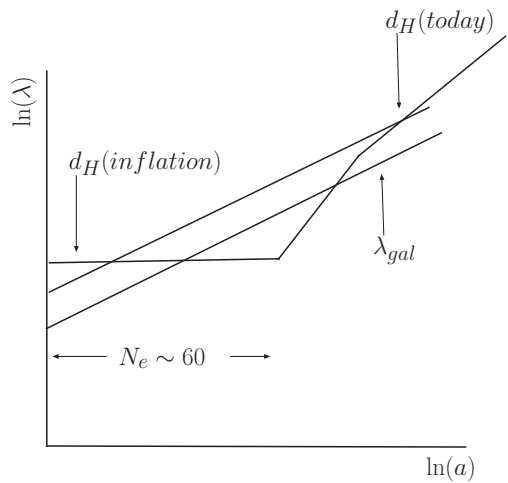,height=2.5in,width=4.5in,keepaspectratio=true}
\caption{Logarithm of physical scales vs. logarithm of the scale
factor. The causal horizon $d_H$ is shown for the inflationary
(De~Sitter), radiation dominated and matter dominated stages. The
physical wavelengths   for today's Hubble radius $d_H(today)$, and
a galactic scale $\lambda_{gal}$ are shown.  }
\label{fig:inflation}
\end{center}
\end{figure}

Inflation gives rise to a remarkable phenomenon: physical
wavelengths grow \emph{faster} than the size of the Hubble radius
$ d_H = a(t)/\dot{a}(t) = 1/H(t) $, indeed \be
\frac{\dot{\lambda}_{phys}}{\lambda_{phys}} = \frac{\dot{a}}{a}=
H(t)= \frac{\dot{d}_H}{d_H}+ d_H \, \frac{\ddot{a}}{a} \; ,
\label{inflad} \ee \noindent Eq.(\ref{inflad}) states that during
inflation \emph{physical wavelengths become larger than the Hubble
radius}. Once a physical wavelength becomes larger than the Hubble
radius, it is causally disconnected from physical processes. The
inflationary era is followed by the radiation dominated and matter
dominated stages where the acceleration of the scale factor
becomes negative since $ p/\rho =1/3 $ in a radiation dominated
era and $ p=0 $ in a matter dominated era [see eq.(\ref{accel})].
With a negative acceleration of the scale factor, the Hubble
radius   grows faster than the scale factor, and wavelengths that
were outside, can now re-enter the Hubble radius. This is depicted
in fig.\ref{fig:inflation}.
This is the main concept behind the inflationary paradigm for the
generation of temperature fluctuations as well as for providing
the seeds for large scale structure formation: quantum
fluctuations generated early in the inflationary stage exit the
Hubble radius during inflation, and eventually re-enter during the
matter dominated era. The basic mechanism for generation of
temperature anisotropies as well as primordial gravitational waves
through inflation is the following\cite{bookkolb}-\cite{muk}: the
energy momentum tensor is split into a the fluid component $
T^{\mu \nu}_{fluid} $ [eq.(\ref{fluido})] that drives the
classical FRW metric plus small quantum fluctuations, namely $
T^{\mu \nu} = T^{\mu \nu}_{fluid}+ \delta \,T^{\mu \nu} $ the
quantum fluctuation of the matter fields induce a quantum
fluctuation in the metric (geometry) $ \delta \, G^{\mu \nu} =
\delta\,T^{\mu \nu}/M^2_{Pl} $. In the linearized approximation
the different wavelengths of the perturbations evolve
independently. After a given wavelength exits the Hubble radius,
the corresponding perturbation becomes \emph{causally
disconnected} from microphysical processes. Perturbations that
re-enter the Hubble radius at the time of photon decoupling, about
$400000$ years after the beginning of the Universe, induce small
fluctuations in the space time metric which induce fluctuations in
the matter distribution driving acoustic oscillations in the
photon-baryon fluid. At the last scattering surface, when photons
decouple from the plasma these oscillations are imprinted in the
power spectrum of the temperature anisotropies of the CMB and seed
the inhomogeneities which generate structure upon gravitational
collapse\cite{dodelson,hu}. The horizon problem, namely why the
temperature of the CMB is nearly homogeneous and isotropic (to one
part in $ 10^5 $) is solved by an inflationary epoch because the
wavelengths corresponding to the Hubble radius at the time of
recombination were \emph{inside} the Hubble radius hence in causal
contact during inflation. This mechanism is depicted in fig.
\ref{fig:inflation}. While there is a great diversity of
inflationary models, they generically predict a gaussian and
nearly scale invariant spectrum of (mostly) adiabatic scalar
(curvature) and tensor (gravitational waves) primordial
fluctuations. These generic predictions of inflationary models
make the inflationary paradigm robust. The gaussian, adiabatic and
nearly scale invariant spectrum of primordial fluctuations provide
an excellent fit to the highly precise wealth of data provided by
the Wilkinson Microwave Anisotropy Probe
(WMAP)\cite{WMAP1,kogut,spergel,peiris}. WMAP has also provided
perhaps the most striking validation of inflation as a mechanism
for generating \emph{superhorizon}  fluctuations, through the
measurement of an anticorrelation peak in the
temperature-polarization (TE) angular power spectrum at $ l \sim
150 $ corresponding to superhorizon
scales\cite{kogut,spergel,peiris}.

\subsection{Inflation and scalar field dynamics}

A simple implementation of the inflationary scenario is based on a single
scalar field, the \emph{inflaton} with a Lagrangian density \be
\mathcal{L} = a^3(t)\left[\frac{\dot{\varphi}^2}{2}-
\frac{(\nabla\varphi)^2}{2a^2(t)}-V(\varphi) \right]\ee with $
V(\varphi) $ the inflationary potential. The energy density and
the pressure for a spatially homogeneous and isotropic inflaton $
\varphi(t) $ field in the early universe [see sec.
\ref{ladrillos}] are given by~\cite{bookkolb}-\cite{liddlerev}
\be\label{enerpres} \rho = \frac{\dot{\varphi}^2}{2}+ V(\varphi)
\quad , \quad p  =\frac{\dot{\varphi}^2}{2}-V(\varphi) \; . \ee
Inflation should last at least $ N_e \gtrsim 60 $ efolds in order
to solve the entropy and horizon problems. This entails a slow
evolution   and small  temporal derivatives for the inflaton (slow
roll), namely $ \dot{\varphi}^2 \ll V(\varphi) $. This implies $
\rho = -p \simeq  V(\varphi) \simeq {\rm constant}$ as   the
equation of state leading to a de Sitter universe with  scale
factor $ a(t)= e^{Ht} , \; H = \sqrt{V(\varphi)/[3 \; M^2_{Pl}]} $
[see eq. (\ref{scalefac})]. This situation is achieved via a
variety of inflationary scenarios, old, new, chaotic, hybrid
inflation etc. (see \cite{bookkolb}-\cite{muk}  for discussions on
different models).

While inflationary dynamics is typically studied in terms of a
\emph{classical} homogeneous inflaton field, such classical field
must be understood as the expectation value of a \emph{quantum
field} in an isotropic and homogeneous quantum state. In
ref.\cite{classi,tsu} the \emph{quantum dynamics} of inflation was
studied for inflation potentials which features a discrete
symmetry breaking, $ V(\varphi) = -m^2 \varphi^2/2 +\lambda
\varphi^4 $ (new inflation) as well as unbroken symmetry
potentials $ V(\varphi)= +m^2 \varphi^2/2 +\lambda \varphi^4 $.
The  initial quantum state was taken to be  a gaussian wave
function(al) with vanishing or non-vanishing expectation value of
the field. This state evolves in time with the full inflationary
potential which features a spinodal region for $ \varphi^2 <
m^2/12\lambda $ in the broken symmetric case. Just as in the case
of Minkowski space time, there is a band of spinodally or
parametrically unstable wave vectors, within this band the
amplitude of the quantum fluctuations grows. Because of the
cosmological expansion wave vectors are redshifted into the
unstable band and when the wavelength of the unstable modes
becomes larger than the Hubble radius these modes become
\emph{classical} with a large amplitude and a frozen phase. These
long wavelength modes assemble into a classical coherent and
homogeneous condensate, which obeys the equations of motion of the
classical inflaton\cite{classi,tsu}.  This phenomenon of
classicalization and the formation of a homogeneous condensate
takes place during the \emph{first} $ 5-10 $ e-folds after the
beginning of the inflationary stage. The full quantum theory
treatment in refs.\cite{classi,tsu} show that this rapid redshift
and classicalization justifies the use of an homogeneous classical
inflaton  leading to the following robust
conclusions\cite{classi,tsu}:
\begin{itemize}
\item{ The quantum fluctuations of the inflaton are of two
different kinds: (a) Large amplitude quantum fluctuations generated
at the begining of inflation through spinodal or parametric
resonance depending on the inflationary scenario chosen. They have
comoving wavenumbers in the range of $ e^{N_T-60} \; 10^{13}
\mbox{GeV} \lesssim  k \lesssim e^{N_T-60} \; 10^{15}$ GeV and they
become superhorizon a few efolds after the begining of inflation.
The phase of these long-wavelength fluctuations freeze out and their
amplitude grows thereby effectively forming a homogeneous
\emph{classical} condensate. The study of more general initial
quantum states featuring highly excited distribution of quanta lead
to similar conclusions\cite{tsu}: during the first few e-folds of
evolution the rapid redshift results in a classicalization of
long-wavelength fluctuations and the emergence of a homogeneous
coherent condensate that obeys the \emph{classical equations of
motion} in terms of the inflaton potential. (b) Cosmological scales
relevant for the observations \emph{today} between $ 1\,\textrm{Mpc}
$ and the Hubble radius  had first crossed (exited) the Hubble
radius about $ \sim 50 $ e-folds before the end of inflation within
a rather narrow window of about $ 8 $ e-folds\cite{bookkolb}. These
correspond to small fluctuations of high comoving wavevectors in the
range\cite{dodelson} $e^{N_T-60} \; 10^{16} \, GeV < k < e^{N_T-60}
\; 10^{20} \, GeV $ where $ N_T \geq 60 $ is the total number of
efolds.}

\item{ During the rest of the inflationary stage the dynamics is described
by this
classical homogeneous condensate that obeys the classical equations
of motion with the inflaton potential. Thus inflation even if
triggered by an initial quantum state or density matrix of the
quantum field, is effectively described in terms of an homogeneous
scalar condensate. }

\end{itemize}

The body of results emerging from these studies provide a
justification for the description of inflationary dynamics in
terms of  \emph{classical} homogeneous scalar field. The conclusion
is that  after a few initial e-folds during which the unstable
wavevectors (a) are redshifted well beyond the Hubble radius, all
that remains for the ensuing dynamics is an homogeneous
condensate, plus small fluctuations corresponding to modes (b).

\subsection{Slow roll inflation}\label{sri}
Amongst the wide variety of inflationary scenarios, \emph{slow roll}
inflation\cite{barrow,stewlyth} provides a simple and generic
description of inflation consistent with the WMAP data\cite{peiris}.
In this scenario, inflation is driven by the dynamics of the
\emph{classical} coherent and homogeneous condensate of the inflaton
field $ \varphi(t) $, which obeys the classical equation of motion
\be\label{eqno} {\ddot \varphi} + 3 \, H \, {\dot \varphi} +
V'(\varphi) = 0 \; . \ee \noindent Its energy density is given by $
\rho(t) = \frac{{\dot \varphi}^2}{2} + V(\varphi) $. The basic
premise of slow roll inflation is that the potential  is fairly flat
during the inflationary stage. This flatness not only leads to a
slowly varying inflaton and Hubble parameter, hence ensuring a
sufficient number of e-folds, but also provides  an explanation for
the gaussianity of the fluctuations as well as for the (almost)
scale invariance of their power spectrum. A flat potential precludes
large non-linearities in the dynamics of the \emph{fluctuations} of
the inflaton, which is therefore determined by a gaussian free field
theory. Furthermore, because the potential is flat the inflaton is
almost massless (compared with the scale of $ V^{\frac14} $), and
modes cross the horizon with an amplitude proportional to the Hubble
parameter. This fact combined with a slowly varying Hubble parameter
yields an almost scale invariant primordial power spectrum.
Departures from scale invariance and gaussianity are determined by
the departures from flatness of the potential, namely by derivatives
of the potential with respect to the inflaton field. These
derivatives are small and can be combined into a hierarchy of
dimensionless slow roll parameters\cite{barrow} that allow an
assessment of the \emph{corrections} to the basic predictions of
gaussianity and scale invariance\cite{peiris}.  The slow roll
expansion introduces a hierarchy of small dimensionless quantities
that are determined by the derivatives of the
potential\cite{barrow,stewlyth}: \be \epsilon_V  =
\frac{M^2_{Pl}}{2} \; \left[\frac{V^{'}(\varphi)}{V(\varphi)}
\right]^2 ~,~ \eta_V   = M^2_{Pl}  \;
\frac{V^{''}(\varphi)}{V(\varphi)}  ~,~ \xi_V = M^4_{Pl} \;
\frac{V'(\varphi) \; V^{'''}(\varphi)}{V^2(\varphi)} \; .
\label{sig} \ee \noindent The slow roll
approximation\cite{barrow,bookliddle,stewlyth} corresponds to $
\epsilon_V \sim \eta_V \ll 1 $  with the hierarchy $ \xi_V \sim
\mathcal{O}(\epsilon^2_V) $, namely $ \epsilon_V $ and $ \eta_V $
are first order in slow roll, $ \xi_V $ second order in slow roll,
etc. The slow roll variable $ \epsilon_V \ll 1 $ implies that the
evolution of the inflaton $ \varphi$ is slow, and to leading order
(neglecting the second derivatives), the equation of motion
(\ref{eqno}) becomes \be 3 H(t)~ \dot{\varphi} + V'(\varphi)=0 \ee
with \be  H^2(t) = \frac{V(\varphi)}{3M^2_{Pl}} \; . \ee \noindent
The second slow roll variable $ \eta_V \ll 1 $ implies that the
inflationary potential is nearly \emph{flat} during the inflationary
stage. During slow roll inflation the number of e-folds, from the
time $ t $ till the end of inflation, at which the value of the
inflaton is $ \varphi_e $, is given by \be N_e[\varphi(t)] =
-\frac{1}{M^2_{Pl}} \int^{\varphi_e}_{\varphi(t)}
\frac{V(\varphi)}{V'(\varphi)} \; d \varphi \; . \ee Small
fluctuations of the scalar (matter) fields around the classical
inflaton lead to small fluctuations in the space-time geometry
through Einstein's equations. There are two types   fluctuations
that are relevant: curvature perturbations and gravitational waves,
both are produced during inflation. The fluctuations of the scalar
field generate directly fluctuations in the curvature of space-time,
while the expansion of the Universe, itself determined by the
dynamics of the scalar field generates gravitational waves (see
refs.\cite{dodelson,hu}). In the linearized approximation different
wavevectors of the fluctuations evolve differently, the power
spectra per logarithmic wave vector interval for curvature ($ R $)
and gravitational wave ($ h $) fluctuations are respectively given
by \be \Delta^2_{R}(k) = \frac{k^3}{2\pi^2} \langle
|R_k|^2\rangle~~,~~ \Delta^2_h(k) = \frac{k^3}{\pi^2} \langle
|h_{+k}|^2 + |h_{\times,k}|^2 \rangle \; , \ee where the $ R_k $ are
the quanta of curvature fluctuations, $h_{+,\times}$ are the two
independent polarizations of the quanta of gravitational waves, and
the expectation value is in the vacuum state during inflation. These
power spectra feature power laws in the slow roll regime, \be
\label{espflu} \Delta^2_{R}(k) = \Delta^2_{R}(k_0)
\left(\frac{k}{k_0}\right)^{n_s-1} \quad , \quad \Delta^2_h(k)=
\Delta^2_h(k_0)\left(\frac{k}{k_0}\right)^{n_T} \; , \ee namely
during slow roll the power spectra of curvature and gravitational
wave perturbations are nearly scale invariant. The amplitude of the
curvature perturbations for wavelengths that re-entered the Hubble
radius at the last scattering surface are directly
related\cite{bookkolb,bookliddle} to the temperature anisotropies
measured by COBE and WMAP\cite{COBE,WMAP1}-\cite{peiris}, \be
\label{ampcmb} \Delta_R = \left. \frac{4}{5}\frac{\Delta
T}{T}\right|_{WMAP}=(4.67\pm 0.27)\times 10^{-5} \ee \noindent The
amplitude and the power laws are determined in slow roll by
\be\label{ampo} \Delta^2_{R}(k_0) = \frac{V(\varphi)}{24 \, \pi^2 \;
M^4_{Pl}\,\epsilon_V} ~~;~~
\frac{\Delta^2_{h}(k_0)}{\Delta^2_{R}(k_0)} = 16 \; \epsilon_V~~;~~
n_s-1 = - 6 \; \epsilon_V + 2\; \eta_V ~~;~~ n_T = -2\; \epsilon_V\;
. \ee \noindent The WMAP values are  for $ k_0 = 0.002/\textrm{Mpc}
$, which yields an upper bound for the scale of the inflationary
potential during slow roll inflation $ V^{1/4} < 3.3 \times
10^{16}\textrm{GeV} $,   \emph{suggesting} a connection between the
scale of grand unification and that of inflation. While inflation
may not be related to phase transitions in the early Universe, the
nearly scale invariant spectrum of gaussian fluctuations
\emph{suggests} a connection with a critical theory. 
Ref. \cite{hectornorma} provided an effective
field theory description of inflation \emph{akin to the Landau-Ginzburg}
description of critical phenomena. Slow roll dynamics can be organized elegantly
in a systematic expansion in powers of $ 1/N_e $ within
a Landau-Ginzburg effective field theory. This is achieved by
introducing a dimensionless inflaton field and rescaled potential as
\be \chi = \frac{\varphi}{\sqrt{N_e} M_{Pl}}\ee and \be w(\chi) =
\frac{V(\varphi)}{N_e \; M^4} \ee where the value of $ \Delta_R $
[eq.(\ref{ampcmb})] and eq.(\ref{ampo}) fix the scale of inflation $
M $ to be $ M \sim 10^{16}$ GeV and $ N_e \sim 50 $ is the number of
e-folds. To emphasize that the slow roll approximation implies a
slow time evolution it is also convenient to introduce a
\emph{stretched}  (slow) dimensionless time variable $ \tau $ and a
rescaled dimensionless Hubble parameter $ \hat h $ as follows
\be\label{time} 
t = \sqrt{N_e} \; \frac{M_{Pl}}{M^2} \; \tau \qquad
; \qquad H = \sqrt{N_e} \;  \frac{M^2}{M_{Pl}}\; \hat{h} 
\ee
\noindent the Einstein-Friedman equation now reads \be \label{efa}
\hat{h}^2(\tau) = \frac13\left[\frac1{2\;N_e} \left(\frac{d\chi}{d
\tau}\right)^2 + w(\chi) \right] \ee \noindent and the evolution
equation for the inflaton field $ \chi $ is given by \be
\label{eqnmot} \frac1{N_e} \; \frac{d^2 \chi}{d \tau^2} + 3 \;
\hat{h} \; \frac{d\chi}{d \tau} + w'(\chi) = 0 \ee The slow-roll
approximation follows by neglecting the $ 1/{N_e} $ terms in
eqs.(\ref{efa}) and (\ref{eqnmot}). Both $ w(\chi) $ and $
\hat{h}(\tau) $ are of order $ N^0_e $ for large $ N_e $. Both
equations make manifest the slow roll expansion as a systematic
expansion in $ 1/N_e $\cite{hectornorma,mangano}. Slow roll
dynamics, the inflationary scale and all of the observational
phenomenology is reproduced with $ \chi \sim \mathcal{O}(1), \;
w(\chi) \sim \mathcal{O}(1) $ during inflation.

\begin{figure}[h]
\begin{center}
\includegraphics[height=2.3in,width=2.2 in,keepaspectratio=true]{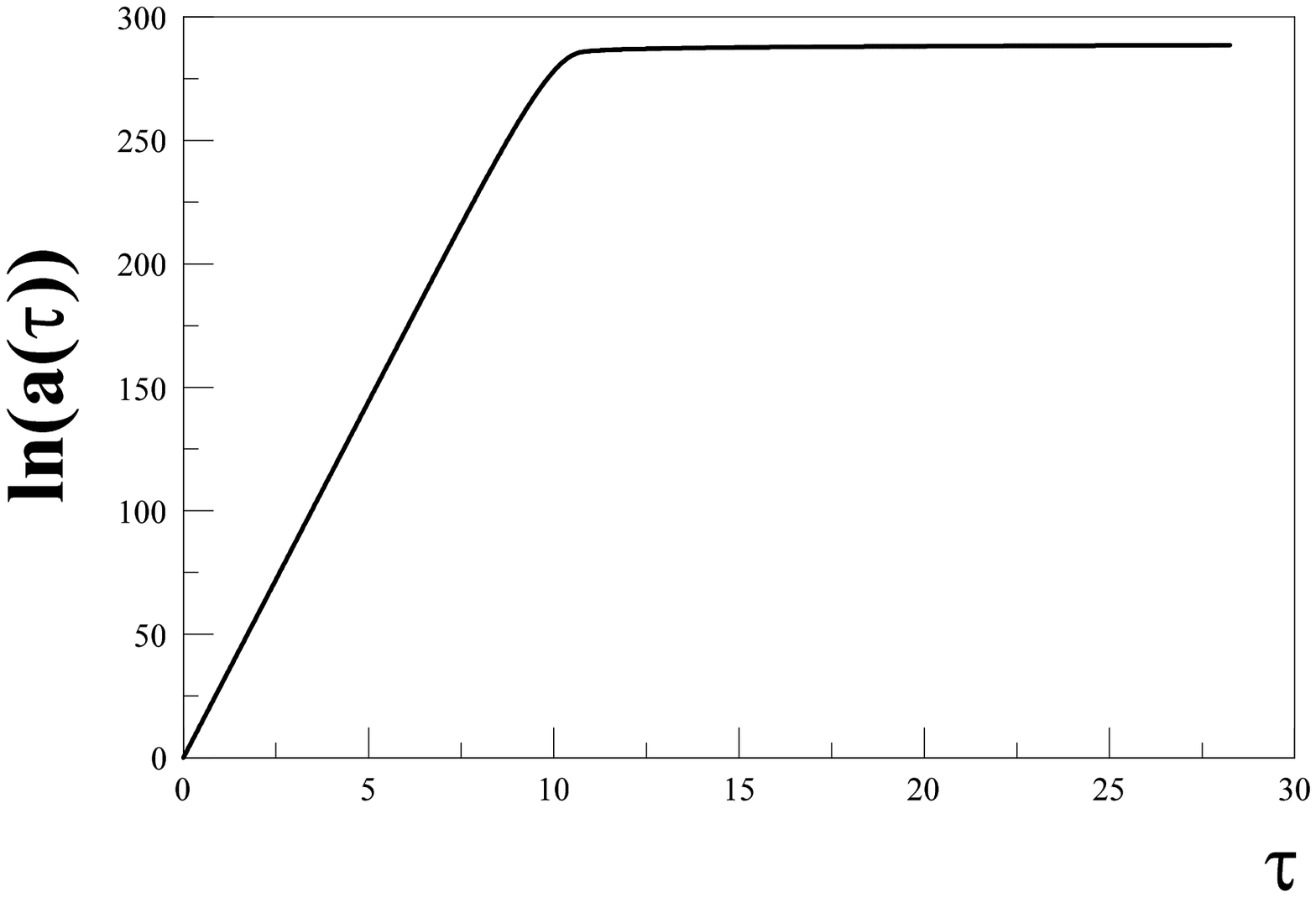}
\includegraphics[height=2.2 in,width=2.2 in,keepaspectratio=true]{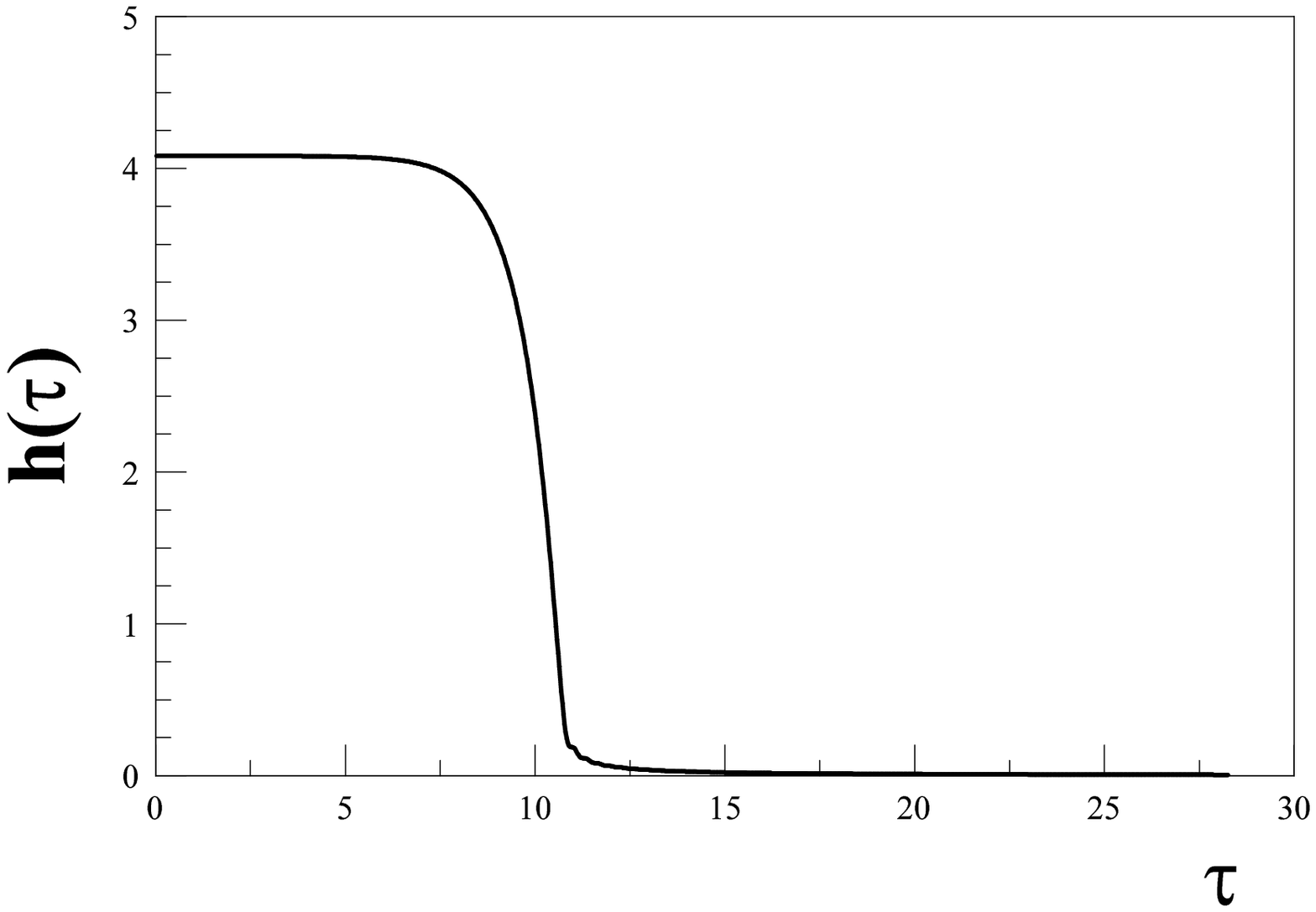}
\includegraphics[height=2.3in,width=2.2 in,keepaspectratio=true]{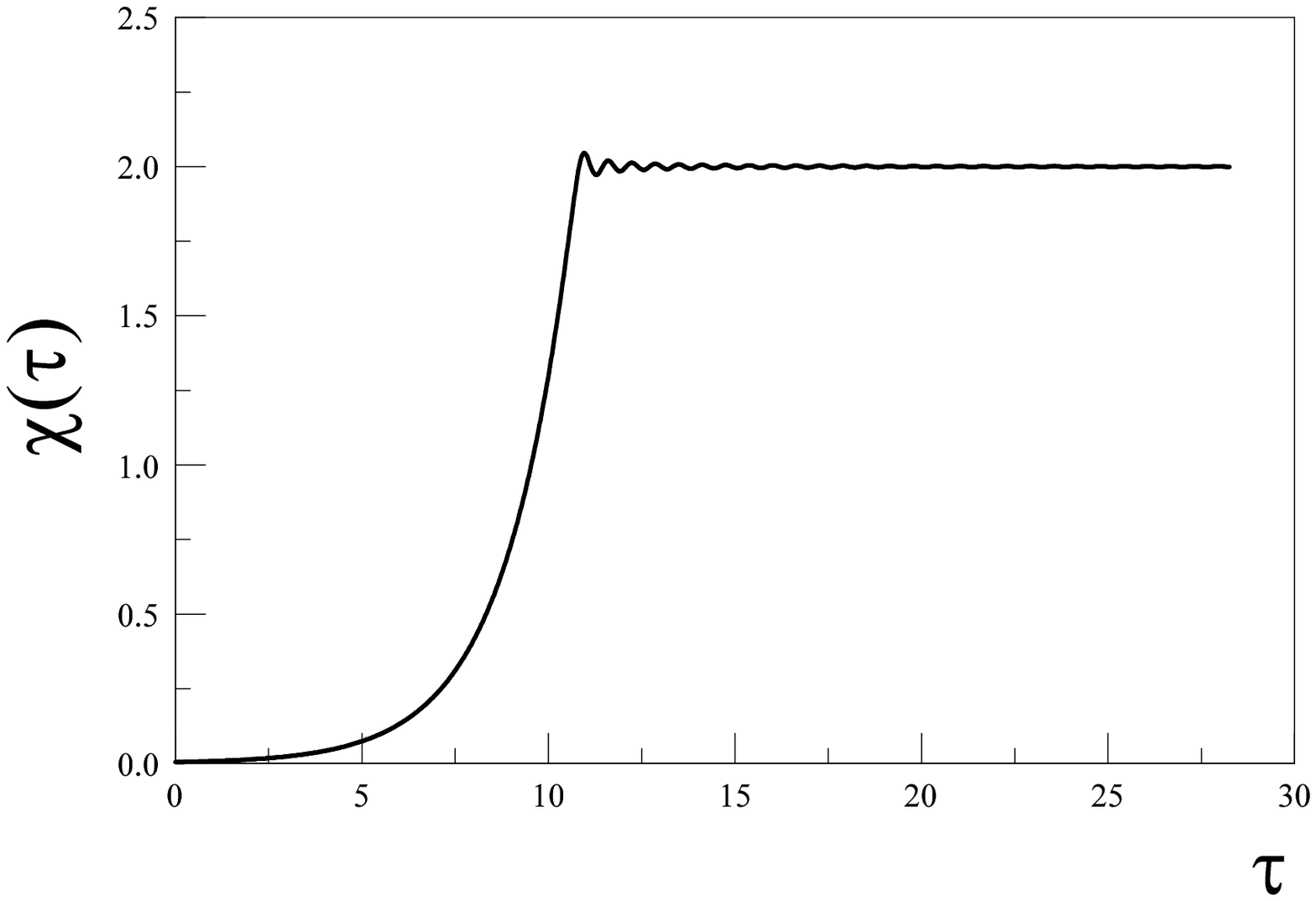}
\caption{Left panel: $ \ln a(\tau) $ which is the number of e-folds
as a function of time. Middle panel $\hat{h}(\tau)$. Right panel:
$ \chi(\tau) $ for $ N_e=60 $. } \label{fig:infla}
\end{center}
\end{figure}

As a pedagogical example for slow roll inflation we display the main
features of the inflationary dynamics for $ w(\chi) =
(\chi^2-4)^2/16 $ in fig.\ref{fig:infla}. These figures reveal
clearly that inflation ends at $ \tau \sim 10 $ when $ \hat{h}(\tau)
$ begins a rapid decrease   when the field is no longer slowly
coasting along near the maximum of the potential $ \chi \sim 0 $ but
rapidly approaching its equilibrium minimum. The left panel $ \ln
a(\tau) $ depicts the number of e-folds as a function of time [by
definition $ N_e(t)=\ln a(t) $], wavelengths of cosmological
relevance today have crossed the Hubble radius during the last $
\sim 50 $ e-folds before the \emph{end} of inflation within a narrow
window $ \Delta N_e \sim 8 $, corresponding to a very small interval
$ \Delta \tau \sim 1 $. A successful inflationary scenario requires
at least $ N_e\sim 60 $ to solve the horizon and entropy problems of
the standard hot Big Bang\cite{bookkolb} which occur in an interval
$ \Delta \tau \sim 3 $ prior to the end of inflation and during
which the dimensionless field changes about $ \Delta \chi \sim 1 $.
These are general aspects of a wide range of inflationary
scenarii\cite{hectornorma}.
We consider here a translationally and rotationally invariant
cosmology where the only source of inhomogeneities are (small)
quantum fluctuations. Indeed, inhomogeneities cannot be excluded
at the beginning of inflation but the redshift of scales during
inflation by at least $ e^{60} \sim 10^{26} $ effectively erases
all eventual initial inhomogeneities.

Inflation is now an established part of cosmology with several
important aspects, such as the superhorizon origin of density
perturbations, having been spectacularly validated by
WMAP\cite{peiris}. Simple but phenomenologically accurate
descriptions of inflation invoke an \emph{effective} field theory
for a homogeneous scalar field, akin to the Landau-Ginzburg
description of critical phenomena described in section \ref{crit}
above\cite{hectornorma}. Analysis of the WMAP data puts very
strongly pressure   on  the simple monomial $ w(\chi) = G_4 \;
\chi^4 $ at the $ 3 \sigma $ level\cite{peiris,afterWMAP1} and
forthcoming observations of the CMB as the Planck satellite and
others bear the promise of yielding precise information on the
inflaton effective potential corresponding to the stage of inflation
during which wavelengths of cosmological relevance today first
crossed the Hubble radius during inflation (see
\cite{peiris,afterWMAP1,hectornorma} for discussions of the lessons
learned so far).

\section{The Electroweak scale: phase transitions and
baryogenesis}\label{sec:EWPT}

There is a large body of observational evidence that suggests that
the there is more matter than antimatter in the Universe up to
scales of the order of the Hubble radius\cite{dolgovmat,trodden}.
The origin of this \emph{baryon asymmetry} is one of the deep
mysteries in particle physics and cosmology. The value of this
asymmetry is quantified by the ratio \be \eta =
\frac{n_b-n_{\bar{b}}}{n_{\gamma}} \ee where $n_b$ ($n_{\bar{b}}$)
is the baryon (antibaryon) density and $n_{\gamma}$ is the photon
density. This is the \emph{only} free input parameter that enters in
nucleosynthesis calculations of the primordial abundance of light
elements\cite{steigman,turnerBBN}. The agreement between the WMAP
results\cite{WMAP1,kogut,spergel,peiris} and the most recent
analysis of the primordial deuterium abundance\cite{kneller} yields
\be \eta = (6.1 \pm 0.3)\times 10^{-10} \label{bau}\ee What is the
origin of this ratio? namely what is the microscopic mechanism
responsible for \emph{baryogenesis}?. The necessary conditions for
successful baryogenesis were first identified and outlined by
Sakharov\cite{sakharov}: \begin{itemize}\item{Baryon number
violation. }\item{C and CP violation: unless these symmetries are
violated in any process in which baryons are created and
annihilated, the rates for baryon  production equals that of the
reverse reaction and no net baryon can be generated by these
processes.}\item{ Departure from equilibrium: in equilibrium the
density matrix only depends on the Hamiltonian (and simultaneously
commuting operators) which in all microscopic theories is invariant
under CPT. Since the baryon number operator is \emph{odd} under CPT,
the expectation value of the baryon operator must vanish if the
density matrix is that of equilibrium\cite{trodden}. }
\end{itemize}

Notably, the standard model of particle physics has the main
ingredients for baryogenesis\cite{trodden,buch}: \begin{itemize}
\item{{\bf $B+L$ violation in the standard model}: Since the weak
interactions only involve the left handed quark  (baryon) and
lepton currents there is a quantum mechanical anomaly in their
conservation laws\cite{thooft}:
\be
\partial_{\mu}J^{\mu,B}=\partial_\mu J^{\mu,l} =
\frac{N_f}{32 \; \pi^2}\left[-g^2 \;
W^a_{\mu\nu}\widetilde{W}^{a,\mu\nu}+ g'^2 \; B_{\mu
\nu}\widetilde{B}^{\mu \nu}\right] \ee where $ N_f $ is the number
of generations, $W,B$ are the field strength tensors for the $ SU(2)
$ and $ U(1) $ gauge fields and $ \widetilde{W}^{\mu \nu}=
\epsilon^{\mu \nu \alpha \beta} \; W_{\alpha \beta} $ and similarly
for $\widetilde{B}$. As a consequence of this anomaly the change in
the baryon (and lepton) number is related to the change in the
topological charge of the gauge field \be \Delta B = N_f \, \Delta
N_{CS}~~;~~ N_{CS}= \frac{g^2}{32\pi^2} \int d^3 x \;
\epsilon^{ijk}\,\textrm{Tr}\left[W_i \;  \partial_j W_k+
\frac{2\,i\,g}{3} \; W_i \; W_j \; W_k \right] \ee where $ W_i $ is
the $ SU(2) $ gauge field and $ N_{CS} $ is an \emph{integer} that
characterizes the topological structure of the gauge field
configuration. Kuzmin, Rubakov and Shaposhnikov\cite{kuzi} noticed
that in the high temperature medium that prevailed in the early
Universe there are \emph{non-perturbative} field configurations,
called \emph{sphalerons} that induce transitions between gauge field
configurations with different values of $N_{CS}$. The sphalerons
lead to $B+L$ violating processes with  transition rate per unit
volume estimated to be\cite{gamsphal} $\Gamma_{sph} \sim \alpha^5 \;
T^4  \; \ln(1/\alpha) $. This estimate for the transition rate
suggests that sphaleron processes are in thermal equilibrium for $
100\,\textrm{GeV}<T<10^{12}$ GeV. \item{{\bf CP violation in the
standard model} Because only left handed quarks and leptons couple
to the charged and neutral vector bosons that mediate the weak
interactions, the standard model violates P maximally. However, CP
violation is much more subtle and is the result of CP violating
phases in the complex Cabibbo Kobayashi Maskawa mass matrix for
quarks resulting from complex Yukawa couplings to the Higgs field.
For $N_f$ generations of quarks and leptons there are $
(N_f-1)(N_f-2)/2 $ independent phases in the CKM mass matrix, and a
non-zero value for \emph{any} of these phases implies CP violation.
For  $ N_f \geq 3$ there is at least one CP violating phase, hence
the standard model, with $N_f=3$ does indeed have the possibility of
CP violation. Experimentally CP violation is observed in the   $
K_0\,\bar{K}_0 $ and $B\,\bar{B}$ systems. }
\item{{\bf Non-equilibrium:} as discussed above weak interaction
processes are in LTE down to $T \sim 1\,\mathrm{MeV}$, therefore the
only possibility for non-equilibrium is through a phase transition.
Since the expansion rate of the Universe is much smaller  than the
weak interaction rate, it is very likely that a second order phase
transition at the electroweak scale $T \sim 100 \,\textrm{GeV}$
would occur in LTE, hence departure from equilibrium requires a
strong first order phase transition. An estimate of the possibility
and strength of a first order electroweak phase transition in the
standard model is gleaned from a one loop calculation of the
effective potential in the $SU(2)+$ Higgs model (i.e, neglecting the
$U(1)$ gauge group)\cite{trodden} \be V^{(1)}(\varphi;T) =
\left(\frac{3g^2}{32} + \frac{\lambda}{4}+
\frac{m^2_t}{4v^2_0}\right)(T^2-T^2_*) \;
\varphi^2-\frac{3g^2}{32\pi} \,T \,\varphi^3 +\frac{\lambda}{4} \;
\varphi^4, \ee  where $ \varphi= \sqrt{\phi^*\phi}, \; v_0 $ is the
Higgs vacuum expectation value at $ T=0 $ and $ m_t $ is the top
quark mass. The second term proportional to $ \varphi^3 $ arises
from the gauge field contribution and is responsible for a first
order phase transition. Including the $ U(1) $ gauge group changes
the above only quantitatively. The one loop effective potential as a
function of $\varphi$ has a typical shape as in the right panel in
fig. \ref{fig:Veff} and features a global and a local minimum
for\cite{trodden} \be T< T_c =
m_H\left(\frac{3g^2}{8}+\lambda-\frac{9g^6}{256\pi^2}+\frac{m^2_t}{v^2_0}
\right)^{-\frac{1}{2}}\ee where $m_H$ is the Higgs mass. A measure
of the strength the phase transition is the ratio $\Delta
\varphi(T_c)/T_c$\cite{shapo} where $\Delta \varphi(T_c)$ is the
jump in the order parameter between the two (degenerate) minima at
$T_c$. Successful baryogenesis requires this ratio to be
$>1$\cite{shapo}. One important aspect that emerges from this simple
analysis is that \emph{the higher the Higgs mass the weaker the
phase transition}. A study of higher orders\cite{buchfod} reveals
that perturbation theory does not give reliable information about
the electroweak phase transition for Higgs masses beyond $ \sim 70$
GeV. Lattice studies have shown\cite{jansen} that the ratio $ \Delta
\varphi(T_c)/T_c <1$ for $ m_H > 45\,\textrm{GeV} $. \emph{If} the
standard model features a strong first order phase transition this
phase transition occurs via the formation of nucleated bubbles just
as described in section \ref{noneqPT}. Finally, a mechanism for
baryogenesis involves transport of baryons through the bubble walls
as the nucleated bubbles grow and percolate filling the space with
the globally stable phase\cite{kaplan}: CP violating interactions of
quarks and leptons in the thermal medium with the bubble walls leads
to an excess of left handed quarks, which sphaleron transitions
convert into a net baryon asymmetry. For a discussion of these
mechanisms see\cite{trodden} and references therein. }}
\end{itemize}
{\bf Caveats:} While the standard model features the main
ingredients for successful baryogenesis,  a substantial body of work
has revealed that for a Higgs mass larger than about $72$ GeV there
is no first order phase transition but a smooth crossover in the
standard model\cite{laine,csikor}. The current LEP bound for the
standard model Higgs mass $m_H \gtrsim 115$ GeV,   all but rules out
the possibility of a strong first order phase transition and
suggests a smooth crossover from the broken symmetry into the
symmetric phase in the standard model. While it has become clear
that the LEP bound on the Higgs mass precludes baryogenesis in the
standard model, some supersymmetric extensions of the standard model
with a \emph{stop} lighter than the top may be able to explain the
observed baryon asymmetry\cite{laine,csikor,carena}.  An alternative
scenario for baryogenesis proposes that a primordial asymmetry
between \emph{leptons} and antileptons or \emph{leptogenesis} is
responsible for generating the baryon asymmetry\cite{fuku,buch}. The
leptogenesis proposal depends on the details of the origin of
neutrino masses and remains a subject of ongoing study.

\section{The QCD phase transition in the early Universe}

One of the most spectacular epochs in the early Universe is the QCD
transition, when quarks and gluons become confined in hadrons. In
the early Universe the baryon asymmetry is very small [see
eq.~(\ref{bau})], and at RHIC (and soon at LHC) it is expected that
the mid-rapidity region for central collisions is also baryon free
(see the discussion in section \ref{urhic} below). For the purpuse
of cosmology, one would wish to directly measure $T_{QCD}$ and the
equation of state from experiments with relativistic heavy ions, but
it turns out not to be a simple task. The problem arises from
extremely different time scales: $10^{-5}$ s in the cosmological QCD
transition (very close to thermal equilibrium), but only $10^{-23}$
s in the laboratory (out-of-equilibrium effects may be important).

Since QCD is asymptotically free, it is expected that at high
temperature a perturbative evaluation of the equation of state in
terms of a weakly interacting gas of quark and gluons should be
reliable. However near the hadronization phase transition the nature
of the degrees of freedom changes from quarks and gluons to hadrons
and QCD  becomes non-perturbative. The only known first principle
method to study QCD non-perturbatively in a wide temperature range
is lattice gauge theory (LGT). Over the last several years there has
been steady progress in the study of the QCD phase diagram with and
without chemical potential including light and heavy quarks
\cite{karsch2,Laermann:2003cv,Katz:2005br}. As explained above, the
problems to incorporate a finite chemical potential in LGT (see
\cite{Laermann:2003cv}) is of no concern  to cosmology and the
mid-rapidity region of central collisions at RHIC and LHC, because
the relevant baryochemical potentials $\mu_B/T \ll 1$.

For an extended review of the cosmological consequences of the
QCD transition and the physics of the first second of the Universe
see \cite{Schwarz:2003du}.

\subsection{ The QCD transition and equation of state:}
\subsubsection{Lattice gauge theory results}\label{sec:lgt}

It has been established that lattice QCD without dynamical quarks
(quenched approximation) exhibits a thermal first-order phase
transition \cite{Fukugita:1989yb} at a critical temperature of
$T_\star \approx 270$ MeV \cite{Laermann:2003cv}. This is also in
agreement with the expectation of a first-order phase transition
from the simplest bag model \cite{bagEOS,teaney,shuryak} (see
below). Unfortunately, the situation is not  clear for dynamical
quarks and is especially unclear for the physical values of the
up, down and strange quark masses.

For dynamical quarks, lattice QCD calculations provide a range of
estimates for $T_c$. In the case of two-flavour QCD $T_c \approx
175$~MeV \cite{AliKhan:2000iz,Karsch:2000kv}, whereas for
three-flavour QCD $T_c \approx 155$ MeV \cite{Karsch:2000kv},
almost independent of the quark mass. For the most interesting
case of two light quark flavors (up and down) and the more massive
strange quark, a value of $T_c \approx 170$ MeV has been obtained
recently, both from standard \cite{Aoki:1998gi} and improved
\cite{Bernard:2004je} staggered quarks. Accordingly, in the
discussion that  follows  we adopt a transition temperature $T_c =
170$~MeV, keeping in mind that the systematic uncertainty is
probably of the order $10$ MeV.

The order of the phase transition and the value of $T_c$ is still
under investigation. For massless quarks the theoretical
expectation is a second order transition for two quark flavors and
a first-order transition for three and more quark flavors
\cite{Pisarski:1983ms}. On the lattice, for two light quarks the
results are inconclusive. The predicted universality class is not
confirmed so far (see discussions in \cite{Laermann:2003cv}); most
recent studies even claim to find hints for a first-order
transition \cite{D'Elia:2005sy}. For three flavours close to the
chiral limit the lattice results clearly indicate that the phase
transition is of first order (see \cite{Laermann:2003cv}), as
expected from theory. Some older simulations suggested that this
holds true for the physical case as well \cite{Iwasaki:1995yj}.
The latter result was obtained using the Wilson quark action,
whereas results with standard \cite{Aoki:1998gi} and improved
\cite{Bernard:2004je} staggered quarks indicate a crossover for
the physical quark masses.

\begin{figure}[h]
\begin{center}
\epsfig{file=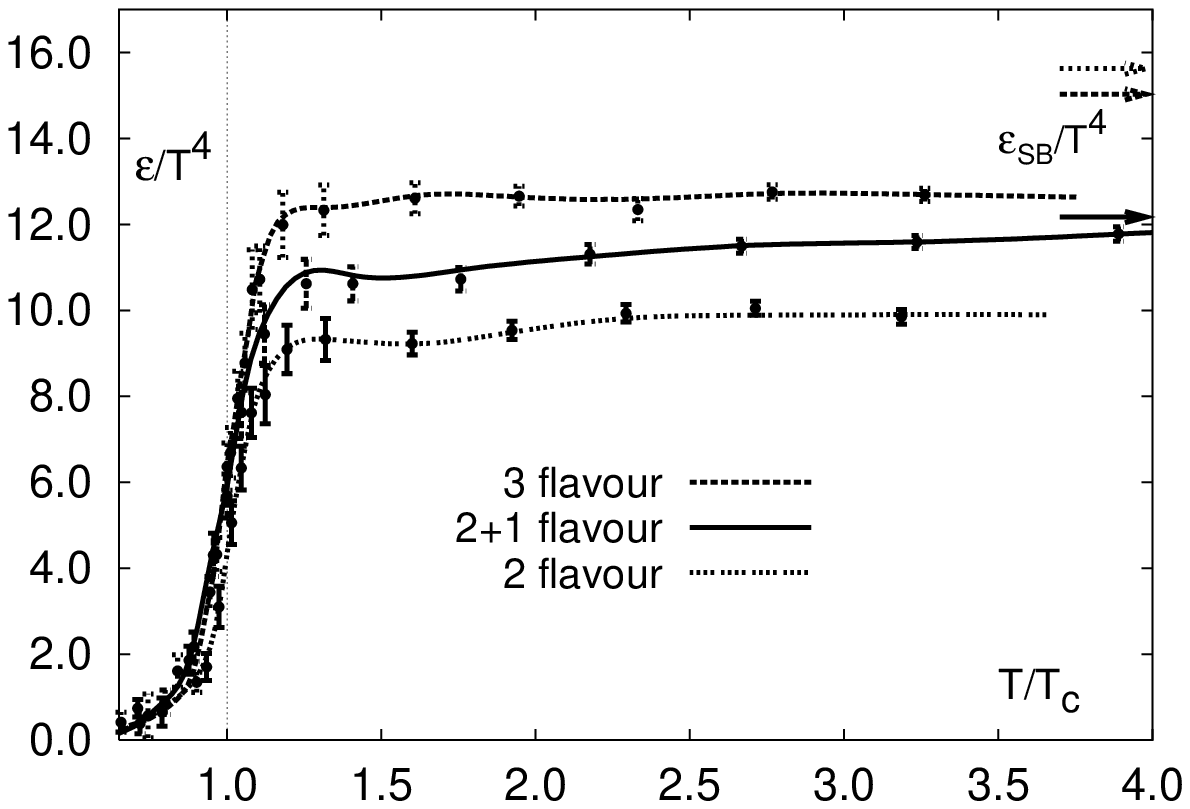,height=2in,width=2in}
\epsfig{file=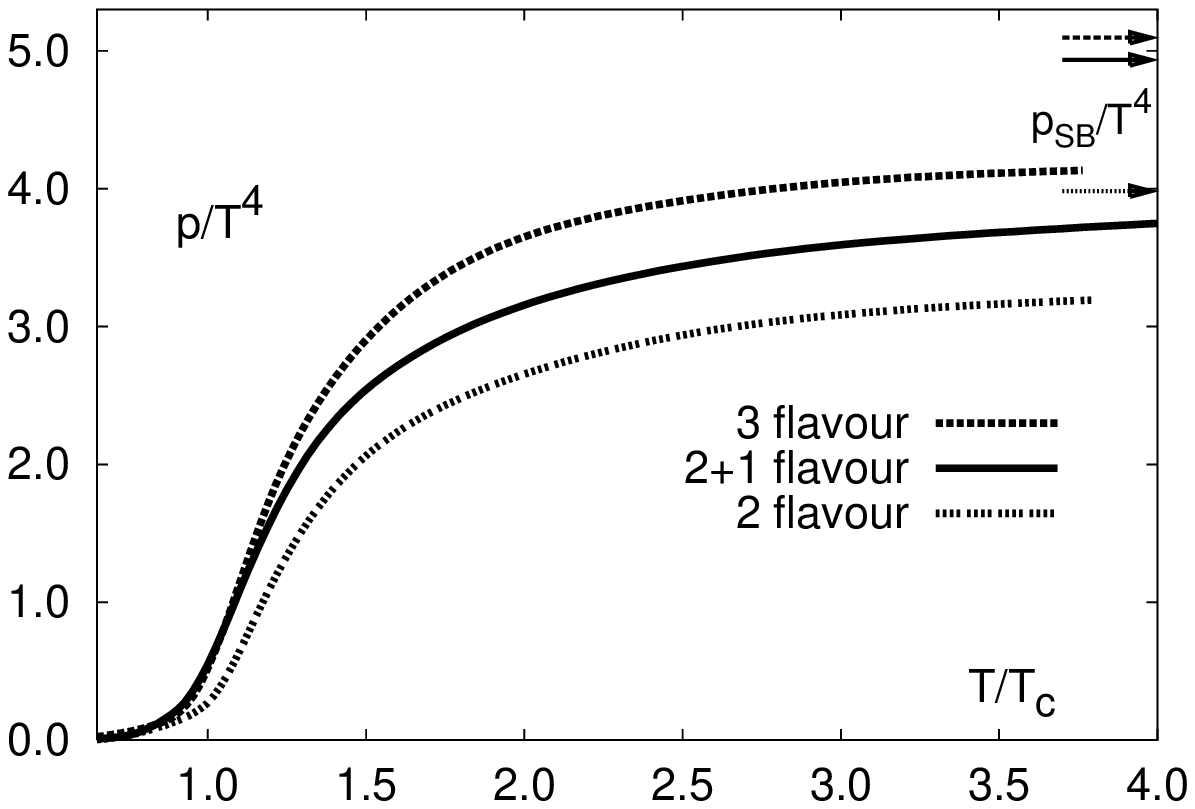,height=2in,width=2in}
\epsfig{file=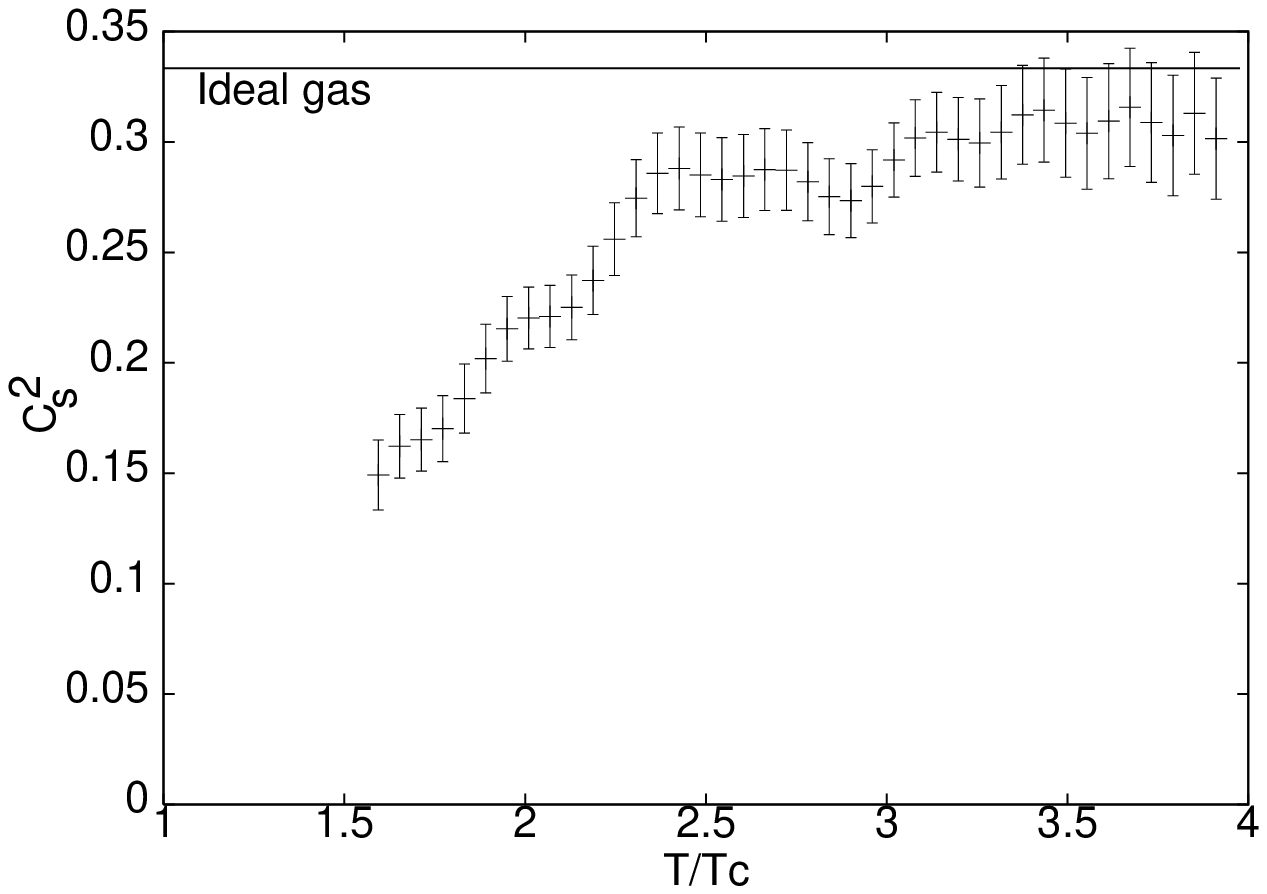,height=2in,width=2in}
 \caption{$\varepsilon/T^4$ and $p/T^4$, vs $T/T_c$ for a
 $16^3\times 4$ lattice  From~\cite{karsch2}. The arrows in the two
 left figures mark the Stephan-Boltzmann result.
  Right figure: speed of sound $C^2_s =  \frac{dP}{d\varepsilon} $.
From~\cite{bernard} } \label{fig:lgt}
\end{center}
\end{figure}

A microscopic description of phase transitions in QCD requires also a
reliable assessment of the equation of state (EoS). Furthermore,
as it will be described below, a hydrodynamic description of the
space-time evolution in relativistic heavy ion collisions also
requires knowledge of the equation of state at high temperature.
The consensus that seems to be emerging is that for the physical
masses of two light (up and down) and one heavier (strange) quark there
is a \emph{sharp crossover} between a high temperature gas of quark and gluon
quasiparticles and a low temperature hadronic phase \emph{without}
any thermodynamic discontinuities. This is displayed in fig.~\ref{fig:lgt}
which summarize results from LGT for the energy
density and pressure (both divided by $ T^4 $ to compare to a free
gas of massless quarks and or gluons) as a function of $ T/T_c $
\cite{karsch2}.
The most recent simulations of the equation of state are reviewed
in \cite{Katz:2005br}.

These figures clearly reveal a sharp decrease in the energy density
and pressure at $ T=T_c $, the value of the energy density is
$\varepsilon_c \sim 6 \;T^4_c$ \cite{karsch2} thus predicting an
energy density $\varepsilon_c \sim 0.7\,\textrm{GeV}/\textrm{fm}^3$.
Furthermore the high temperature behavior is {\em not quite} given
by the Stephan-Boltzmann law, see fig. (\ref{fig:lgt}), suggesting
that even at large temperatures the plasma is not described by free
quarks and gluons up to temperatures $ T \sim 4.0 \; T_c \sim 700
~{\rm Mev}$.

In the hydrodynamic limit, the speed of sound
\begin{equation}
c_s \equiv \left(\frac{\partial p}{\partial \epsilon} \right)_S^{1/2}
\end{equation}
is a quantity of central interest. A strong decrease in the speed
of sound, already above $T_c$, has been observed in lattice QCD,
see \cite{bernard,Aoki:2005vt} for the most recent data.
Consistent with previous results from quenched QCD and two-flavor
QCD, $c_s^2(T_c) \approx 0.1$ when approaching the critical
temperature from above (in the QGP). The right panel of
fig.~\ref{fig:lgt} displays the speed of sound $c^2_s$
\cite{bernard} clearly showing a dramatic decrease for $T \lesssim
2 \; T_c$ and approaching $1/3$ for $T \gg T_c$ in agreement with
an ultrarelativistic gas of quarks and gluons.

\subsubsection{The bag EOS and a first order phase transition}

\begin{figure}[h]
\begin{center}
\epsfig{file=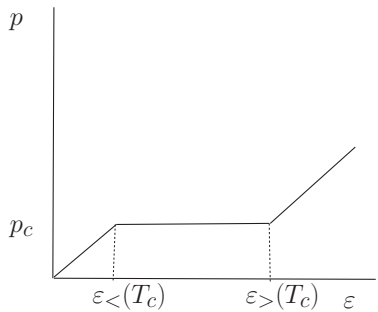,height=2 in,width=2
in,keepaspectratio=true} \caption{Bag (EoS) for a massless pion
gas for $T<T_c$ and quarks and gluons in a bag for $T>T_c$. }
\label{fig:bageos}
\end{center}
\end{figure}

If the transition from $ T>T_c $ to $ T<T_c $ is continuous but
sharp as evidenced by the lattice data in fig. \ref{fig:lgt} the
behavior may not be too different from an actual transition which
may be modelled by a simpler (EoS) which would allow an analytic
treatment and to be included in a hydrodynamic description. The
bag model\cite{bagEOS} provides a semi-phenomenological
description of an (EoS) that features a  quark-hadron transition.
The simplest version considers the thermodynamics in two different
regions: for $ T>T_c $ a gas of massless quarks and gluons and
confinement is accounted for in terms of a bag constant
$B$\cite{bagEOS}, for $ T<T_c $ a gas of free massless pions (in
the chiral limit). At $ T=T_c $ quarks, gluons and pions coexist
in equilibrium at   constant pressure and temperature $p_c,T_c$.
In this model the pressure and energy density for vanishing
chemical potential are given by \bea && p_> = g_> \;
\frac{\pi^2}{90} \; T^4 - B ~~;~~ \varepsilon_> = g_> \;
\frac{\pi^2}{30} \; T^4 + B~~;~~\textrm{for} \; \,T>T_c  \;  ,
\label{QGP}\\
&& p_< = g_<  \; \frac{\pi^2}{90} \; T^4   ~~;~~ \varepsilon_< =
g_< \; \frac{\pi^2}{30} \; T^4  ~~;~~\textrm{for}\,T<T_c  \;
,\label{HG} \eea where $ B $ is the bag constant with a typical
value $ B^{\frac{1}{4}}\sim 200$ MeV\cite{bagEOS}. For a
description of a quark-hadron transition $ g_>= 37,47.5 $ for $
N_f=2,3 $ respectively and $ g_<=3 $ for a gas massless pions. The
two regions are joined together by a \emph{flat} coexistence curve
with constant $ p_c, \; T_c $ according to a Maxwell construction
(see sec.\ref{sec:liqgas}). The value of $ T_c $ is determined by
this coexistence condition and is given by \be T^4_c = \frac{90 \;
B}{(g_>-g_<) \; \pi^2} \; \;  . \label{Tbag} \ee For $ N_f=2 $ and
three massless pions $ T_c \sim 145$ Mev which is not too far from
the lattice results $ T_c \sim 170$ MeV. The flat coexistence
region describes a \emph{mixed} phase in which a quark-gluon
plasma is in coexistence with a hadron gas with a vanishing
isothermal sound speed, just like in the liquid-gas transition.
The proportion of each phase in coexistence is determined by the
lever-rule described in section \ref{sec:liqgas}.  The bag
equation of state is therefore given by \be p=
\frac{\varepsilon}{3}-\frac43  \; B ~~;~~\textrm{for}~~
\varepsilon > \varepsilon_>(T_c) \quad , \quad p =
\frac{\varepsilon}{3} ~~;~~\textrm{for}~~  \varepsilon <
\varepsilon_<(T_c) \; , \label{bagi} \ee with a flat coexistence
line for $ \varepsilon_<(T_c) \leq p \leq \varepsilon_>(T_c) $.
This (EoS) is displayed in fig. \ref{fig:bageos}, compare to the
liquid-gas equation of state in the left panel of fig.
\ref{fig:liqgas}.

Since the Helmholtz free energy density is $ f(T)=
\varepsilon-t\,s=-p $ with $ s $ the entropy density, it is clear
that for $ T>T_c $ the free energy of the quark gluon phase is
smaller and for $ T<T_c $ the hadron phase has a smaller free
energy. Since on the coexistence line $ p=p_c, \; T=T_c $ are constant,
there is a jump in the entropy, namely a latent heat between the
QGP and the hadronic phase, hence the bag model (EoS) leads to a
\emph{first order phase transition} with a latent heat
\be\label{latheat} \textit{l} = T_c \; \Delta s=\varepsilon_>(T_c) -
\varepsilon_<(T_c) = 4 \;  B \; ,
\ee independent of the number of degrees of freedom.
In the quenched approximation on the lattice, the latent heat was determined
to be $ l \approx 1.4 \; T_c^4 $
\cite{Iwasaki:1992ik,Beinlich:1996xg}, which is
a factor of $ 5 $ smaller than expected from a quenched bag model.

More realistic versions of the bag equation of state have
been constructed in ref.\cite{teaney,shuryak}. The importance of
the bag (EoS) not only stems from its simplicity and the fact that
it models a sharp crossover as evidenced by the lattice data but
also because it is widely used in the hydrodynamic approach to
ultrarelativistic heavy ion collisions [see the discussion in
section \ref{sec:hydro} below].

Besides the latent heat, the surface tension is the crucial
parameter for the nucleation of bubbles in a first-order phase
transition. The surface tension $\sigma \equiv   \left({{\rm d}
W}/{{\rm d} A}\right)_V$  is the work $ {\rm d} W $ that has to be
done per area $ {\rm d} A $ to change the phase interface at fixed
volume. The absence of surface excitations in hadronic spectra
suggests that $ \sigma^{1/3} \ll B^{1/4} $ \cite{Farhi:1984qu} in
the bag model. A self-consistent calculation of the surface
tension within the bag model shows that the surface tension
vanishes for massless quarks and gluons if no interactions besides
the bag constant (i.e.~$ \alpha_s = 0 $) are taken into account
\cite{Farhi:1984qu}. This can be cured by introducing short-range
interactions \cite{Farhi:1984qu} or by including the strange quark
mass \cite{Berger:1986ps,Madsen:1994vp}. For the surface tension
rather small values are found from quenched lattice QCD, $ \sigma
\approx 0.015  \; T_c^3$ \cite{Iwasaki:1993qu,Beinlich:1996xg}.

\subsection{Cosmological consequences of the QCD transition}

 In the radiation dominated Universe for $T>T_c$ the particle
 content is: quarks, gluons,   leptons and photons. It is a good
approximation   to treat all particles with $ m \ll 3 \;  T $ as
massless. Above the QCD transition $ g_{\rm quarks} = (7/8) 12 N_f $
(number of quark flavours) and $ g_{\rm gluons} = 8 $, below the QCD
transition $ g_{\rm hadrons} $ quickly drops to zero, i.e.~pions
disappear at $ T \sim 40 $ MeV. Therefore at the QCD epoch  $ g_> =
51.25\, (61.75)~,~   g_< = 17.25 (21.25)$ without (with) strange
quarks and without (with) kaons, respectively. At the QCD transition
the Hubble radius is about $10$ km, corresponding to scales of $ 1 $
pc or $ 3 $ light-years today. The mass inside a Hubble volume is $
(4\pi/3) \; \epsilon(T_c) \; R_H^3 \sim 1 M_\odot $. This mass is
redshifted $ \propto (1+z) $ as the Universe expands, because it is
made up of radiation.   The mass of cold dark matter in a comoving
volume is invariant, $ M_{\rm CDM} \equiv [(1+z_{\rm eq})/(1+z)] \;
M(z) $. At the QCD transition $ (1+z_{\rm eq})/(1+z) \sim 10^{-8} $
(recall $ z_{\rm eq} \sim 10^4 $), and thus
\begin{equation}
M_{\rm H}^{\rm CDM} \sim 10^{-8} M_\odot \ .
\end{equation} The Hubble time at the QCD transition, $ t_{QCD} \sim 10^{-5} $ s, is
much longer than the relaxation time scale of particle interactions.
Thus, leptons, photons,   the QGP  and the hadron gas (HG) are in
thermal and chemical equilibrium during cosmological time scales.
All components are  in LTE  and form a radiation fluid. There
conserved quantum numbers for this radiation fluid,   lepton and
baryon numbers (in the standard model). However, the corresponding
chemical potentials are much smaller than the temperature at the QCD
epoch. In this situation we can   use   the equation of state
calculated in thermal lattice QCD   and apply it directly to study
the equilibrium aspects of cosmology. In standard cosmology, all
lepton chemical potentials are assumed to vanish exactly. The baryon
number density $ n_{\rm B} $ follows from the ratio of baryons to
photons at Big Bang Nucleosynthesis (BBN), $ \eta \equiv (n_{\rm
B}/n_\gamma)_{\rm BBN} $, and the conservation of baryon number and
entropy, $ n_{\rm B}/s = $ constant:
\begin{equation}
n_{\rm B}(T_c) = \eta\left(\frac{n_{\gamma}}{s}\right)_{\rm BBN} \;
s(T_c)  = (8.2 \pm 0.4) \times 10^{-11} \;  s(T_c) \;  , \label{nB}
\end{equation}
using the value of $ \eta $ given in eq.(\ref{bau}). Finally, the
baryon number inside a Hubble volume is given by
\begin{equation}
B_{\rm H} \approx  \left( \frac{61.75}{ g} \right)^\frac12 \;
             \left(\frac{170 \mbox{\ MeV}}{ T_c} \right)^3 \;
  \left(\frac{\eta}{6.1\times  10^{-10}}\right) \;  2.0 \times 10^{48}
\end{equation} at the beginning of the transition and about twice that value at the
end\footnote{This formula is correct if no black holes
are formed during the QCD transition and if the quark nuggets
that might have formed evaporate before the BBN epoch.}.

\subsubsection{A first-order QCD transition \label{bubbles}}

If the cosmological QCD transition is of first order, it proceeds
via bubble nucleation \cite{Hogan83,DeGrand}. Its typical duration
is $ 0.1 t_{QCD} $. From the small values of surface tension and
latent heat found in lattice QCD calculations
\cite{Iwasaki:1993qu}, the amount of supercooling is found to be
small \cite{Ignatius}. Hadronic bubbles nucleate during a short
period of supercooling, $\Delta t_{\rm sc} \sim 10^{-3} \;
t_{QCD}$. In a homogeneous Universe without `dirt' the bubbles
nucleate owing to thermal fluctuations (homogeneous nucleation),
with a typical bubble nucleation distance of \cite{Christiansen}
\begin{equation}
d_{\rm nuc} \sim 1 \mbox{\ cm\ } \sim 10^{-6} \;  R_{\rm H}\ .
\end{equation}
The hadronic bubbles grow very fast, within $10^{-6} t_{QCD}$,
until the released latent heat has reheated the Universe to $T_c$.
By that time, just a small fraction of volume has gone through the
transition. For the remaining $99\%$ of the transition, the HG and
the QGP coexist at the pressure $p_{\rm HG}(T_c) = p_{\rm
QGP}(T_c)$. During this time the hadronic bubbles grow slowly and
the released latent heat keeps the temperature constant until the
transition is completed. The energy density decreases continuously
from $\epsilon_{\rm QGP}(T_c)$ at the beginning of the transition
to $\epsilon_{\rm HG}(T_c)$ when the transition is completed.

In first-order phase transitions that we know from our everyday
experience, like the condensation of water drops in clouds, the
drops (bubbles) are nucleated at impurities (`dirt'). This could
happen in the early Universe as well. Candidates for cosmic `dirt'
are primordial black holes, monopoles, strings, and other kinds of
defects. Of course, the existence of any of these objects has not
been verified so far. Under these circumstances the typical
nucleation distance may differ significantly from the scenario of
homogeneous nucleation. For the QCD transition, heterogeneous
nucleation has been studied in some detail in
ref.\cite{Christiansen}.

When the magnitude of primordial temperature fluctuations is of the same
order or larger than the typical supercooling,
the transition proceeds via inhomogeneous bubble nucleation. The
mean nucleation distance results from the scale and amplitude of the
temperature fluctuations.

\textbf{Homogeneous nucleation \label{homn}:} The probability to
nucleate a bubble with critical radius (i.e.~the minimal bubble size
that can grow after its formation) by a thermal fluctuation per unit
volume and unit time is given by (see discussion in Sec. III B.2)
\begin{equation} \label{I}
I(T) = I_0(T) \;  \exp\left( - \frac{\Delta W_{\rm c}}{T}\right) \ ,
\end{equation}
with $ \Delta W_{\rm c} = 16 \; \pi \; \sigma^3/[3 \; (p_{\rm HG} -
p_{\rm QGP})^2] $. For dimensional reasons   $ I_0 \sim C \;  T_c^4
$, with $ C = {\cal O}(1) $. A more detailed calculation of $ I_0 $
within the bag model has been provided in \cite{Csernai}. It was
shown in Ref.~\cite{Christiansen} that the temperature dependence of
the prefactor $ I_0 $ can be neglected for the calculation of the
supercooling temperature $ T_{\rm sc} $ in the cosmological QCD
transition.

For small supercooling $\Delta \equiv 1 -  T/T_c \ll 1$ we may evaluate
$(p_{\rm HG} - p_{\rm QGP})(T)$ by using the second law of thermodynamics,
i.e.\ $p_{\rm HG} - p_{\rm QGP} \approx l \; \Delta$, and thus
\begin{equation} \label{A}
I(\Delta) \approx I_0(T_c) \;  \exp\left(-A/\Delta^2\right),
\end{equation}
with $A \equiv 16 \;  \pi \;  \sigma^3/(3 \;  l^2  \; T_c)$ and
$ I_0(T_c) \approx T_c^4 $. Note that this result does not depend on the
details of the QCD equation of state. For the values of $ l=1.4 \;  T_c^4 $ and
$ \sigma=0.015 \; T_c^3 $ from quenched lattice QCD \cite{Iwasaki:1993qu}
$ A \approx 3 \times 10^{-5} $. In the bag model $ A \approx 5 \times 10^{-2}
 \; (\sigma/T_c^3)^3 $.

The amount of supercooling that is necessary to complete the transition,
$ \Delta_{\rm sc} $, can be estimated from the schematic case of one single
bubble nucleated per Hubble volume per Hubble time, which is
\begin{equation} \label{Delta_sc}
{\cal O}(\Delta_{\rm sc}) = \left[\frac{A}{4 \; \ln(T_c/H_{QCD})} \right]^{1/2}
\approx 4 \times 10^{-4}
\end{equation}
for the values of $l$ and $\sigma$ from quenched lattice QCD.
For the bag model we assume  $ \sigma < 0.1 \;  T_c^3 $, which implies that
$ \Delta_{\rm sc} < 6 \times 10^{-4} $.

The time lapse during the supercooling period follows from the conservation
of entropy and reads
\begin{equation}
\Delta t_{\rm sc}/t_{QCD} = \Delta_{\rm sc}/(3 \;  c_s^2) =
{\cal O}(10^{-3}) \ .
\end{equation}
Here we used the relation $ c_s^2 = {\rm d} \ln s/{\rm d} \ln T $
for the speed of sound in the supercooled phase. For realistic
models $0 < c_s(\Delta) < 1/\sqrt{3}$, and in  the bag model
$c_s(\Delta) = 1/\sqrt{3}$.

\begin{figure}[t]
\centerline{\includegraphics[width=0.35\textwidth]{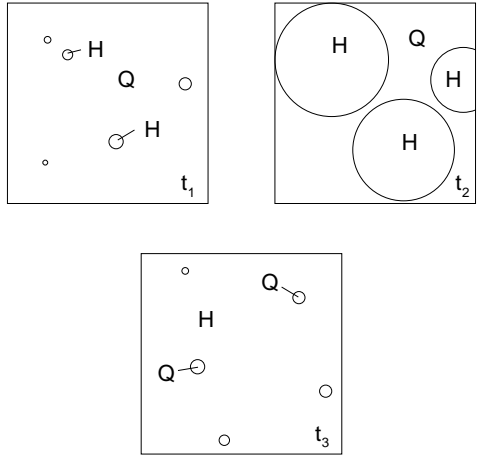}}
\caption{\label{fig7}
Sketch of a first-order QCD transition via homogeneous bubble nucleation:
above the critical temperature the Universe is filled with a quark--gluon
plasma (Q). After a small amount of supercooling the first
hadronic bubbles (H) nucleate at $ t_1 $, with mean separation $d_{\rm nuc}$.
At $ t_2 > t_1 $ these bubbles have grown and have released enough
latent heat to quench the formation of new bubbles. The supercooling,
bubble nucleation, and  quenching takes just $ 1\% $ of the full transition
time. In the remaining $ 99\% $ of the transition time the bubbles grow
following the adiabatic expansion of the Universe. At $ t_3 $ the transition
is almost finished. The shrinking QGP drops are separated by the typical
distance $ d_{\rm nuc} $.}
\end{figure}

After the first bubbles have been nucleated, they grow most probably
by weak deflagration \cite{DeGrand,Kurki-Suonio,Kajantie,Ignatius}.
The deflagration front (the bubble wall) moves with the velocity
$ v_{\rm defl} \ll 1/\sqrt{3} $ \cite{Kajantie92}. The energy that is
released from the bubbles is distributed into the surrounding QGP by a
supersonic shock wave and by neutrino radiation. This reheats the QGP
to $ T_c $ and prohibits further bubble formation. Since the amplitude of
the shock is very small \cite{Kurki-Suonio}, on scales smaller than the
neutrino mean free path, heat transport by neutrinos is the most efficient.
Neutrinos have a mean free path of $ 10^{-6} R_{\rm H} $ at $ T_c $.
When they do most of the heat transport, heat goes with
$ v_{\rm heat}= {\cal O}(c) $. For larger scales, heat transport is much
slower. Figure \ref{fig7} shows a sketch of the homogeneous bubble nucleation
scenario.

Let us now calculate the mean bubble separation, $d_{\rm nuc}$, and the
final supercooling, $\Delta_{\rm sc}$, for a scenario with weak deflagration.
Bubbles present at a given time have typically been nucleated during the
preceding time interval
\begin{equation}
\label{tnucl}
\Delta t_{\rm nuc} \equiv  I/({\rm d}I/{\rm d} t) \ .
\end{equation}
Using the relation between time and supercooling,
$ {\rm d}\Delta/{\rm d}t = 3 \;  c_s^2 /t_{QCD} $, we find
\begin{equation}\label{Delta-nucl}
\Delta t_{\rm nuc}/t_{QCD}= \Delta_{\rm sc}^3/(6 \;  A \;  c_s^2) =
{\cal O}(10^{-5}) \quad \rm{and} \quad
\Delta_{\rm nuc} = \frac{\Delta t_{\rm nuc}}{\Delta t_{\rm sc}} \;
\Delta_{\rm sc} = {\cal O}(10^{-2}) \;  \Delta_{\rm sc} \ .
\end{equation}
During the time interval $ \Delta t_{\rm nuc} $ each bubble
releases latent heat, which is distributed over a typical
distance $ \approx  2 \;  v_{\rm heat} \;  \Delta t_{\rm nuc} $. This distance
has a weak dependence on the precise value of $ \Delta_{\rm sc} $, but the
bubble nucleation rate increases strongly with $ \Delta $ until
one bubble per volume $ \sim (\Delta t_{\rm nuc} \;  v_{\rm heat})^3 $ is
nucleated. Therefore the mean bubble separation is \begin{equation}
\label{dnucl} d_{\rm nuc} \approx 2 \;  v_{\rm heat} \;
\Delta t_{\rm nuc} \approx
\frac{v_{\rm heat}}{3 \; c_s^2} \; \frac{\Delta_{\rm sc}^3}{A}  \;  R_{\rm H} =
{\cal O}(10^{-6} \;  R_{\rm H})= {\cal O}(1 {\rm cm}),
\end{equation}
where we used $ v_{\rm heat} = {\cal O}(0.1), 3 \; c_s^2 = {\cal O}(0.1) $,
which gives a typical value for the nucleation distance. The suppression of
bubble nucleation due to already existing bubbles is neglected.

The estimate eq.(\ref{dnucl}) of the mean bubble separation applies if the
released latent heat by means of sound waves and by neutrino free streaming is
sufficient to reheat the QGP to $ T_c $, i.e. to quench the nucleation of
new bubbles. On the other hand the typical bubble separation could be
given by the rate of release of latent heat, i.e. by the bubble wall velocity
$ v_{\rm defl} $. Since the period of supercooling lasts about $ 1 \% $ of the
time needed for completing the entire first-order phase transition, $ 1 \% $ of
the QGP must be converted to HG in the process of sudden reheating to
$ T_c $; the bubble radius at quenching must therefore reach $ 0.2 $ of the
bubble separation, $ R_{\rm bubble}\approx 0.2 \; d_{\rm nuc} $.  With $ R_{\rm
bubble} \approx v_{\rm defl} \; \Delta t_{\rm nuc} $, and using the above
relation $ d_{\rm nuc} \approx 2 \; v_{\rm heat} \; \Delta t_{\rm nuc} $,
we require $ v_{\rm defl} \ge 0.4 v_{\rm heat}$ for consistency.
If $ v_{\rm defl} $ is smaller than this, the limiting factor for quenching
is the rate of release of latent heat by bubble growth, and the bubble
separation is \begin{equation}
d_{\rm nuc} \approx 2 \;  v_{\rm defl} \;  \Delta t_{\rm nuc} \approx
\frac{v_{\rm defl}}{3 \; c_s^2} \; \frac{\Delta_{\rm sc}^3}{A} \; R_{\rm H} \; ,
\end{equation}
i.e. the bubble separation will be smaller than the estimate
in eq.~(\ref{dnucl}).

We are now in a position to improve the estimate of $\Delta_{\rm sc}$:
one bubble nucleates in the volume $(v_{\rm heat}\Delta t_{\rm nuc})^3$
during $\Delta t_{\rm nuc}$.
This can be written as
\begin{equation}
\label{Delta-exact}
1 \approx  (v_{\rm heat} \;  \Delta t_{\rm nuc})^3 \;  \Delta t_{\rm nuc}
I(t_{\rm sc}) \ ,
\end{equation}
which in terms of the supercooling parameter $\Delta_{\rm sc}$ is
given by: \be 1 \approx  \frac{v_{\rm heat}^3}{ (3 \;  c_s^2 \;
A)^4} \; \left(\frac{T_c}{H_{QCD}}\right)^4 \; \Delta_{\rm sc}^{12}
\;  \exp\left(-\frac{A}{\Delta_{\rm sc}^2}\right) \approx 10^{94}
\;  \Delta_{\rm sc}^{12} \; \exp\left(-\frac{2.89 \times
10^{-5}}{\Delta_{\rm sc}^2}\right)  \;  . \ee While the
pre-exponential factor is smaller   than the naive estimate
eq.(\ref{Delta_sc}) by a factor of $10^{20}$, the amount of
supercooling is just $20\%$ larger than in eq.(\ref{Delta_sc}),
i.e.\ $ \Delta_{\rm sc} = 5 \times 10^{-4} $, confirming, that
that numerical prefactors in eq.(\ref{Delta-exact}) are irrelevant
in the calculation of $ \Delta_{\rm sc} $.

In summary, the scales on which non-equilibrium phenomena occur
are given by the mean bubble separation, which is about $ 10^{-6}
R_{\rm H} $. The entropy production is tiny, i.e. $ \Delta S/ S
\sim 10^{-6} $, since the supercooling is small $ \sim 10^{-3} $.
After supercooling, which lasts $ 10^{-3} \; t_{QCD} $, the
Universe reheats in $ \Delta t_{\rm nuc} \approx 10^{-6} \;
t_{QCD} $. After reheating, the thermodynamic variables follow
their equilibrium values and bubbles grow only because of the
expansion of the Universe.

\textbf{Inhomogeneous nucleation\label{ihn}:} The local
temperature $ T(t,{\bf x}) $ of the radiation fluid fluctuates,
because cosmological perturbations have been generated during
cosmological inflation\cite{hu}. Let us denote the temperature
fluctuation by $ \Delta_T \equiv \delta T/T $. Inflation predicts
a Gaussian distribution of perturbations (see
section\ref{inflation}):
\begin{equation}
P(\Delta_T) \; {\rm d}\Delta_T = \frac1{\sqrt{2\pi}\Delta_T^{\rm rms}}
\exp\left[ - \frac12 \frac{\Delta_T^2}{(\Delta_T^{\rm rms})^2}\right]
{\rm d}\Delta_T \ .
\end{equation}
If one allows for a tilt in the power spectrum
of density fluctuations, the rms temperature fluctuation reads [see
eq.(\ref{espflu}) and ref.\cite{IS}], \begin{equation}
\Delta_T^{\rm rms} \approx 10^{-4} \;  (3 \; c_s^2)^{3/4} \;
\left(\frac{k}{k_0}\right)^{(n_s - 1)/2} \ ,
\end{equation} where $ k_0 $ is the wave number of the mode that crosses
the Hubble radius today. For the scale-invariant spectrum ($ n_s=1
$) it is found\cite{IS} $ \Delta_T^{\rm rms}(k_{\rm QCD}) \approx
2 \times 10^{-5} $.

The detailed analysis in ref.\cite{IS} leads to the conclusion
that the picture of homogeneous bubble nucleation, where bubbles
form from statistical fluctuations, is false for the most probable
cosmological scenarios.

\begin{figure}[t]
\centerline{\includegraphics[width=0.35\textwidth]{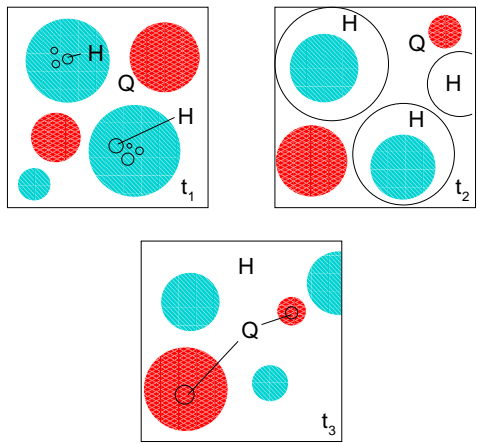}}
\caption{\label{fig8}
Sketch of a first-order QCD transition in the inhomogeneous
Universe (from \protect\cite{IS}):
at $t_1$ the coldest spots (dark grey) are cold enough to render the
nucleation of hadronic bubbles (H) possible, while most of the
Universe remains in the quark--gluon phase (Q). At $t_2 > t_1$ the bubbles
from the cold spots have merged and have grown to bubbles as large
as the fluctuation scale. Only the hot spots (light grey) are still in the
QGP phase. At $t_3$ the transition is almost finished. The last
QGP drops are found in the hottest spots of the Universe. The mean
separation of these hot spots can be much larger than the homogeneous
bubble nucleation separation.}
\end{figure}
A new scenario for the cosmological QCD transition was developed
in ref.\cite{IS} in which a small scale cut-off in the spectrum of
primordial temperature fluctuations  from collisional damping by
neutrinos \cite{Weinberg,SSW2} is included. This study revealed
that at the QCD transition neutrinos travel freely on scales $
\lambda_{\nu-{\rm mfp}} \approx 10^{-6} \;  R_{\rm H} $, and that
fluctuations on the diffusion scale of neutrinos are washed out by
the time of the QCD transition:
\begin{equation}
\lambda_{\nu-{\rm diff}} =
\frac13 \;  \sqrt{\lambda_{\nu-{\rm mfp}} \;  c \;  t_{QCD}}
\approx 10^{-4} \;  R_{\rm H} \ .
\end{equation}
Thus the old picture of homogeneous bubble nucleation still applies
within the small homogeneous patches of $\lambda_{\rm smooth} =
10^{-4} \; R_{\rm H}$.

The compression time scale for a homogeneous patch is $ \delta t =
\lambda_{\rm smooth}/c_s \sim 10^{-3} \;  t_{QCD} $. If the compression
time scale is larger than $ \Delta t_{\rm nuc} $ the temperature fluctuations
are frozen with respect to the time scale of nucleations.

A sketch of inhomogeneous bubble nucleation is shown in fig.~\ref{fig8}. The
basic idea is that temperature inhomogeneities determine the location of
bubble nucleation. In cold regions, bubbles nucleate first.
In general we have two possible situations:
\begin{enumerate}
\item If $\Delta_{\rm nuc} > \Delta_T^{\rm rms}$, the temperature
      inhomogeneities are negligible and the phase transition proceeds
      via homogeneous nucleation (see section \ref{homn}).
\item If $\Delta_{\rm nuc} < \Delta_T^{\rm rms}$, the nucleation rate
      is inhomogeneous and we have to consider the scenario sketched in
      fig.~\ref{fig8}.
\end{enumerate}
A first attempt to analyse inhomogeneous nucleation has been given in
\cite{IS}. According to \cite{IS}, the nucleation distance $ d_{\rm nuc} $
exceeds the scale $ \lambda_{\rm smooth} $, if
\begin{equation}
\lambda_{\rm smooth} < 2 \;  \frac{v_{\rm heat}}{3 \;  c_s^2} \Delta_T^{\rm rms}
 \; R_{\rm H}.
\end{equation}
If $ \Delta_T^{\rm rms} > 5 \times 10^{-5} $, it is quite likely that this
condition is met. In that case
we can conclude that the typical inhomogeneity scale in the baryon
distribution is inherited from the scale of density inhomogeneities in
the radiation fluid at the end of the QCD transition. The effect in terms
of length scales is at least two orders of magnitude larger than the
nucleation distance in homogeneous nucleation and is
$ {\cal O}(1 \mbox{\ m}) $, which is of interest for inhomogeneous BBN.

\subsubsection{Effects from a first-order QCD
transition}\label{effects}

Let us now briefly summarize the effects that have been suggested to
emerge from the cosmological QCD transition. There are two kinds of effects:
the effects that have been found in the mid 80s and early 90s
stem from the bubble scale and they thus affect scales
$\lambda \leq d_{\rm nuc}$. The formation of quark nuggets, the generation
of isothermal baryon fluctuations, the generation of magnetic fields and
gravitational waves belong to the effects from the bubble scale.

In recent years it was found that there is another class of possible
consequences from the QCD transition, which are connected to the Hubble
scale and therefore affect scales $ \lambda \leq R_{\rm H} $. Among these
effects are the amplification of inhomogeneities and later formation of cold
dark matter clumps, the modification of primordial gravitational waves, and the
enhanced probability of black hole formation during the QCD transition.

\textbf{Quark nuggets/Strangelets:} In 1971 Bodmer \cite{Bodmer}
suggested the possibility that strange quark matter might be the
ground state of bulk matter, instead of ${}^{56}$Fe. Later Witten
\cite{Witten} rediscovered this idea. Strange quark matter was
further studied by Farhi and Jaffe \cite{Farhi:1984qu}. The idea
of strange quark matter is based on the observation that the Pauli
principle allows more quarks to be packed into a fixed volume in
phase space if three instead of two flavours are available. Thus
the energy per baryon would be lower in strange quark matter than
in nuclei. However, the strange quark is heavy compared with up
and down quarks, and this mass counteracts the advantage from the
Pauli principle. No strange quark matter has been found
experimentally so far \cite{smexp}. The issue of stability of
strange quark matter has not been settled yet; for a recent review
see \cite{Madsen}.

Witten \cite{Witten} pointed out that a separation of phases during the
coexistence of the hadronic and the quark phase could gather a large number
of baryons in strange quark nuggets \cite{Witten}.
These quark nuggets could contribute to the dark matter today \cite{Witten}
or affect BBN \cite{steigman,turnerBBN}.
At the end of the transition the baryon number
in the quark droplets could exceed the baryon number in the hadron phase
by several orders of magnitude, $ n_{\rm B}^{\rm QGP} $ could be close to
nuclear density. However, it was realized
that the quark nuggets, while cooling, lose baryons. The quark nuggets
evaporate as long as the temperature is above $ \sim 50$~MeV \cite{Alcock}.
Quark nuggets may survive this evaporation if they contain much more
than $ \sim 10^{44} $ baryons initially \cite{evap}. This number should be
compared with the number of baryons inside a Hubble volume at the QCD
transition, which is $ 10^{48} $. Thus, the mean
bubble nucleation distance should be $ > 3 \times 10^{-2} R_{\rm H}
\sim 300 $ m so as to collect enough baryons. This seems impossible from
todays perspective, as explained above.

\textbf{Inhomogeneous nucleosynthesis:} Applegate and Hogan
\cite{Applegate} found that a strong first-order QCD phase
transition induces inhomogeneous nucleosynthesis. It is extremely
important to understand the initial conditions for BBN, because
many of our ideas about the early Universe rely on the validity of
the standard (homogeneous) BBN scenario. This is in good agreement
with observations \cite{steigman,turnerBBN}. In inhomogeneous
nucleosynthesis \cite{IBBN}, large isothermal fluctuations of the
baryon number (the remnants of the quark droplets at the end of
the QCD transition) could lead to different yields of light
elements. As a minimal requirement for an inhomogeneous scenario
of nucleosynthesis, the mean bubble nucleation distance has to be
larger than the proton diffusion length, which corresponds to
$\sim 3$ m \cite{Mathews} at the QCD transition. This is two
orders of magnitude above recent estimates of the typical
nucleation distance \cite{Christiansen}.

Although values for $ \eta $ dramatically different from those in
the standard BBN are excluded both from measurements of the light element
abundances and from the CMB, it might be possible to alter
the primordial abundance of heavy elements ($ A > 7 $) in
inhomogeneous scenarios \cite{JedamzikHE}.

\textbf{Cold dark matter clumps:} Scales $\lambda$ that are of the
order of the Hubble radius $R_{\rm H}$ are not sensitive to
details of the bubbles. It was reported in Refs.\ \cite{SSW,SSW2}
that the evolution of cosmological density perturbations (see
Sec.~III) is strongly affected by a first-order QCD transition for
subhorizon scales, $ \lambda < R_{\rm H} $.

In the radiation-dominated Universe subhorizon density perturbations
perform acoustic oscillations. The restoring force is provided by
pressure gradients. These, and therefore the speed of sound
$ c_s = \left(\partial p/\partial \epsilon \right)_S^{1/2} $
(on scales much larger than the bubble separation scale) drop to
zero at a first-order QCD transition \cite{SSW}, because both phases
coexist at the pressure $p_c$ only ($a$ is the scale factor of
the Universe):
\begin{equation}
c_s^2 = \frac{{\rm d} p_c/{\rm d} a}{{\rm d} \epsilon(a)/{\rm d} a} = 0 \ .
\end{equation}
It stays zero during the entire transition and suddenly rises back to
the radiation value $ c_s=1/\sqrt{3} $ after the transition. A significant
decrease in the effective speed of sound $c_s$ during the cosmological QCD
transition was also pointed out by Jedamzik \cite{Jedamzik}.

As the speed of sound drops to zero, the restoring force for acoustic
oscillations vanishes and density perturbations for subhorizon modes
fall freely. The fluid velocity stays constant during this free fall.
Perturbations of shorter wavelengths have higher velocities
at the beginning of the transition, and thus grow proportional to the wave
number $ k $ during the phase transition. The primordial Harrison--Zel'dovich
spectrum of density perturbations is amplified on subhorizon scales.
The spectrum of density perturbations on superhorizon scales,
$ \lambda > R_{\rm H} $, is unaffected. At $ T\sim 1 $ MeV the neutrinos
decouple from the radiation fluid. During this decoupling the large peaks
in the radiation spectrum are wiped out by collisional damping \cite{Weinberg}.

Today a major component of the Universe is dark matter, most likely CDM.
If CDM is kinetically decoupled from the radiation fluid
at the QCD transition, the density perturbations in CDM do not suffer
from the neutrino damping. This is the case for primordial black
holes or axions, but not for supersymmetric dark matter.
At the time of the QCD transition the energy density of CDM is small, i.e.\
$ \epsilon_{\rm cdm}(T_c) \sim 10^{-8} \;  \epsilon_{\rm rad}(T_c) $.
CDM falls into the potential wells provided by the dominant
radiation fluid. Thus, the CDM spectrum is amplified on subhorizon
scales. The peaks in the CDM spectrum go non-linear
shortly after radiation--matter equality. This leads to the
formation of CDM clumps with
mass $ < 10^{-10} M_\odot $. Especially the clumping of axions has important
implications for axion searches \cite{Sikivie}.
If the QCD transition is strong enough, these
clumps could be detected by gravitational femtolensing \cite{femtolensing}.

\subsubsection{Damping of gravitational waves at the QCD
transition}\label{gw}

In principle, primordial gravitational waves (e.g.~from cosmological inflation)
present a clean probe of the dynamics of the early Universe, since
they know only about the Hubble expansion. As was shown in \cite{Schwarz}
a step is imprinted in the spectrum of primordial gravitational waves
by the cosmological QCD transition. This step does not allow us to tell
the difference between a first-order transition and a crossover,
but an estimate of the transition temperature and a measurement of the
drop in effective number of relativistic degrees of freedom would be
possible.

\begin{figure}[h]
\begin{center}
\epsfig{file=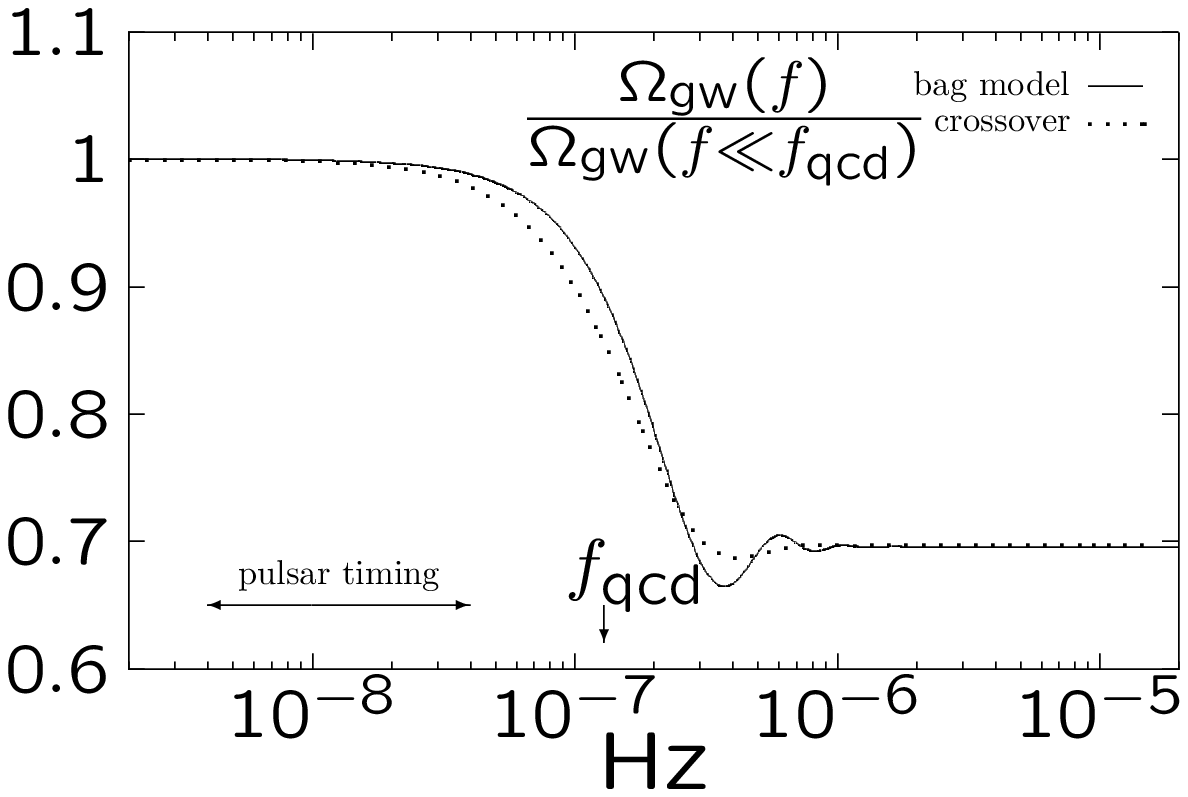,height=2in,width=4in,keepaspectratio=true}
\caption{The modification of the energy density, per logarithmic
frequency interval, for primordial gravitational waves from the QCD
transition (from \protect\cite{Schwarz}). \label{fig6}}
\end{center}
\end{figure}

Primordial gravitational waves are predicted to be generated during
inflation \cite{Starobinskii} and could be detected by observing the
so-called B-mode (parity odd patterns) polarisation of the CMB.
Inflation predicts an almost scale-invariant energy density per
logarithmic frequency
interval for the most interesting frequencies ($ \sim 10^{-8} $ Hz for pulsar
timing, $ \sim 10^{-3} $ Hz for LISA, and $\sim 100$ Hz for LIGO and VIRGO) of
the gravitational waves.
The energy fraction in gravitational waves, per logarithmic interval in
frequency $ f $,
is defined by
\begin{equation}
\label{omegagw}
\Omega_{\rm gw} (f) \equiv  f \;  {{\rm d} \epsilon_{\rm gw}\over {\rm d} f}
 \; {1\over \epsilon_{\rm c}} \ .
\end{equation}
Figure \ref{fig6} shows the transfer function $\Omega_{\rm gw}(f)/
\Omega_{\rm gw}(f \ll f_{QCD})$ from the cosmological QCD transition.
The typical frequency scale is
\begin{equation}
f_{QCD} \approx 1.5 \left(g\over 17.25\right)^{\frac12} \;
{T_c \over 170 \mbox{\ MeV}} \;  10^{-7} \mbox{\ Hz} \ ,
\end{equation}
which corresponds to the mode that crosses the Hubble horizon at the end of
the bag model QCD transition. Scales that cross into the horizon after the
transition (l.h.s.~of the figure) are unaffected, whereas modes that
cross the horizon before the transition are damped by an additional factor
$\approx 0.7$. The modification of the differential spectrum has been
calculated for a first-order (bag model) and a crossover QCD transition.
In both cases the step extends over one decade in frequency. The detailed
form of the step is almost independent from the order of the transition.

The size of the step can be calculated analytically \cite{Schwarz}.
Comparing the differential energy spectrum for modes that cross into the
horizon before and after the transition gives the ratio
\begin{equation}
\label{ratio}
{\Omega_{\rm gw}(f \gg f_{QCD}) \over \Omega_{\rm gw}(f\ll f_{QCD})} =
\left(g_{\rm a}\over g_{\rm b}\right)^{\frac13} \approx 0.696 \ ,
\end{equation}
for the QCD transition.

In fig.~\ref{fig6} we indicated the frequency range ($\sim 1$ yr$^{-1}$)
in which limits on $\Omega_{\rm gw}$ have been reported
from pulsar timing residuals \cite{pulsar}. Unfortunately, today's technology
does not enable us to detect primordial gravitational waves at frequencies
around $10^{-7}$ Hz, because their expected amplitude is too small.


\section{Observing the early Universe: the earliest is the latest.}

As discussed above, during the inflationary stage the acceleration
of the scale factor is positive, while in the radiation or matter
dominated eras it is negative. Eq.(\ref{inflad}) shows that
wavelengths grow faster than the Hubble radius during inflation
but slower during radiation and matter dominated eras.
Perturbations of cosmological relevance with wavelengths between
our Hubble radius today to galactic scales exited the Hubble
radius during a window of $ \sim 5-8 $ e-folds about $55$ e-folds
before the end of inflation thus becoming causally disconnected
from microphysics. Wavelengths relevant to CMB anisotropies
re-entered at the time of recombination prior to photon
decoupling. Therefore CMB anisotropies provide \emph{direct}
information about the inflationary stage. Contrary to this case,
the wavelengths of quantum fluctuations produced inside the Hubble
radius during the radiation or matter dominated eras are
\emph{always} inside the Hubble radius and are causally influenced
by microphysical processes. Hence observable consequences of
processes that occurred during these eras are generally
\emph{indirect}. The baryon asymmetry is \emph{perhaps} a remnant
of the the microphysics of the standard model (or extensions
thereof)  and \emph{perhaps} the QCD phase transition left an
observable footprint as discussed in secs.\ref{effects} and
\ref{gw}. The time scale between the QCD phase transition $\sim 10
\mu\textrm{secs}$ and that of nucleosynthesis $\sim 200\,
\textrm{secs}$ is very large on the scale of strong interactions
$\sim 10^{-23}\,\textrm{secs}$ which tend to erase any potentially
interesting signature. Thus CMB observations, $400000$ years after
the Big Bang,   provide a window into the \emph{earliest}
phenomena that took place during inflation, while microphysical
processes within the standard model of particle physics occur much
later, are causal all throughout the evolution, and interactions
blur their observable signals. An important \emph{possible}
remnant of these phase transitions are primordial magnetic fields.

A variety of astrophysical observations including Zeeman splitting,
synchrotron emission, Faraday rotation measurements (RM) combined
with pulsar dispersion measurements (DM) and polarization
measurements suggest the presence of large scale magnetic
fields\cite{cmp,han}. The  strength of typical galactic magnetic
fields is of the order $\sim \mu~G$\cite{cmp,han} and they are
correlated on very large scales up to galactic or even larger
reaching to scales of cluster of galaxies $\sim
1~\mbox{Mpc}$\cite{cmp}. The origin of these large scale magnetic
fields is still a subject of discussion. It is currently agreed that
a variety of dynamo mechanisms are efficient in {\bf amplifying}
seed magnetic fields with typical growth rates $\Gamma \sim
\mbox{Gyr}^{-1}$ over time scales $\sim 10-12 $ Gyr \cite{cmp}.
There are different  proposals for the origin of the initial seed at
different stages in the history of the early Universe, in particular
during inflation,  and or phase transitions\cite{cmp}. Primordial
(hyper) magnetic fields may have important consequences in
electroweak baryogenesis\cite{grassoEW}, Big Bang nucleosynthesis
(see\cite{cmp}), the polarization of the CMB\cite{varios} via the
same physical processes as Faraday rotation, and structure
formation\cite{cmp},\cite{otros}. If the electroweak and/or the
chiral phase transitions occurred out of equilibrium they could be a
significant source of primordial magnetic fields\cite{magne}. Thus
cosmic magnetic fields may be one of the few  observational relics
of primordial phase transitions.  The study of the origin of cosmic
magnetic fields is one of the key science projects of the
forthcoming Square Kilometer Array (SKA)\cite{SKA}.

\section{Studying phase transitions   with accelerators}

\subsection{Ultrarelativistic heavy ion collisions: seeking the Quark Gluon
Plasma }\label{urhic} The program of ultra relativistic heavy ion
collisions (URHIC) whose primary goal is to study the phase diagram
of QCD began almost two decades ago with the fixed target heavy ion
programs at the AGS at Brookhaven and the SPS at CERN. A summary of
the results  of these efforts mainly through the $Pb+Pb$ experiments
at SPS-CERN provided compelling evidence in favor of the existence
of a new state of matter\cite{heinz}. The program continues at the
Relativistic Heavy Ion Collider (RHIC) at BNL. Unlike the fixed
target experiment SPS at CERN, RHIC is a collider experiment which
currently studies $Au+Au$ collisions with center of mass energy
$\sqrt{s} \sim 200 {\rm AGev}$ and luminosity $\sim 10^{26}
cm^{-2}s^{-1}$. The future ALICE (A Large Ion Collider Experiment)
heavy ion program at LHC is expected to study $Pb+Pb$ collisions
with c.m. energies up to $\sqrt{s}\sim 5 {\rm ATev}$ and
luminosities $\sim 10^{27} cm^{-2}s^{-1}$. In these collisions the
heavy nuclei can be pictured in the CM frame as two Lorentz
contracted pancakes. For $Au+Au$ collisions the size of each pancake
in the direction transverse to the beam axis is about 7 fm. At RHIC
and LHC energies most of the baryons are expected to be carried away
by the receding pancakes (the fragmentation region) while in the
region of the collision a large energy (density) is deposited in the
form of quark pairs and gluons. At least two important mechanisms
for energy deposition in the collision region are at
work~\cite{muharris,blaizot,shuryak,qgpbooks}: i) the establishment
of a strong color electric field (flux tube) that eventually breaks
up into quark-antiquark pairs when the energy in the field is larger
than the pair production threshold and ii) the partons (quarks and
gluons) inside the colliding nuclei interact and redistribute their
energy~\cite{geiger}. An estimate of the energy deposited in the
collision region has been provided by Bjorken~\cite{bjorken}
\begin{equation}\label{enercen}
\epsilon = \frac{1}{\tau_0 \pi R^2_A} \frac{dE_T}{dy}
\end{equation}
\noindent with $\tau_0 \sim 1 {\rm fm}/c$, $R_A \sim 7 {\rm fm}$
for Au and $dE_T/dy$ is the transverse energy per unit rapidity
which is {\em measured}. Parton-parton scattering is expected to
lead to thermalization on time scales $\sim 1\, {\rm fm}/c$
~\cite{muharris,blaizot}. After the quark-gluon plasma achieves
LTE, the evolution is conjectured to be described by hydrodynamic
expansion~\cite{hydro1}. As the QGP expands and cools, the
temperature falls near the critical temperature and the
confinement and chiral phase transitions occur\cite{meyer}. Upon
further cooling the quark-gluon plasma hadronizes, the hadrons
rescatter until the hadron gas is dilute enough that the mean free
path is larger than the mean distance between hadrons. At this
point hadrons freeze-out and stream out freely to the detectors
from this {\em last scattering or freeze-out surface}. This
picture is summarized in fig. \ref{fig:uhic} below.
\begin{figure}[h]
\begin{center}
\epsfig{file=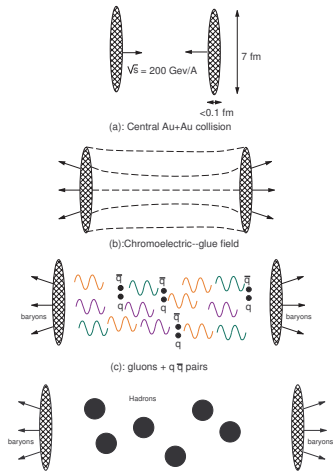,height=2.5in,width=4in,keepaspectratio=true}
\caption{Heavy Ion Collisions. } \label{fig:uhic}
\end{center}
\end{figure}
Estimates based on this picture and on detailed numerical
evolution~\cite{evolutionqgp} suggest that at RHIC the quark gluon
plasma lifetime is of order $\sim 10 fm/c$ while the total
evolution until freeze-out is $\sim 50-100 fm/c$.

\subsubsection{Hydrodynamics, lattice gauge theory and the equation of state.}\label{sec:hydro}
Although the evolution of the quark gluon plasma from the initial
state described by the parton distribution of the colliding nuclei
until freeze-out of hadrons  clearly requires a non-equilibrium
description, a hydrodynamic picture of the evolution is both
useful and experimentally relevant~\cite{hydro1,hydro2}. In the
approximation in which quarks and gluons are strongly coupled in
the sense that their mean free paths are much smaller than the
typical wavelength for the variation of collective phenomena, the
QGP can be described as a {\em fluid} in LTE, for which the energy
momentum tensor is of the form
\begin{equation}\label{tmunu} T^{\mu \nu} =
(\varepsilon+p)\; u^{\mu}\; u^{\nu}-p \; g^{\mu \nu} \; ,
\end{equation}
\noindent with $ \varepsilon, \; p $ the energy density and pressure
respectively, $ g^{\mu \nu} $ is the  metric and
\begin{equation}\label{velo}
u^{\mu}=\gamma(x)\; (1,{\vec v}(x))
\end{equation}
\noindent is a local 4-velocity vector with $ \gamma(x) $ the local
Lorentz contraction factor.

In this description the dynamical
evolution of the fluid is obtained from the conservation laws of
energy-momentum, baryon number and
entropy~\cite{shuryak,qgpbooks,meyer,hydro1} with an equation of
state $p=p(\varepsilon)$   to close the set of equations. A simple
and phenomenologically useful description is provided by Bjorken's
model for longitudinal
expansion~\cite{shuryak,qgpbooks,bjorken,hydro1}. There are three
important ingredients in Bjorken's models: i) the central
rapidity region where the energy deposited and the formation of
the QGP takes place well separated from the fragmentation
region which is the region of the receding pancakes where most
of the baryons are (see fig. \ref{fig:uhic}). ii) The
thermodynamic variables are invariant under boosts along the beam
(longitudinal) axis, this is based on the observation that the
particle distributions are invariant under these boosts in the
central rapidity region and  iii) the expansion only occurs
along the beam axis, i.e, longitudinal expansion. In the baryon
free region, local thermodynamics implies that the entropy density $ s $
in the local rest frame of the fluid is related to the energy
density $ \varepsilon $ and pressure $ p $ as
\be
\varepsilon+p=T\; s \; .\label{entropy}
\ee It is convenient to introduce the space time rapidity $\eta$ and
proper-time $\tau$ as
\begin{eqnarray}\label{etatau} \eta =
\frac{1}{2}\ln\left[\frac{t+z}{t-z} \right] ~~;~~\tau =\sqrt{t^2-z^2}\; ,
\end{eqnarray}
\noindent with $z$ the coordinate along the beam axis
(longitudinal) and the fluid rapidity $ \theta $ by writing the
local velocity eq.(\ref{velo}) as
$ u^{\mu} = (\cosh\theta,0,0,\sinh\theta) $.
 Bjorken's model assumes the fluid to be composed of
free streaming particles for which $ v_z=z/t $ in which case the
fluid rapidity $ \theta $  becomes the space-time rapidity $ \eta $.
Boost invariance along the longitudinal direction entails that
$ \varepsilon,p,s,T $ are all functions of proper time only\cite{bjorken,hydro1}.

The energy-momentum conservation equation $ \partial_{\mu}T^{\mu
\nu} $ leads to two equations by projecting along the direction
$ u^{\mu} $ and perpendicular to it using the projector
$ g^{\mu\nu}-u^{\mu}u^{\nu} $. Under the assumption that the central
region is baryon free, the conservation of entropy and energy and
momentum lead to the following equations (for details
see~\cite{hydro1,meyer,shuryak,qgpbooks,bjorken})
\be
\frac{d\varepsilon(\tau)}{d\tau}+\frac{1}{\tau}(\varepsilon+p)=0
\label{enercons} \quad , \quad
\frac{ds(\tau)}{d\tau}+\frac{s}{\tau}=0 \; .
\ee Assuming an equation of state $ p(\tau)= p(\varepsilon(\tau)) $ and
combining eqs.(\ref{enercons}) with the
thermodynamic relation  (\ref{entropy}) (for baryon free plasmas)
one finds that the evolution of the temperature is given by
\begin{equation}\label{temp}
\left. \frac{\tau}{T}\frac{dT}{d\tau}=-c^2_s  \quad , \quad c^2_s=
\frac{dP}{d\varepsilon} \right|_s \; .
\end{equation}
Which for constant speed of sound results in the cooling law
\begin{equation}\label{coollaw}
T(\tau) = T_0 \left(\frac{\tau_0}{\tau} \right)^{c^2_s}
\end{equation}
Obviously, the form of eq.(\ref{enercons}) is similar to
the energy conservation equation (\ref{conener}) in cosmology
with the expansion rate $ \dot{a}/a =1/(3\tau) $ when the proper time
$ \tau $ is identified with the
comoving time $ t $ in the cosmological setting. The similarity
becomes even more remarkable for the case of the QGP being modelled
as a radiation fluid (which is expected to be a good approximation
at high temperature) since in this case $ c^2_s=1/3 $ and the
connection with a radiation dominated cosmology with scale factor
$ a(t)\propto t^{\frac{1}{3}} $ is evident.

To find the general evolution equations for the QGP, an equation of state is needed.
 It is at this stage where the connection with lattice gauge
theory (LGT) is made. LGT obtains the thermodynamic
functions {\em in equilibrium} and these are input in the
hydrodynamic description as local functions of space-time under
the assumption of LTE. However, \emph{practically} hydrodynamic
simulations\cite{hydro2} use a bag (EoS). For a discussion of the
parameters in the bag (EoS) as well as a justification for using
this equation of state which features a strong first order phase
transition instead of the lattice (EoS) which features a crossover
see\cite{hydro2}.

\subsection{Predictions and observations pre-RHIC}

Several experimental signatures had been associated with the
formation of the QGP~\cite{muharris,signatures}, and the heavy ion
program  at SPS  focused on several of them. We summarize the
observables that have been proposed as telltales of a QGP and the
data gathered by the SPS from $Pb+Pb$  and $Pb+Au$ collisions,
which taken together provide a hint of
evidence~\cite{heinz,blaizot,phystoday,gyul} for  a QGP, although
many   of them could have alternative explanations.

\subsubsection{$J/\Psi$ suppression and strangeness enhancement}
\begin{figure}[h]
\begin{center}
 \epsfig{file=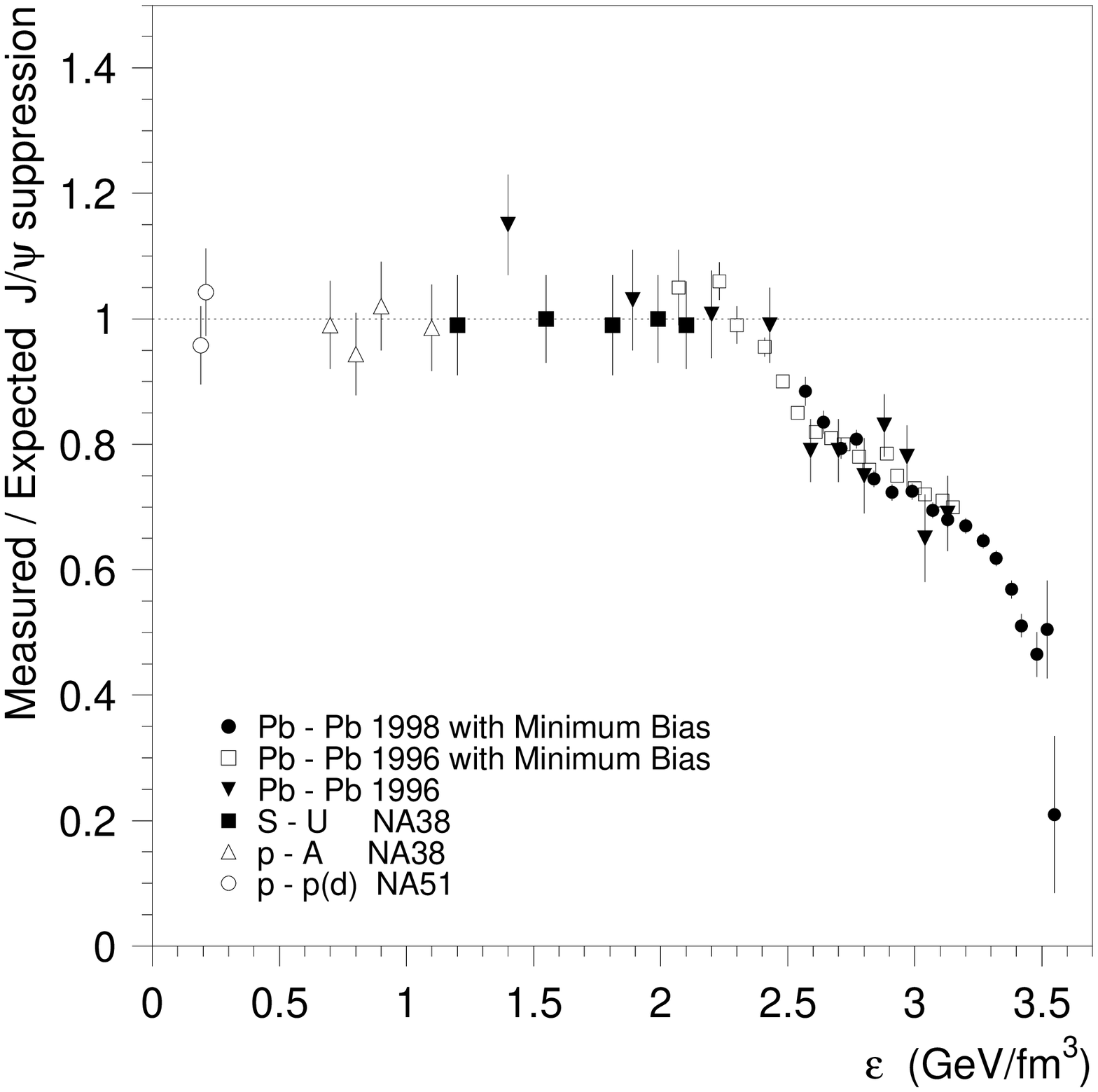,height=2
in,width=2 in,keepaspectratio=true} \epsfig{file=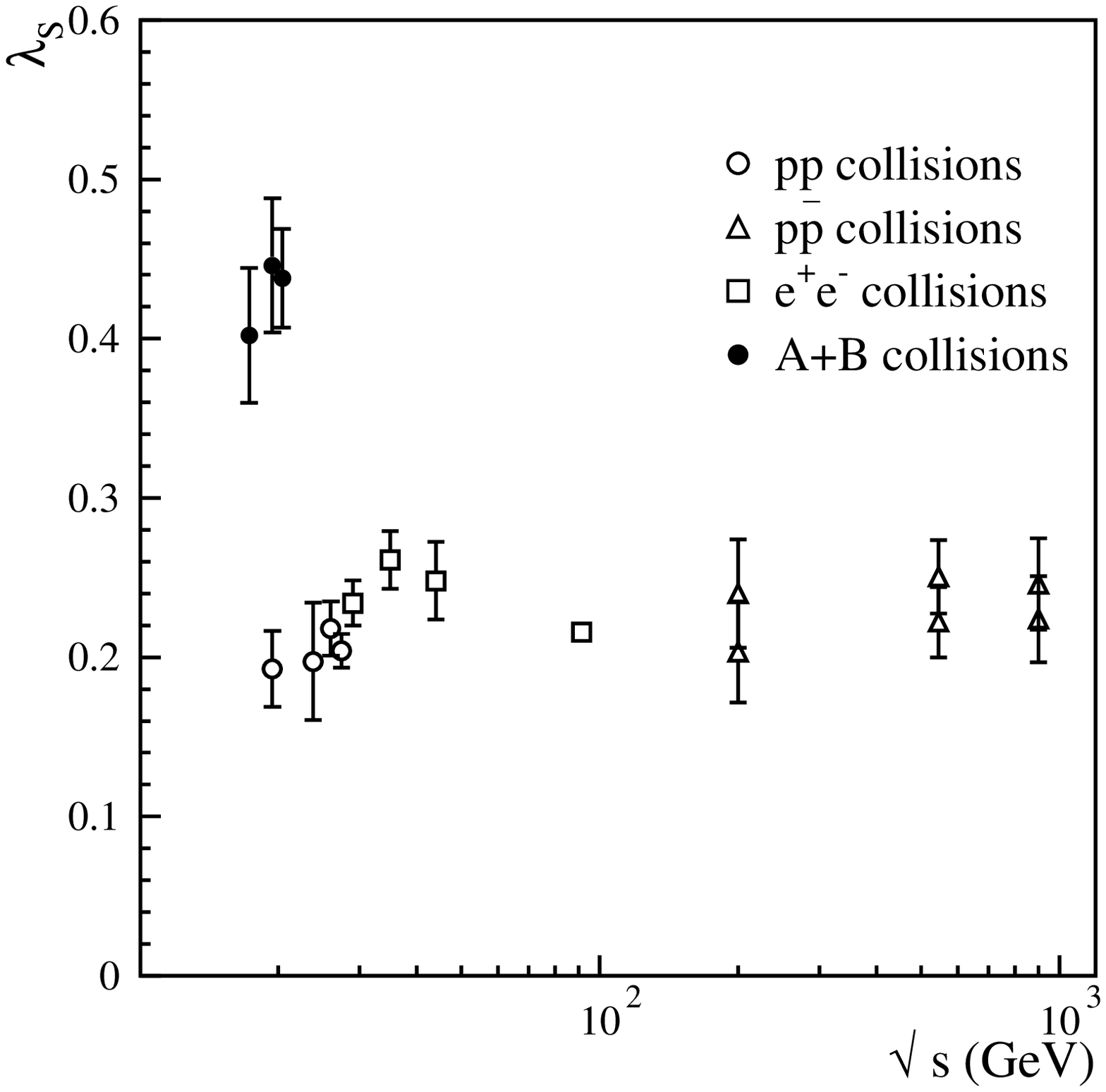,height=
2 in,width=2 in,keepaspectratio=true}
\caption{Left panel: anomalous $J/\psi$ suppression as a function
of the initial energy density. From~\cite{jpsi}. Right panel: The
ratio $\lambda_s$ as a function of $\sqrt{s}$, the energy of the
collision for nucleon-nucleon and nucleus-nucleus collisions.
From~\cite{soll}}   \label{fig:jpsisup}
\end{center}
\end{figure}

The $J/\Psi$ is a narrow $\bar{c}c$ bound state, which, \emph{if}
produced in the early stages of the collision can probe the QGP
because its lifetime is longer than that of the QGP and decays
into dilepton pairs which leave the plasma without scattering. The
original suggestion~\cite{satz} is that when the screening length
of the color force is smaller than the size of the $\bar{c}c$
bound state, this narrow resonance will melt. This argument
suggests that charmonium suppression could provide evidence for a
QGP. Alternatively a suppression could result from scattering with
hard gluons in the plasma and the dissociation of the bound
state~\cite{satz}. A normal suppression of charmonium is expected
on the grounds that once formed the $\bar{c}c$ bound state
interact with other nucleons inside the nucleus. This expected
suppression is studied in proton nucleon collisions and
extrapolated to nucleon-nucleon collisions. This is considered the
normal suppression in contrast to the abnormal suppression
expected from the presence of a plasma. The left panel of fig.
\ref{fig:jpsisup}  shows the data gathered by the NA50
collaboration at the SPS-CERN~\cite{jpsi}.

This   figure   reveals an abnormal suppression when the energy
density is $\varepsilon > 2.5 \, {\rm GeV}/{\rm fm}^3$. The energy
density in this figure has been computed with Bjorken's formula
eq.(\ref{enercen}). The NA50 collaboration at
CERN-SPS~\cite{jpsi2}   combined data for $J/\Psi$ suppression
from the NA38 and NA50 experiments. The analysis reveals that
while for the most peripheral (largest impact parameter)
collisions the suppression can be accounted for by nuclear
absorption, there is no saturation in the suppression in the  most
central $Pb+Pb$ collisions and that the observed suppression
pattern can be   understood in a deconfinement scenario. This
report concluded that {\em the $J/\Psi$ suppression pattern
observed in the NA50 data provides significant evidence for
deconfinement of quarks and gluons in $Pb+Pb$ collisions}.
Strangeness enhancement along with chemical equilibration are some
of the earliest proposals for clear signatures of the formation of
a QGP~\cite{strangeness}. The main idea is based on the estimate
that the strangeness equilibration time in a hot QGP is of the
same order as the expected lifetime of the QGP ($\sim 10 fm/c$)
produced in nucleus-nucleus collisions. Two important aspects of
this estimate make strangeness enhancement a prime candidate: if
strangeness attains chemical equilibrium in the QGP, this
equilibrium value is significantly higher than the strangeness
production in nucleon-nucleon collisions. Also strangeness
production through hadronic rescattering or final state
interactions was estimated to be negligibly
small~\cite{strangeness}. In the QGP, color deconfinement leads to
a large gluon density that leads to the creation of $s\bar{s}$
pairs, furthermore chiral symmetry makes the strange quark lighter
thus lowering the production threshold. This situation is in
contrast to the case of hadronic rescattering or final state
interactions where the production of pairs of strange quarks has
large thresholds and small cross sections~\cite{heinz}. The usual
measure of strangeness enhancement is through the ratio $
\lambda_s =  {2 \langle \bar{s}s \rangle}/{\langle
\bar{u}u+\bar{d}d \rangle}$. The right panel in fig.
\ref{fig:jpsisup}  displays $\lambda_s$ as a function of
$\sqrt{s}$  for nucleon-nucleon as well as nucleus-nucleus
collisions ($S+S,S+Ag,Pb+Pb$) at SPS.
The data displayed in fig. \ref{fig:jpsisup} is taken as
evidence that nucleus-nucleus collisions at SPS are creating a hot
state of matter.


\subsubsection{Electromagnetic probes: dileptons and direct
photons}

\begin{figure}[h]
\begin{center}
\epsfig{file=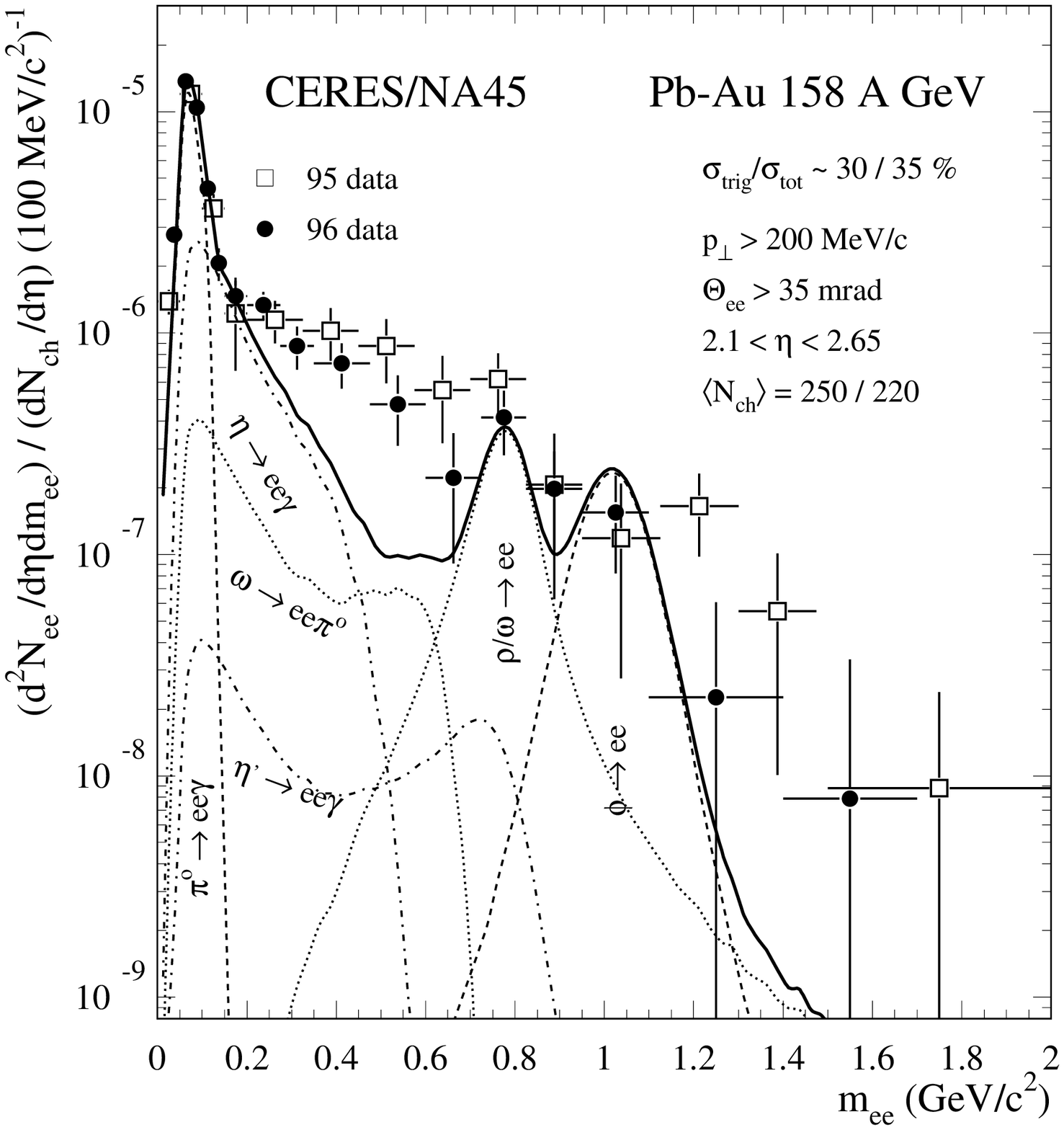,height=2.1in,width=2.1in}
\epsfig{file=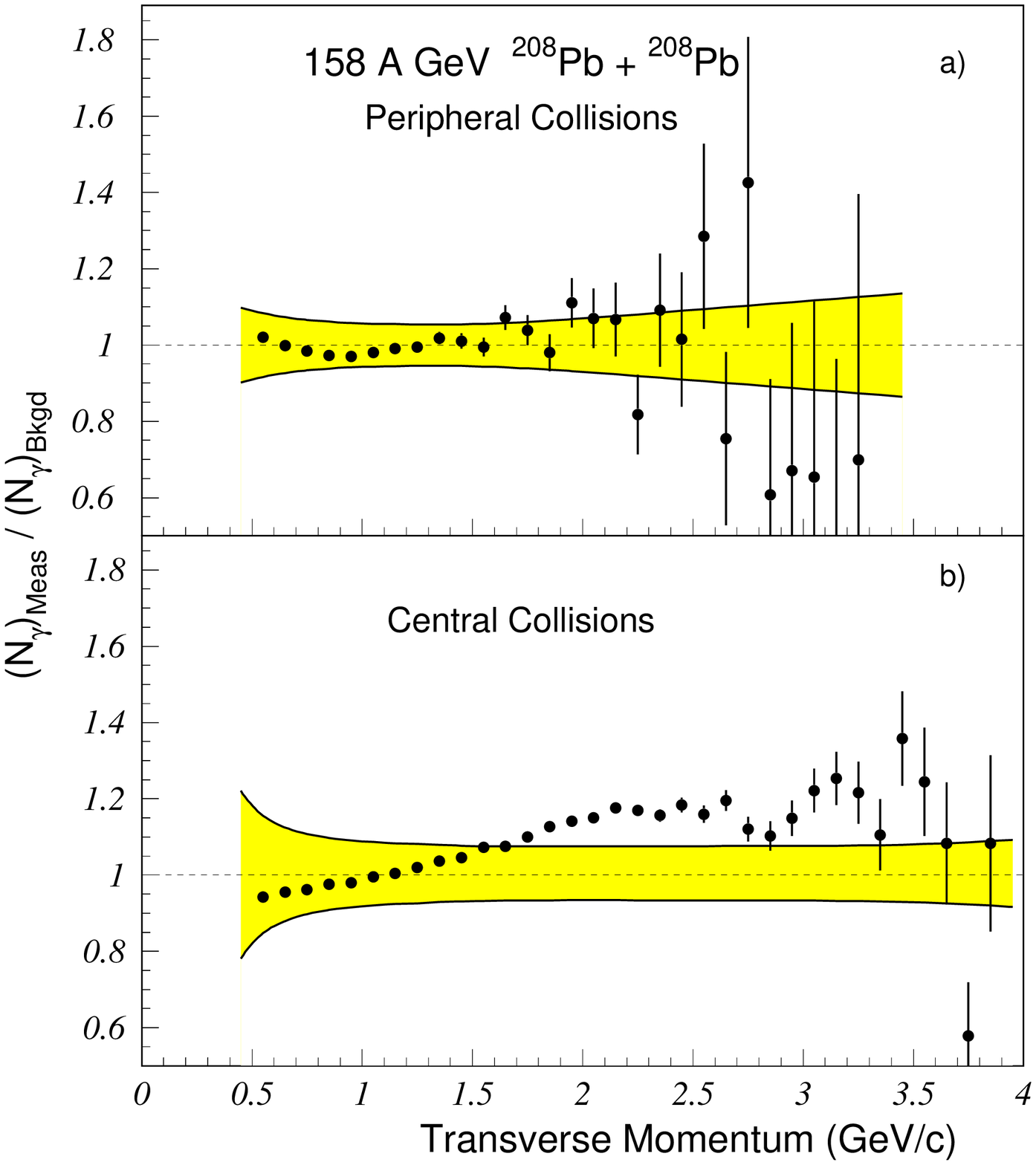,height=2in,width=2in}
\caption{Left figure: invariant mass spectrum of $e^+e^-$ pairs from
$Pb+Au$ collisions at 158 AGev at the SPS-CERN (CERES/NA45).
From~\cite{ceres}. Right figure: ratio of total measured yield of
direct photons to hadronic background vs. transverse momentum from
$Pb+Pb$ at 158 AGev for peripheral and central collisions.
From~\cite{WA98}. } \label{fig:ceres}
\end{center}
\end{figure}

Electromagnetic probes: $e^+e^-$  or $\mu^+\mu^-$ dilepton pairs
and direct (prompt) photons are prime probes of the hot
plasma~\cite{wam,electro}, since once produced they leave the
plasma without further interactions because their mean free path
is {\em much} larger than the typical size of the plasma. The
$\rho$ vector meson is important because it decays into dileptons
and its lifetime is $\sim 1 fm/c$. Therefore once it is produced
in the hadron gas it decays {\em within} the hot hadronic plasma
and the produced dileptons carry direct    information from the
plasma. Thus while dileptons produced from the decay of the $\rho$
meson do not yield evidence of the earlier stages in the QGP, they
do
  offer information on the hadronic stage. The left
panel in fig. \ref{fig:ceres} presents the data gathered by the
CERES-NA45 collaboration for the invariant mass spectrum for
electron-positron pairs from 158 AGeV $Pb+Au$ collisions at the
SPS-CERN. The solid line represents the expected spectrum from the
decays of hadrons produced in proton-nucleon and proton-proton
collisions extrapolated to $Pb+Au$ collisions and is the sum of the
contributions shown in the graph.  There are two remarkable features
in this graph: a clear enhancement of dileptons in the region
$2~{\rm MeV} \lesssim M_{e^+e^-} \lesssim 700~{\rm MeV}$ and  that
instead of the $\rho$ meson peak at $m_{\rho}=770 ~{\rm MeV}$ there
is a broad distribution. The excess of dileptons in the small
invariant mass region cannot be explained by charged pion
annihilation~\cite{heinz}. It is remarkable   that the dilepton
enhancement is {\em below} the putative $\rho$ peak and that there
is no hint of the $\rho$ at $770~{\rm MeV}$!. The current
understanding of these features is that the medium effects result in
a shift in the $\rho$ meson mass as well as a change in its
width\cite{wam,heinz,blaizot}. Thus while this interpretation does
not directly yield information on the QGP, it does support the
picture of a hot gas of hadrons, mainly pions which is the main
interaction channel of the $\rho$ vector meson. Direct photons are
conceptually a clean direct probe of the early stages of the QGP.
Photons are produced in the QGP by several processes:
gluon-to-photon Compton scattering off (anti)quark
$q(\bar{q})g\rightarrow q(\bar{q})\gamma$ and quark-antiquark
annihilation to photon and gluon $q\bar{q}\rightarrow g\gamma $ and
to the {\em same order} (see Kapusta et. al. and Aurenche et. al.
in~\cite{electro}) (anti)quark bremsstrahlung $qq(g)\rightarrow
qq(g)\gamma$ and quark-antiquark annihilation with scattering
$q\bar{q}q(g)\rightarrow q(g)\gamma$. Detailed calculations
including screening corrections~\cite{electro} reveal that direct
photons from the QGP could provide a signal that could be
discriminated against the signal from the hadronic background. This
work indicates the theoretical feasibility of direct photons as
direct probes of the early stages of the QGP. The WA98 collaboration
at SPS-CERN reported their analysis for the {\em first} observation
of direct photons from $Pb+Pb$ collisions with $\sqrt{s}= 158 \,
\textrm{AGeV}$~\cite{WA98}. Their data is summarized in the right
panel of  fig. \ref{fig:ceres}. The transverse momentum distribution
of direct photons is determined on a statistical basis and compared
to the background photon yield predicted from a calculation of the
radiative decays of hadrons. The most interesting result is that a
significant excess of direct photons beyond that expected from
proton-induced reaction at the same $\sqrt{s}\,$ is observed in the
range of transverse momentum greater than about $1.5\;{\rm GeV}/c$
in central collisions. A comparison of  the data to the theoretical
predictions for direct photons from an equilibrated QGP was
performed in ref~\cite{hatsuda}. The conclusions are that while it
is not clear if SPS   reached the energy density to form the QGP,
the data indicates a hot and dense phase that could be its
precursor. It has recently  been shown~\cite{wangnon} that
\emph{non-equilibrium} effects in an expanding QGP formed in RHIC
collisions lead to an enhancement of direct photons in the region of
transverse momentum $p_T \gtrsim 2~ {\rm GeV}$.


\subsubsection{Quantum spinodal decomposition and pion domains} For
massless up and down quarks, QCD has an $ SU(2)_L\otimes SU(2)_R $
symmetry which is spontaneously broken down to $ SU(2)_{L+R} $,
this is the chiral phase transition. It has been argued that for
massless quarks the equilibrium aspects of this transition is
described by the universality class of the $ O(4) $ Heisenberg
ferromagnet\cite{raja}. The Landau-Ginzburg effective field theory
for this universality class is given by the effective potential
\be V(\vec{\Phi}) = a \;  (T-T_c) \; \vec{\Phi}^2 + g \;
(\vec{\Phi}^2)^2~~;~~\vec{\Phi}= (\sigma,
\vec{\pi})\label{dccveff}  \; , \ee where $ \sigma\sim \bar{\psi}
\gamma^5 \psi  $ is the pseudoscalar chiral order parameter field
that acquires an expectation value below the critical temperature,
and $ \vec{\pi}\sim \bar{\psi} \gamma^5 \vec{\tau}\psi $ ($
\tau^a=\textrm{isospin~matrices} $) is the triplet of pions, which
are the Goldstone bosons associated with the symmetry breaking. In
an URHIC, chiral symmetry is expected to be restored if a QGP is
formed with $ T>T_c $. Upon expansion and rapid cooling chiral
symmetry is again broken.  \emph{If} the chiral phase transition
occurs out of equilibrium, the ensuing quench can bring the
hadronic phase into the spinodal region of the effective potential
(\ref{dccveff}). As a result, quantum spinodal instabilities would
lead to the formation of correlated domains inside which the order
parameter is \emph{disoriented} resulting in a \emph{disoriented
chiral condensate}(DCC)\cite{dcc,raja}. Spinodal instabilities
would result in the formation and growth of \emph{pion
domains}\cite{boydcc} with distinct observational signatures in
the pion distribution. A systematic study by the WA98
collaboration at SPS-CERN\cite{dccexp} measured charged particle
multiplicities in $ 158  \,\textrm{AGeV}\,Pb+Pb $ collisions and
ruled out the formation of (DCC) domains at $ 90\%\,CL $. The
Minimax collaboration studied the possibility of (DCC)-like events
at  Fermilab's Tevatron in $p\bar{p}$ collisions\cite{minimax} and
found no statistically significant signals of large isospin
fluctuations.

\subsection{News from RHIC:}
The four experiments at RHIC have presented their analysis of the
first three years of operation at the top energy $\sqrt{s} = 200
A\,\textrm{GeV}$ for $Au+Au$ collisions\cite{phenix}. The combined
body of results not only confirms most of the findings at SPS but
also reveal novel and unexpected phenomena\cite{gyul,phystoday}.
To begin with, the results on particle multiplicity when combined
with Bjorken's estimate for the value of the energy density
eq.(\ref{enercen}) yield $\varepsilon \sim 5.5
\,\textrm{GeV}/\textrm{fm}^3$ for  $\tau_0 \sim 1\,\textrm{fm}/c$.
This energy density is almost one order of magnitude larger than
the critical value obtained from (LGT), $\varepsilon_{LGT}\sim 0.7
\,\textrm{GeV}/\textrm{fm}^3$ (see section (\ref{sec:lgt}))
therefore these experiments probe QCD in the deconfined regime. At
least two important results stand out:

\subsubsection{Elliptic flow, ideal hydrodynamics and early
thermalization}
\begin{figure}[h]
\begin{center}
\epsfig{file=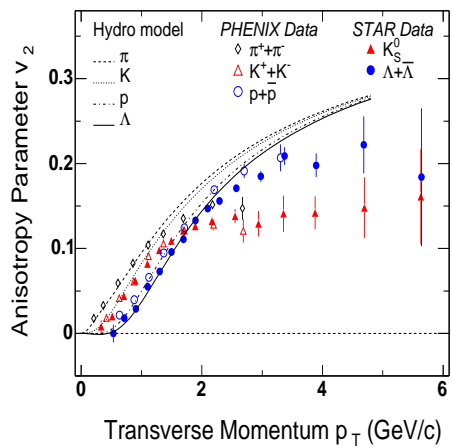,height=2.5in,width=2.5in} \caption{Combined
STAR and PHENIX data on $v_2$. From\cite{phystoday}. }
\label{fig:v2}
\end{center}
\end{figure}

If a QGP is formed early in the collision the
pressure drives a collective expansion or \emph{flow}. An
important measure of this collective flow is \emph{elliptic
flow}\cite{olli} in collisions with non-vanishing impact parameter
(non-central). If the reaction plane is taken to be the $x-z$
plane, with $z$ the beam axis, the triple differential momentum
distribution for hadron $h$ is expanded in a Fourier series in
terms of the angle $\phi$ in the $x-y$ plane as \be \frac{dN_h
}{dydp_T^2d\phi} =\frac{dN_h}{2\pi\,dydp_T^2} \left[1 + 2
v_1(y,p_T ) \cos\phi   + 2  v_2(y,p_T ) \cos 2 \phi + \cdots
\right] \;\;. \label{floweq} \ee The first coefficient $v_1$
vanishes at midrapidity\cite{hydro2}, the second $ v_2 $, the
\emph{elliptic flow} parameter is a measure of the hydrodynamic
flow in peripheral collisions driven by pressure
gradients\cite{hydro2}. A remarkable result from RHIC experiments
is that ideal hydrodynamic calculations of $v_2$ are in excellent
agreement with the combined experimental results from
STAR\cite{starv2} and PHENIX\cite{phenixv2} for a many hadron
species, up to $p_T \sim 2\,\textrm{GeV}$ as shown in fig.
\ref{fig:v2}.
A hydrodynamic calculation requires two inputs: an initialization
(proper) time $ \tau_0 $ at which LTE is assumed to hold and an
equation of state. The initial stage that leads to LTE requires
knowledge beyond the realm of applicability of hydrodynamics.
Hydrodynamic calculations just \emph{input} initial data at a
given (proper) time hypersurface. In most hydrodynamic approaches
a bag-like (EoS) is used\cite{hydro2} with typical values for the
bag constant $ B^{1/4}\sim 200-250\,\textrm{MeV} $ which yield a
critical temperature $ T_c \sim 160 \textrm{MeV} $. Such equation of
state yields a strong first order transition, with a latent heat
$ l \sim 1\,\textrm{GeV}/\textrm{fm}^3 $. While such strong first
order phase transition is  in contradiction with the (EoS) from
(LGT), it is argued in ref.\cite{hydro2} that the latent heat is
consistent with the (gradual) change in entropy density across the
transition in the lattice data.  Agreement between the
hydrodynamical calculations and the experimental values of $ v_2 $
requires a \emph{very short} initialization time scale $\tau_0
\sim 0.6\,\textrm{fm}/c$\cite{hydro2} and systematic studies of
different equations of state in ref.\cite{teaney} points out that
the data is reproduced if the (EoS)   includes a soft point at
which the speed of sound becomes anomalously small as featured by
a quark-hadron mixed phase. As the plasma enters the mixed phase
pressure gradients become very small and anisotropies in the
transverse flow stall\cite{hydro2}. Furthermore an important
aspect of the hydrodynamic results is that \emph{ideal}
hydrodynamics yields a remarkable agreement with the experimental
data on collective flow. Taken together these results suggest the
following:
\begin{itemize}\item{Early thermalization: the data is reproduced
if the initialization proper time at which LTE is assumed to hold
is $\tau_0 \sim 0.6 \textrm{fm}/c$. } \item{Ideal hydrodynamics:
the data is reproduced by \emph{ideal} hydrodynamics, namely a
strongly coupled fluid with a short mean free path and small
viscosity coefficients.} \item{Phase transition: the study of
different (EoS) in ref.\cite{teaney} conclude that hydrodynamics
with an (EoS) that feature a phase transition is consistent with
the elliptic flow data, while an (EoS) without a phase transition
is not. }
\end{itemize} The emerging consensus is that the new state of
matter formed in URHIC is a strongly correlated state or strongly
coupled quark gluon plasma, a strikingly different situation from
the original conjecture of a QGP of weakly interacting quark and
gluon \emph{quasiparticles}. Ideal hydrodynamics combined with
such early thermalization scale leading to a rapid equilibration
can only result from  strong correlations. However, and
alternative interpretation has been  recently argued:  that the
RHIC data, in particular the elliptic flow parameter $v_2$ can be
interpreted from  an \emph{incomplete equilibration}\cite{bhale}.
Clearly the understanding of the new findings at RHIC is still
evolving.

\subsubsection{A new initial state?} A new picture of the initial
state prior to the onset of LTE  and hydrodynamical evolution is
emerging\cite{phystoday}. Current theoretical
understanding\cite{cgc} suggests that the ground  state of   high
energy nuclei prior to the collision is akin to a non-perturbative
Bose-Einstein condensate of gluons, a \emph{color glass condensate}.
Evidence for   a rapid growth in the gluon distribution function as
a function of $x$,   the fraction of the longitudinal momentum
carried by the gluon was first revealed by  deep inelastic
scattering experiments at HERA\cite{hera}.  Gluon self-interaction
results in a characteristic momentum scale below which the gluon
distribution function saturates, at RHIC this scale is $Q_s \sim 1
\textrm{GeV}$. The density of gluons below this scale is
$dN_g/d\ln(1/x)\sim 1/\alpha_s$ which is very large because the
strong interaction coupling $\alpha_s$ becomes small at this scale.
The evidence for this novel initial state at RHIC emerges from the
suppression or quenching of jets and has been observed in comparing
$d+Au$ and $Au+Au$ collisions\cite{phystoday,cgc}. This high gluon
density initial state is conjectured to be the \emph{precursor} of a
QGP\cite{phystoday} and it is possible\cite{roma} that this highly
correlated and non-perturbative initial state is responsible for the
early thermalization required by the hydrodynamic approach. These
novel aspects of URHIC will continue to be the focus of the
experimental and theoretical programs.

\subsection{Little Bang vs. Big Bang}\label{PUN}
One of the stated goals of the URHIC program was to open a window
to the early Universe by creating the conditions that prevailed
when it was about $10\,\mu\textrm{secs}$ old. While there are
similarities between URHIC and early Universe cosmology, there
also are differences which are even more pronounced by the recent
results from RHIC.

\textbf{Similarities}: i) The early Universe is almost baryon
free, and the entropy is dominated by radiation, in central
collisions at RHIC (and the forthcoming LHC) the central rapidity
region is nearly baryon free. ii) Ideal hydrodynamics and
longitudinal expansion are very similar to the Hubble expansion of
the cosmic fluid.

\textbf{Differences}: i) In the early Universe at the time of the
QCD phase transition, the expansion time scale $1/H \sim 10^{-5}
\,\textrm{secs}$  was much \emph{longer} than the time scale of
strong interactions $\sim 10^{-23}\,\textrm{secs}$, hence the
crossover behavior revealed by the lattice data implies that the QCD
phase transition(s) (deconfinement-confinement and chiral symmetry
breaking) occurred in LTE. In URHIC, the expansion time scale is of
order $ 10^{-22}$secs,  ten times the time
scale of strong interactions,  and non-equilibrium effects are more
likely. ii) Since in the early Universe cooling occurs via Hubble
expansion from an initial temperature or energy density \emph{many}
orders of magnitude larger than the energy density required for the
QCD phase transition $\sim 1\,\textrm{GeV}/\textrm{fm}^3$,
undoubtedly a \emph{weakly interacting} gas of quarks and gluons,
namely a weakly interacting QGP was present as a consequence of
asymptotic freedom. Contrary to this expectation, the results from
RHIC seem to suggest a \emph{strongly} interacting \emph{liquid}, a
very different state of matter. Furthermore, the initial
condition in the early Universe, prior to the QCD scale is that of a
radiation dominated fluid in LTE slowly cooling via adiabatic
expansion. RHIC (and HERA) reveal a novel initial state of strongly
correlated gluons with a non-perturbative `color glass
condensate'. Whether the strongly coupled quark gluon liquid is
a consequence of such strongly correlated initial state and whether
we can learn aspects of QCD during the first few
$\mu\,\textrm{secs}$ after the Big Bang will continue to be the
focus of the experimental programs at RHIC and LHC.

\subsection{ The nuclear liquid gas phase transition: spinodal
decomposition and nuclear fragmentation.}\label{nuclearpt}

The interaction between nucleons in nuclei is similar to the Van
der Waals potential which acts between molecules: a short distance
repulsive core and a long distance attractive tail. For this
reason is was suggested that the nuclear interaction should lead
to a liquid-gas phase transition in nuclei\cite{siemens}. This
original work suggested that if the equation of state of nuclear
matter is of the Van der Walls type, collision experiments may
bring excited nuclei into the spinodal region of the phase diagram
in which spinodal instabilities towards phase separation would
lead to the spectacular fragmentation of nuclei. This observation
launched considerable theoretical efforts to yield a better
understanding of the nuclear equation of state\cite{chomaz} and
several experimental facilities to study this phenomenon. The most
commonly used nuclear effective interaction to study the
thermodynamics of nuclear matter is of the Skyrme
type\cite{skyrme}. More modern approaches to the nuclear equation
of state are in terms of effective nuclear field
theories\cite{serot,serotreview}  of hadrons  and mesons where the
hadronic interaction results from  the exchange of $\pi,
\sigma,\rho$ and $\omega$ mesons. At the mean field
level\cite{serotreview} the hadronic interaction does feature a
short range repulsion and a long range attractive tail that yield
an equation of state with a flat region of coexistence akin to the
Maxwell construction in a Van der Walls equation of state. These
theoretical studies suggest that the critical temperature for
nuclear matter is $T_c \sim 18-20\,\textrm{MeV}$.
  From the experimental point of view a number of
accelerator facilities: GANIL, GSI, LNL, MSU, Dubna have been
developed to study the physics of \emph{nuclear
multifragmentation} as a consequence of the \emph{nuclear liquid
gas phase transition}. These experimental facilities study low
energy heavy ion collisions involving several types of nuclei. For
example     heavy ion reactions of $Xe+Sn$ at
$32\,\textrm{MeV}/\textrm{nucleon}$\cite{borderiex} are studied at
GANIL,  and $25-35\,\textrm{MeV}/\textrm{nucleon}$ $Au+Cu$
collisions at the K1200-NSCL Cyclotron at Michigan State
University (MSU)\cite{agostino}. The main experimental signature
of the nuclear liquid-gas phase transition is expected to be the
decay of highly excited nuclei through the process of
\emph{multifragmentation} via spinodal
instabilities\cite{moretto}. Experimentally the process of
multifragmentation leads to abnormally large fluctuations in the
distribution of nuclei\cite{chomazrev}, these fluctuations
becoming more pronounced as the critical temperature is
approached. The distribution of fragments should favor a
particular size, associated with the most spinodally unstable
modes\cite{chomaz,chomazrev}. In finite nuclei the transition is
smoothed by the finite size and the signatures are not expected to
be as sharp as predicted by bulk thermodynamics in an infinite
medium. Theoretical modelling of the dynamics is   based on
transport or quantum molecular dynamics simulations\cite{chomaz}.
There is now a substantial body of experimental
data\cite{agostino,borderiex} that seems to confirm the presence
of such large fluctuations with an impressive identification
capability thanks to high performance $4\pi$ detectors. A summary
of the experimental results from the INDRA detector at GANIL has
been recently presented\cite{rivet}. Some of the remarkable
experimental evidence reported by the INDRA collaboration
are\cite{rivet}: \begin{itemize} \item{Statistically significant
evidence for spinodal fragmentation in $Xe+Sn$ collisions at
energies  $32-45\,\textrm{MeV}/\textrm{nucleon} $. }
\item{Bimodality: in coexistence the order parameter features
bimodality from the two coexisting phases. Such bimodality was
reported in the distribution of fragments in $Au+Au$ collisions at
energies in the range $60-100\,\textrm{MeV}/\textrm{nucleon}$.}
\item{Negative specific heat: the \emph{microcanonical heat
capacity} is a thermodynamic quantity that features a negative
branch in the coexistence region, signaling again an instability.
The results of the experiments are in agreement with such a
negative branch of the microcanonical heat capacity for finite
nuclei in a region of coexistence. }

\end{itemize}

These experimental findings    seem to validate the picture of the
nuclear liquid-gas phase transition and pave the way towards a
detailed understanding of the nuclear equation of state in a regime
of temperatures and density of relevance for understanding nuclear
matter in supernovae explosions and the emerging neutron stars. As
more high precision experiments are carried out at these facilities
a more detailed picture of the equation of state of nuclear matter
will emerge, as well as a first, spectacular confirmation of the
\emph{dynamics} of a phase transition in nuclear matter.

\section{Summary,  conclusions and outlook:}

This article presents an excursion of the early and the present
Universe exploring the standard models of cosmology and particle
physics. The  focus is on the equilibrium and non-equilibrium
physical aspects  and potential observational consequences of the
phase transitions predicted by particle physics. General
relativity, statistical mechanics and the standard model of
particle physics lead to a detailed understanding of the evolution
of the Universe and to time-marks at which phase transitions may
have occurred.  A detailed introduction to static and dynamical
aspects of phase transitions with relevance to cosmology and the
experimental program that studies the phase diagram of quark and
nuclear matter is provided.

An important aspect of the connection between particle physics and
cosmology is the theoretical prediction of an early stage of
inflation during which the visible Universe expands almost
exponentially. This inflationary stage solves outstanding problems
of the standard Big Bang model and has been spectacularly
confirmed by CMB measurements. Inflation is now an integral part
of the standard model of cosmology and inflationary dynamics has a
prominent role in the evolution of the Universe. We have discussed
the potential observable consequences of the electroweak and the
QCD phase transitions in the early Universe and highlighted the
fact that while anisotropies in the CMB provide \emph{direct}
information of the inflationary phase which is the
\emph{earliest}, observable consequences of the electroweak and
QCD phase transition, which took place much \emph{later} are
indirect. The URHIC program initiated with the AGS at BNL,
continued with SPS and CERN and currently with RHIC at BNL  has
provided a wealth of evidence in support of a novel state of
matter, the QGP. We summarize the experimental evidence from SPS
and RHIC and discuss some of the most recent discoveries at RHIC.
While the results from all four experiments at RHIC provide
compelling evidence for a new state of matter, the evidence
suggests a strongly correlated quark-gluon \emph{liquid} instead
of a weakly interacting gas of quarks and gluons which presumably
existed in the early Universe prior to the QCD phase transition.
We have also discussed theoretical and experimental studies of the
liquid-gas phase transition in cold nuclear matter. Recent results
from several experimental programs offer compelling evidence for a
liquid-gas transition and spinodal instabilities in nuclear matter
leading to \emph{multifragmentation} of excited nuclei. These
studies yield a detailed knowledge of the equation of state of
nuclear matter in a regime of temperature and density relevant of
supernovae and neutron stars.

\begin{acknowledgments} D.B. thanks the US NSF for  support under
grant PHY-0242134.
\end{acknowledgments}



\end{document}